\documentclass[prd, twocolumn, showpacs,aps,nofootinbib,floatfix,amsmath,amssymb]{revtex4}
\usepackage{graphicx}
\usepackage{soul}
\usepackage{color}
\usepackage[mathscr]{euscript}
\usepackage{slashed}
\usepackage[usenames,dvipsnames]{xcolor}
\usepackage{amsmath}
\begin{document}

\makeatletter
\newbox\slashbox \setbox\slashbox=\hbox{$/$}
\newbox\Slashbox \setbox\Slashbox=\hbox{\large$/$}
\def\pFMslash#1{\setbox\@tempboxa=\hbox{$#1$}
  \@tempdima=0.5\wd\slashbox \advance\@tempdima 0.5\wd\@tempboxa
  \copy\slashbox \kern-\@tempdima \box\@tempboxa}
\def\pFMSlash#1{\setbox\@tempboxa=\hbox{$#1$}
  \@tempdima=0.5\wd\Slashbox \advance\@tempdima 0.5\wd\@tempboxa
  \copy\Slashbox \kern-\@tempdima \box\@tempboxa}
\def\FMslash{\protect\pFMslash}
\def\FMSlash{\protect\pFMSlash}
\def\miss#1{\ifmmode{/\mkern-11mu #1}\else{${/\mkern-11mu #1}$}\fi}

\newcommand{\psum}[1]{{\sum_{ #1}\!\!\!}'\,}
\makeatother

\title{Bounds on Lorentz-violating Yukawa couplings via lepton electromagnetic moments}

\author{J. Alfonso Ahuatzi-Avenda\~no$^a$, Javier Monta\~no$^b$, H\'ector Novales-S\'anchez$^a$, M\'onica Salinas$^a$, and J. Jes\'us Toscano$^a$}
\affiliation{$^a$Facultad de Ciencias F\'isico Matem\'aticas, Benem\'erita Universidad Aut\'onoma de Puebla, Apartado Postal 1152 Puebla, Puebla, M\'exico.\\$^b$CONACYT-Facultad de Ciencias F\'isico Matem\'aticas, Universidad Michoacana de San Nicol\'as de Hidalgo, Av. Francisco J. M\'ugica s/n, C.~P. 58060, Morelia, Michoac\'an, M\'exico.}

\begin{abstract}
The effective-Lagrangian description of Lorentz-invariance violation provided by the so-called Standard-Model Extension covers all the sectors of the Standard Model, allowing for model-independent studies of high-energy phenomena that might leave traces at relatively-low energies. In this context, the quantification of the large set of parameters characterizing Lorentz-violating effects is well motivated. In the present work, effects from the Lorentz-nonconserving Yukawa sector on the electromagnetic moments of charged leptons are calculated, estimated, and discussed. Following a perturbative approach, explicit expressions of leading contributions are derived and upper bounds on Lorentz violation are estimated from current data on electromagnetic moments. Scenarios regarding the coefficients of Lorentz violation are considered. In a scenario of two-point insertions preserving lepton flavor, the bound on the electron electric dipole moment yields limits as stringent as $10^{-28}$, whereas muon and tau-lepton electromagnetic moments determine bounds as restrictive as $10^{-14}$ and $10^{-6}$, respectively. Another scenario, defined by the assumption that Lorentz-violating Yukawa couplings are Hermitian, leads to less stringent bounds, provided by the muon anomalous magnetic moment, which turn out to be as restrictive as $10^{-14}$.
\end{abstract}


\maketitle


\section{Introduction}
\label{Int}
It has been half a century since its formulation~\cite{Glashow,Weinberg,Salam}, and yet the Standard Model (SM) remains our best theoretical description of fundamental physics~\cite{PDG}. Even so, the SM is nowadays considered, by the scientific community, to be a low-energy manifestation of an underlying theory operating at some very-high energy scale, perhaps of the order of the Planck scale. In general, the two main ingredients behind the definition of field theories are their dynamic variables and symmetries~\cite{Wudka}. Regarding the latter aspect, invariance under spacetime and gauge transformations have traditionally received much attention in model building. While Lorentz symmetry is a conventional assumption in beyond-SM contexts, Planck-scale physical formulations, such as string theory and noncommutative field theory, are able to spontaneously brake it~\cite{KoSa,KoPo,KoPo1,CHKLO}, thus yielding Lorentz-nonconserving physical phenomena which, at current experimental sensitivity, may manifest as measurable tiny effects. Since no compelling evidence of Lorentz violation has been ever observed~\cite{KoRu}, thus leaving us blind as to which place is the best to introduce this kind of new physics, the effective-Lagrangian approach~\cite{Wudka,DGMP}, distinguished for being model independent, seems to be suitable. \\

A couple decades ago, an effective-Lagrangian description of Lorentz-symmetry nonconservation, known as the Lorentz- and CPT-violating SM Extension (SME), was devised~\cite{CoKo1,CoKo2}. The SME has since become a useful tool to comprehensively study this sort of new physics\footnote{The SME is nicely and succinctly reviewed in Ref.~\cite{Tasson}.}, which induces unconventional phenomena such as vacuum birefringence~\cite{KoMe1,KoMe2}, vacuum \v{C}erenkov radiation~\cite{LePo1,LePo2,Altschul1,Altschul2,Altschul3,Altschul4}, oscillations of massless neutrinos~\cite{KoMe3,KoMe4,DiKo,DKSC}, exotic electromagnetic properties of SM particles~\cite{MNTT1,MNTT2}, and violations of standard theorems~\cite{KLP,CNTT}. The dynamic variables of the SME and its gauge-symmetry group are the same as those of the sole SM, the key element being a large set of coefficients, characterized by fully-contracted spacetime indices within Lagrangian terms and which transform as tensors under observer Lorentz transformations~\cite{CoKo1,CoKo2}, thus implying that Lorentz-violating physics is observer independent. Nonetheless, these tensor coefficients, which define preferred directions in spacetime, are invariant under particle Lorentz transformations~\cite{CoKo1,CoKo2}, so they do not preserve Lorentz symmetry. The estimation of the quite vast set of SME coefficients has become the main objective of several phenomenological investigations on Lorentz violation. The present paper is one of such works. A comprehensive catalog of SME-coefficients constraints is provided in Ref.~\cite{KoRu}, which, moreover, is updated every year. Lorentz-violating Lagrangian terms constituting the SME are classified into two types, according to whether they are power-counting renormalizable or not~\cite{KoMe2,MNTT1,MNTT2,CNTT,KoMe5,KoMe6,MNPB,CFLdSS,FLMS}. The full set of renormalizable SME terms define the so-called minimal SME (mSME). Within the framework of the mSME, the present paper is a phenomenological investigation of effects of Lorentz violation on the anomalous magnetic moments (AMMs), $a_A$, and electric dipole moments (EDMs), $d_A$, of charged leptons, $l_A$, with $A=e,\mu,\tau$ labeling lepton flavors. The emergence of contributions to these electromagnetic moments as byproducts of Lorentz-invariance nonconservation, has been addressed by the authors of Refs.~\cite{MNTT1,MNTT2,CheKu,CSV,ACF1,ACF2,ACF3,ABSF}. Under the assumption of Lorentz invariance, contributions to AMMs and EDMs are identified from the well-known parametrization of the electromagnetic vertex $A_\mu l_Al_A$, given by $\overline{u_A}(p')\,\Gamma_\mu u_A(p)$, with $u_A$ the momentum-space Dirac spinor for a charged lepton $l_A$ with mass $m_A$, and $\Gamma_\mu$ given by~\cite{HIRSS,NPR,BGS} 
\begin{equation}
\Gamma_\mu=ie\Big[ \gamma^\mu(f^{\rm V}_A-f^{\rm A}_A\gamma_5)-\sigma_{\mu\nu}q^\nu\Big( i\frac{f^m_A}{2m_A}-\frac{f^d_A}{e}\gamma_5 \Big) \Big],
\label{emvpar}
\end{equation}
for on-shell external fermions and off-shell photon field, in which case all form factors, particularly the magnetic form factor $f^a_A(q^2)$ and the electric form factor $f^d_A(q^2)$, are functions of squared transfered momentum $q^2$ only. The on-shell-photon case, in which $q^2=0$, defines the AMM and the EDM by $a_A\equiv f^a_A(q^2{\textstyle =}\,0)$ and $d_A\equiv f^d_A(q^2{\textstyle =}\,0)$, respectively. While in the presence of Lorentz violation the structure of the corresponding electromagnetic-interaction parametrization is expected to be far richer~\cite{MNTT1,MNTT2}, AMMs and EDMs are still identified from the aforementioned Lorentz-preserving parametrization. Therefore, AMMs and EDMs from Lorentz violation necessarily originate in second-order SME-coefficients contributions, since in such a case form-factor contributions with fully-contracted spacetime indices may emerge. \\

Lorentz-violating effects addressed in the present work emerge from the Yukawa sector of leptons in the mSME, where both flavor and spacetime indices characterize Yukawa-like Lorentz-violating constants. Lorentz violation from these interactions is introduced, in the Feynman-diagrams approach, through three-point vertices and from two-point insertions as well, with both elements including lepton-flavor change. We emphasize, though, that transition electromagnetic moments~\cite{NPR,BGS} are not within the scope of the present work, so external fermion lines are always taken to preserve lepton flavor. Leading contributions to the electromagnetic form factors of interest are generated by Feynman diagrams with a virtual photon line, which dominate over contributions from diagrams in which either Higgs or $Z$ bosons participate. In this context, current AMMs and EDMs data are utilized to estimate upper bounds on SME coefficients. Since the resulting contributions involve a plethora of Lorentz-violation parameters, assumptions, aiming at the  reduction of the number of SME quantities, are made, for which two scenarios are considered. In one of these scenarios some SME parameters, introduced by lepton-flavor-nonconserving two-point insertions, are assumed to be quite small, thus being disregarded and then leaving appropriate conditions to bound Lorentz-violation coefficients to be as small as $10^{-28}$, from the electric dipole moment of the electron, and limits as restrictive as $10^{-14}$ and $10^{-4}$ if constraints on the muon and the tau-lepton electromagnetic moments are taken into account, respectively. Another scenario, relying on the the assumption that Yukawa-like related couplings are Hermitian, also gives rise to bounds on SME coefficients. In this scenario, the analysis of mSME contributions and their comparison with current bounds on electromagnetic moments of charged leptons determine upper limits on the impact of Lorentz-violating coefficients as stringent as $10^{-15}$, which, specifically, are imposed by the anomalous magnetic moment of the muon. In the same scenario, the electric dipole moment of the muon also establishes upper bounds on SME coefficients of order $10^{-12}$. A summary of these bounds is provided by Table~\ref{finaltab}.
\\

The remainder of the paper has been organized as follows. In Sec.~\ref{theor}, a brief discussion on the theoretical framework, necessary for the phenomenological calculation, is performed. We present, in Sec.~\ref{calc}, the analytical calculation of the electromagnetic vertex $A_\mu l_Al_A$ at one loop. Numerical estmations and a discussion on our results are provided in Sec.~\ref{numb}. Finally, we give a summary of the present investigation in Sec.~\ref{conc}.


\section{Lorentz violation in the Yukawa sector}
\label{theor}
Since the mSME is an effective field theory parametrizing heavy physics at SM energy scales, its Lagrangian terms are exclusively defined in terms of the fields of such a low-energy description. In this context, Lorentz-nonconserving interactions are introduced in all the SM sectors, among which we consider, for the phenomenological objectives of the present investigation, Lagrangian terms from the Yukawa sector. Meanwhile, ${\rm SU}(3)_C\times{\rm SU}(2)_L\times{\rm U}(1)_Y$ gauge symmetry is still assumed and then spontaneously broken through implementation of the Brout-Englert-Higgs mechanism~\cite{EnBr,PWHiggs1,PWHiggs2} as usual, in order to define the full set of mass eigenfields within the theory governed by the electromagnetic gauge group~\cite{PeSch,CheLi,GiKi,Langacker,Schwartzbuch}. Of course, this procedure affects Lorentz-violating interactions, for which a discussion on the resultant terms of the mSME Yukawa sector is pertinent.
\\

In the mSME, the Yukawa sector is {\it CPT}-even and is given by~\cite{CoKo2}
\begin{eqnarray}
\label{YL}
{\cal L}_{\rm Y}&=&- (Y_L)^{AB}\bar{L}_A\phi R_B-\frac{1}{2}(H_L)^{AB}_{\mu \nu}\bar{L}_A\phi \sigma^{\mu \nu} R_B \nonumber \\
&&- (Y_U)^{AB}\bar{Q}_A\tilde{\phi} U_B-\frac{1}{2}(H_U)^{AB}_{\mu \nu}\bar{Q}_A\tilde{\phi} \sigma^{\mu \nu} U_B \nonumber \\
&&- (Y_D)^{AB}\bar{Q}_A\phi D_B-\frac{1}{2}(H_D)^{AB}_{\mu \nu}\bar{Q}_A\phi \sigma^{\mu \nu} D_B
\nonumber \\ &&
+{\rm H.c.}
\label{Ybssb}
\end{eqnarray}
Here, $\phi$ is the Higgs doublet. Moreover, $L_A$ and $R_A$ are the SM ${\rm SU}(2)_L$ left-handed lepton doublets and right-handed lepton singlets, respectively. $Q_A$, on the other hand, are the right-handed quark doublets, whereas $U_A$ are the $u$-type quark singlets, and $D_A$ are the $d$-type quark singlets, both right-handed. In all cases, capital-letter indices $A,B$ label fermion flavor.
The matrices with entries $(H_{L})^{AB}_{\mu \nu}$, $(H_{U})^{AB}_{\mu \nu}$, $(H_{D})^{AB}_{\mu \nu}$ are dimensionless, but, as it happens with the SM Yukawa matrices $Y_{L}$, $Y_{U}$, $Y_{D}$, they are not restricted to be Hermitian in flavor space. This opens a window to look for flavor-violation effects mediated by the Higgs boson. A thorough discussion on this Lorentz-violating Yukawa sector, which includes its application to the photon propagator within the scheme of nonlinear covariant gauges~\cite{MRS,MoRe,RoBa,MeTo}, was recently carried out in Ref.~\cite{HMNSTV}. Notation and conventions utilized in the present paper have been taken from that reference.
\\

After spontaneous symmetry breaking, and once implementation of the standard unitary transformations to pass from the gauge basis to the basis of mass eigenstates has been performed, the Lagrangian given in Eq.~(\ref{YL}) can be written, in the unitary gauge, as
\begin{eqnarray}
&&
{\cal L}_{\rm Y}=-\sum_{A}\left(m_{f_A}+\frac{g\,m_{f_A}}{2m_W}\,H\right)\bar{f}_{A}f_{A}
\nonumber \\ &&  
-\frac{1}{2}\sum_{A,B}\left(v+H\right)\bar{f}_A
\left(V^{AB}_{\alpha \beta}+A^{BA*}_{\alpha \beta}\, \gamma^5\right)\sigma^{\alpha \beta}\,f_B,
\label{Yassb}
\end{eqnarray}
where
\begin{eqnarray}
V^{AB}_{\alpha \beta}&=&\frac{1}{2}\left(Y^{AB}_{\alpha \beta}+Y^{BA*}_{\alpha \beta}\right),
\label{VABab}
\\
A^{AB}_{\alpha \beta}&=&\frac{1}{2}\left(Y^{AB}_{\alpha \beta}-Y^{BA*}_{\alpha \beta}\right).
\label{AABab}
\end{eqnarray}
In the above expressions,  $Y_{\alpha \beta}=V^\dag_L H_{\alpha \beta} V_R$, with $V_{L}$ and $V_{R}$ the unitary matrices connecting the gauge and mass-eigenfields bases of chiral spinors. Though not explicitly indicated by the notation of matrices $Y_{\alpha \beta}$, three types of fermions correspond to each of them, namely, $Y^{L}_{\alpha \beta}$, $Y^{D}_{\alpha \beta}$, and $Y^{U}_{\alpha \beta}$, which stand for charged leptons, $u$-type quarks, and $d$-type quarks, respectively. With respect to flavor space, matrices $V_{\alpha\beta}$ are Hermitian, whereas matrices $A_{\alpha\beta}$ are antiHermitian. On the other hand, matrices $V^{AB}$ and $A^{AB}$, given in spacetime group, are both antisymmetric. In the perturbative approach, which is adopted here, these Lorentz-volating Yukawa couplings yield two types of physical couplings: the bilinear insertion $-(v/2)\bar{f_A}\left(V^{AB}_{\alpha \beta}+A^{BA*}_{\alpha \beta}\, \gamma^5\right)\sigma^{\alpha \beta}f_B$ and the trilinear vertex $-(1/2)H\bar{f_A}\left(V^{AB}_{\alpha \beta}+A^{BA*}_{\alpha \beta}\, \gamma^5\right)\sigma^{\alpha \beta}f_B$. For instance, the one-loop contribution from this sort of Lorentz violation to the photon propagator is determined by the bilinear term, whose Feynman rule is $-i(v/2)\left(V^{AB}_{\alpha \beta}+A^{BA*}_{\alpha \beta}\, \gamma^5\right)\sigma^{\alpha \beta}$~\cite{HMNSTV}. Note, from Eq.~(\ref{AABab}), that $A_{\alpha \beta}$ vanishes for Hermitian matrix $Y^\dag_{\alpha \beta}=Y_{\alpha \beta}$, whereas the antiHermitian-matrix condition $Y^\dag_{\alpha \beta}=-Y_{\alpha \beta}$, eliminates $V_{\alpha\beta}$.


\section{One-loop contributions to lepton electromagnetic interactions}
\label{calc}
Next we calculate contributions from Lorentz violation in the Yukawa sector ${\cal L}_{\rm Y}$, Eq.~(\ref{Yassb}), to the electromagnetic vertex $A_\mu f_A f_A$. To execute this task, we follow a perturbative approach, in which the effects from Lorentz-violating Lagrangian terms that are quadratic in fields are taken into account by placing two-point vertex insertions in propagator lines of Feynman diagrams. This {\it modus operandi} has been of profit in previous phenomenological investigations~\cite{MNTT1,MNTT2,CNTT,HMNSTV,CheKu,KLP,KoPi,AST,CSV,BCT,LMT}.\\

As explicitly shown in Refs.~\cite{MNTT1,MNTT2}, electromagnetic interactions at the loop level are modified by the occurrence of Lorentz violation, resulting in a larger number of electromagnetic form factors than those in Eq.~(\ref{emvpar}), constructed under the assumption of Lorentz invariance~\cite{HIRSS,NPR,BGS}. Nevertheless, even if Lorentz symmetry is violated, the form factors defining the AMM and the EDM are to be identified from the Lorentz-invariant contributions of Eq.~(\ref{emvpar}), where Lorentz-violating background fields can only be contracted with themselves. According to Eq.~(\ref{Yassb}), the coefficients $V^{AB}_{\mu\nu}$ and $A^{AB}_{\mu\nu}$ are antisymmetric with respect to spacetime indices, so they are traceless in this sense. Consequently, any first-order contribution to AMMs and EDMs vanish. But note that nonzero Lorentz-invariant contributions may emerge as long as diagrams at the second order in $V^{AB}_{\mu\nu}$ or $A^{AB}_{\mu\nu}$ are considered. 
\\


\subsection{Contributing Feynman diagrams}
\begin{figure}[ht]
\center
\includegraphics[width=3cm]{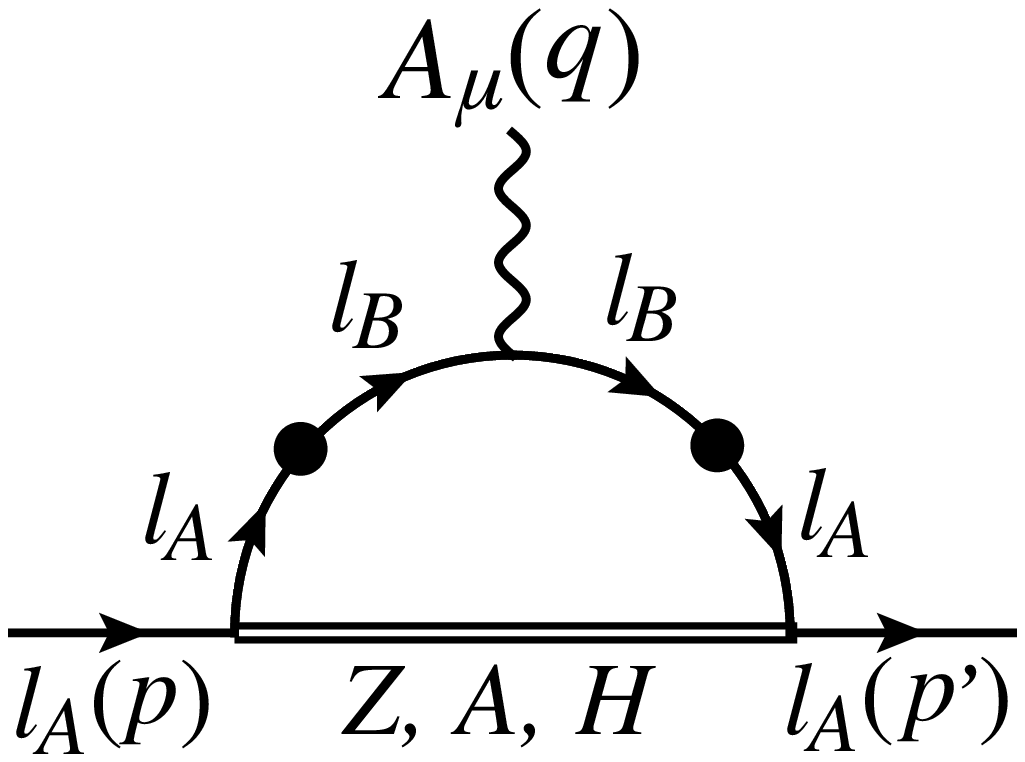}
\hspace{0.8cm}
\includegraphics[width=3.7cm]{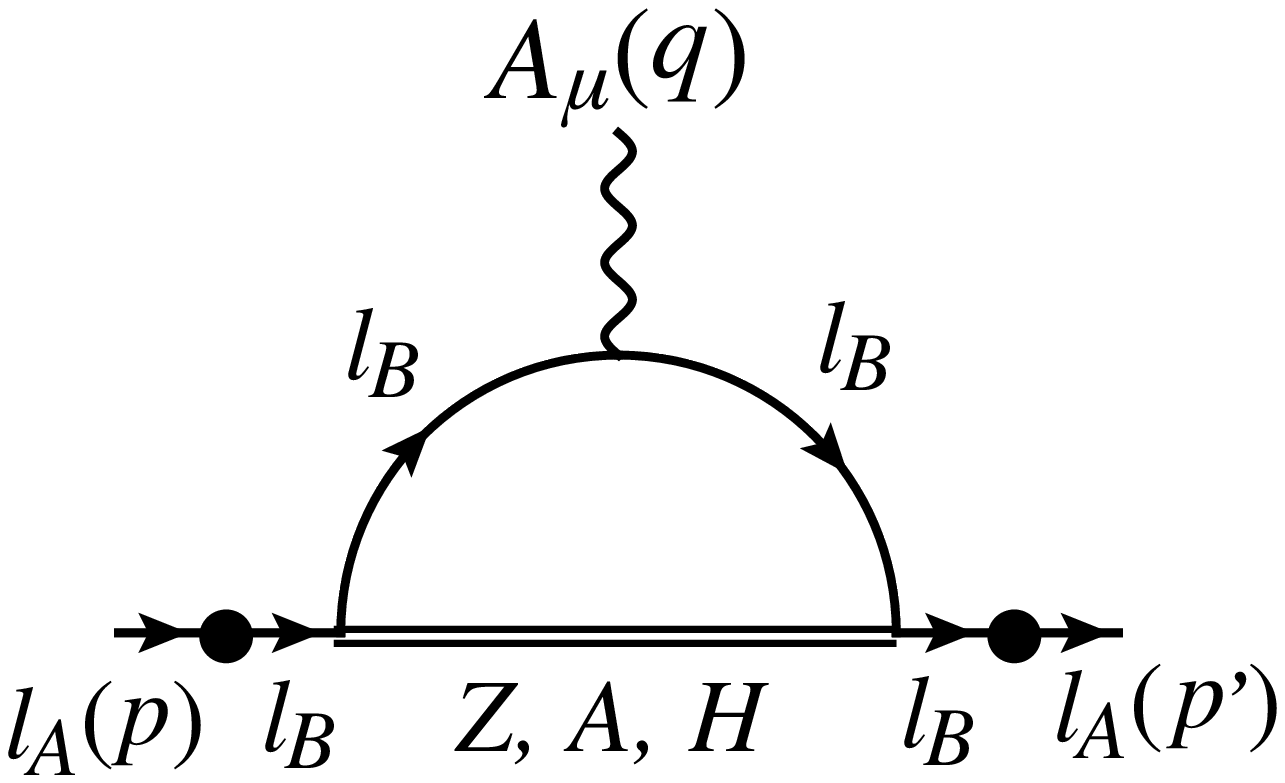}
\vspace{0.3cm}
\\
\center
\includegraphics[width=3cm]{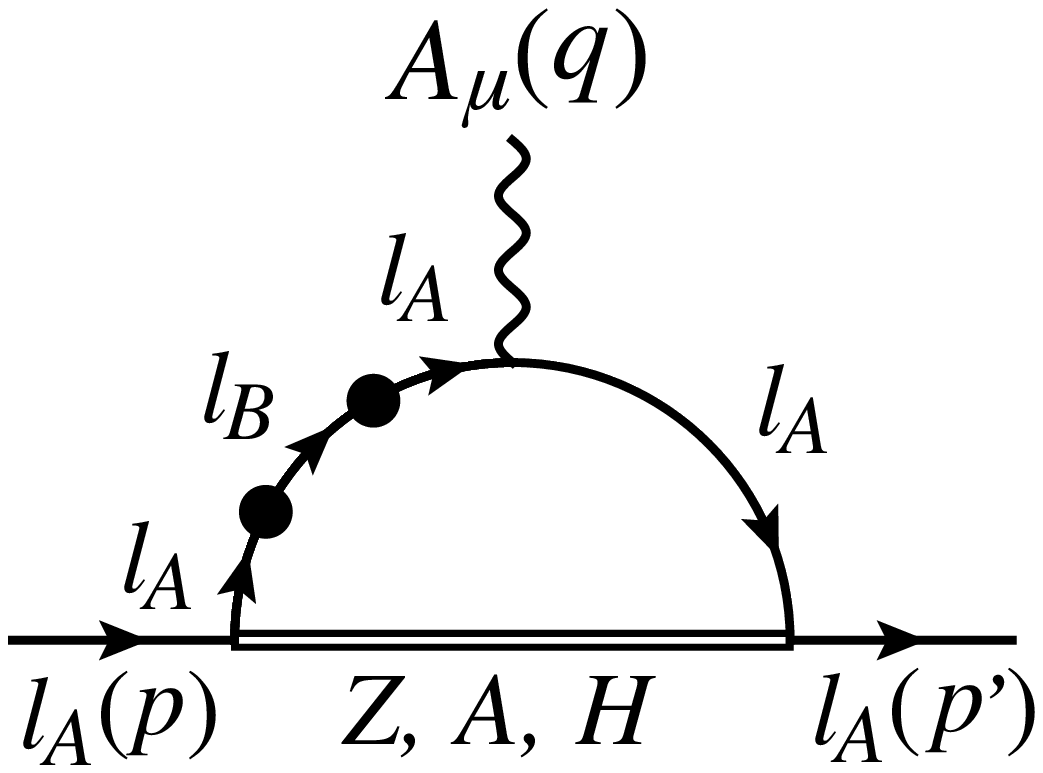}
\hspace{1cm}
\includegraphics[width=3cm]{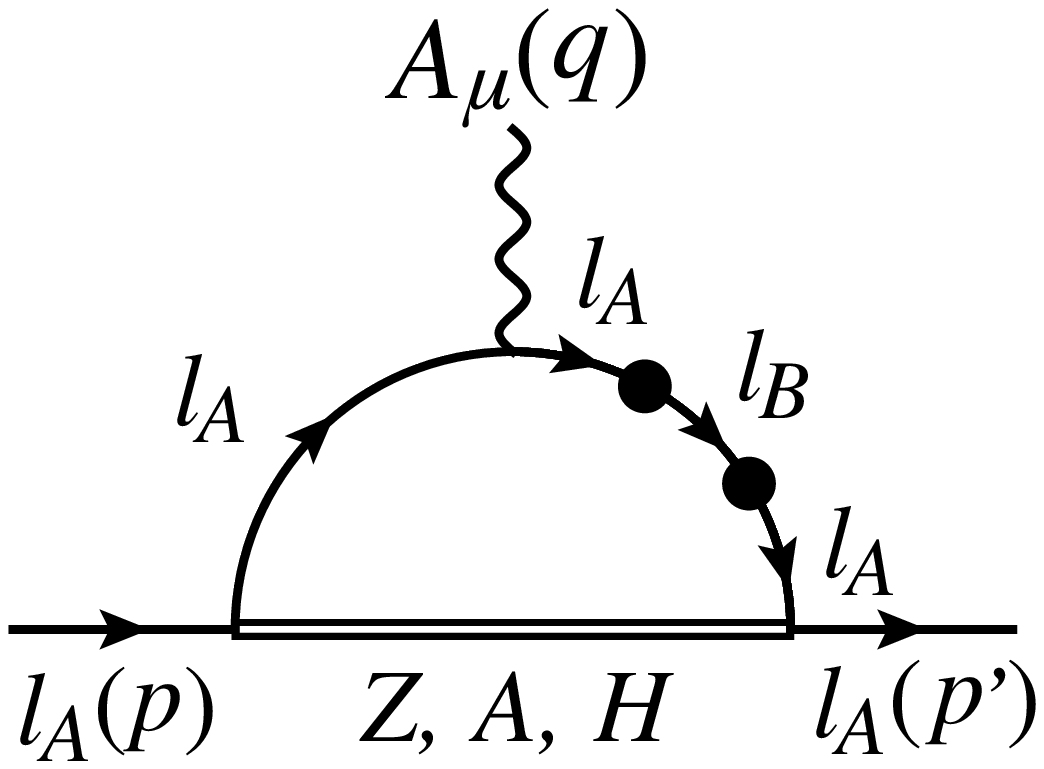}
\vspace{0.3cm}
\\
\center
\includegraphics[width=3.3cm]{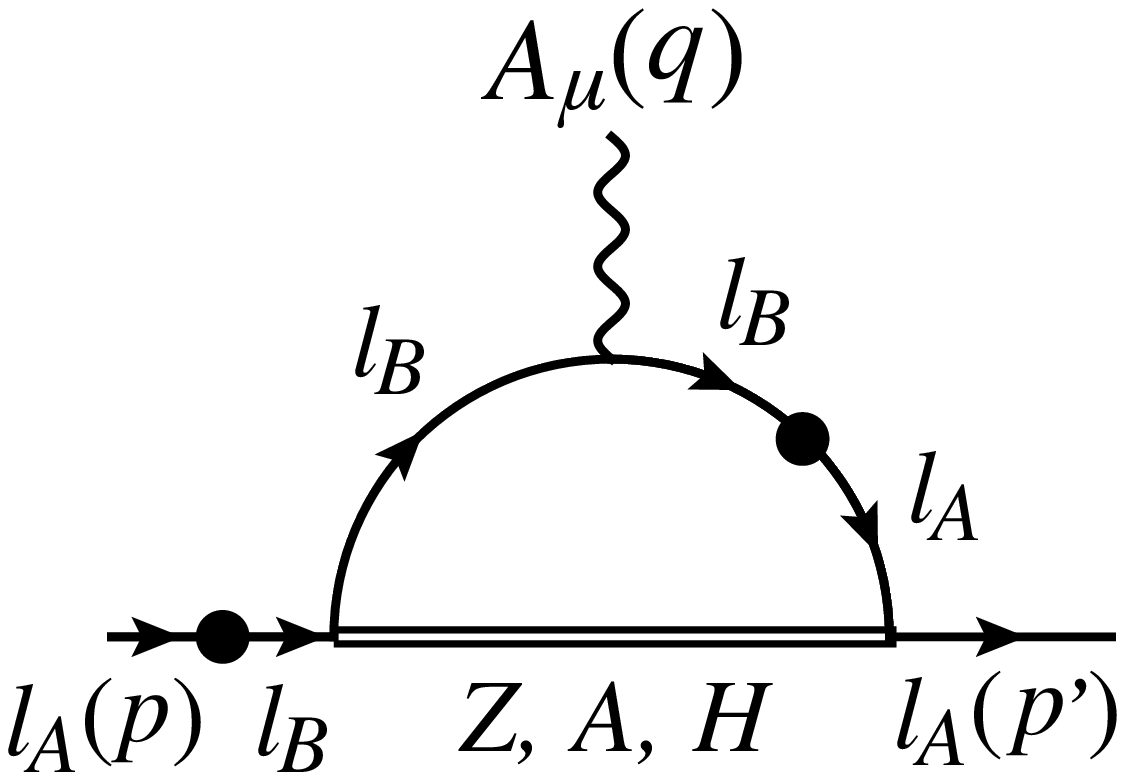}
\hspace{1cm}
\includegraphics[width=3.3cm]{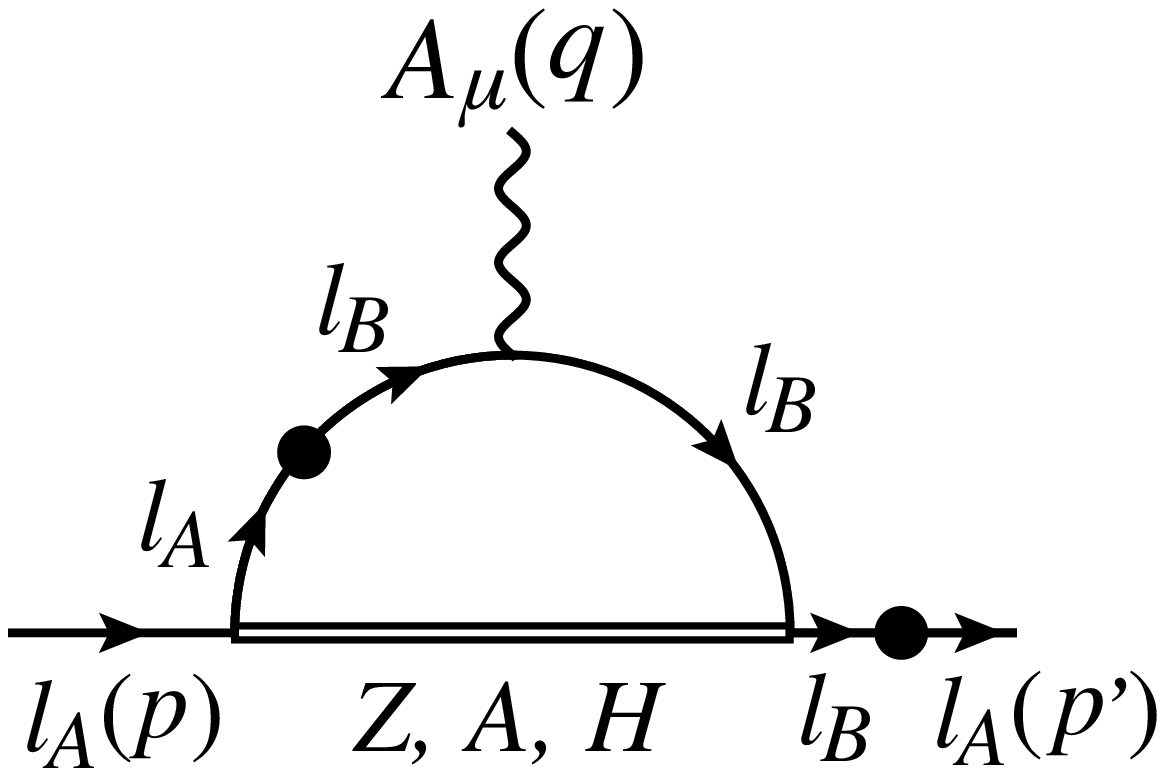}
\vspace{0.3cm}
\\
\center
\includegraphics[width=3.3cm]{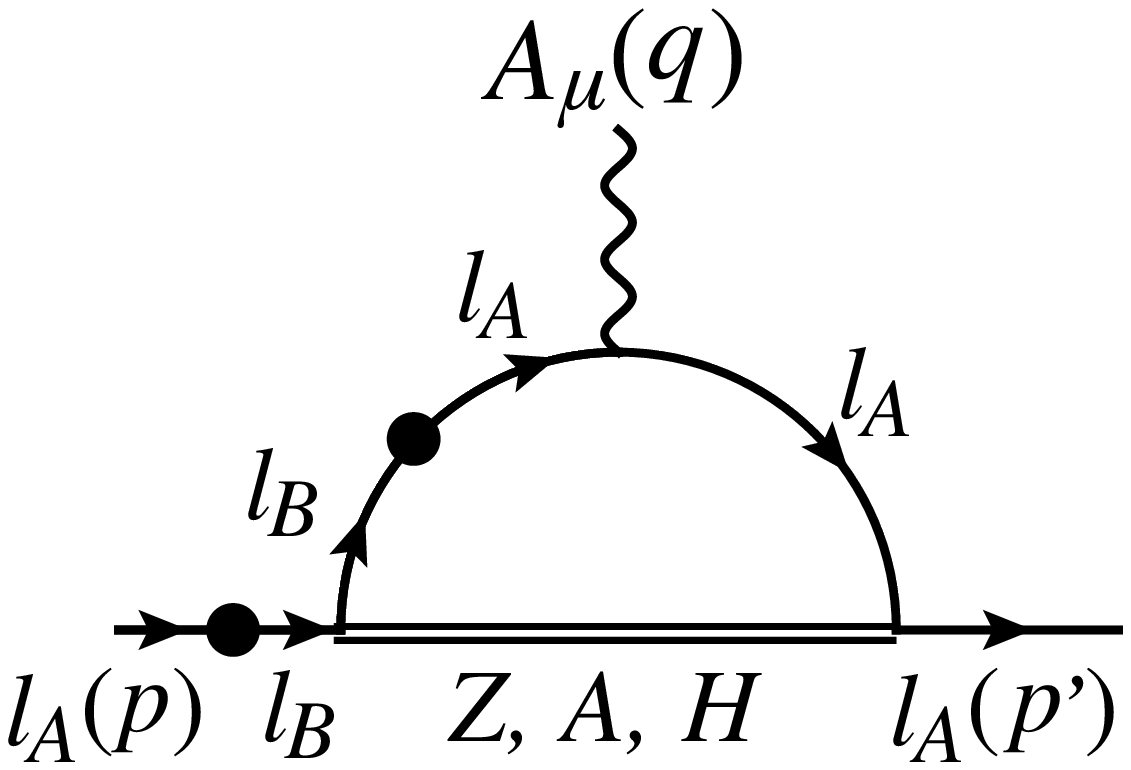}
\hspace{1cm}
\includegraphics[width=3.3cm]{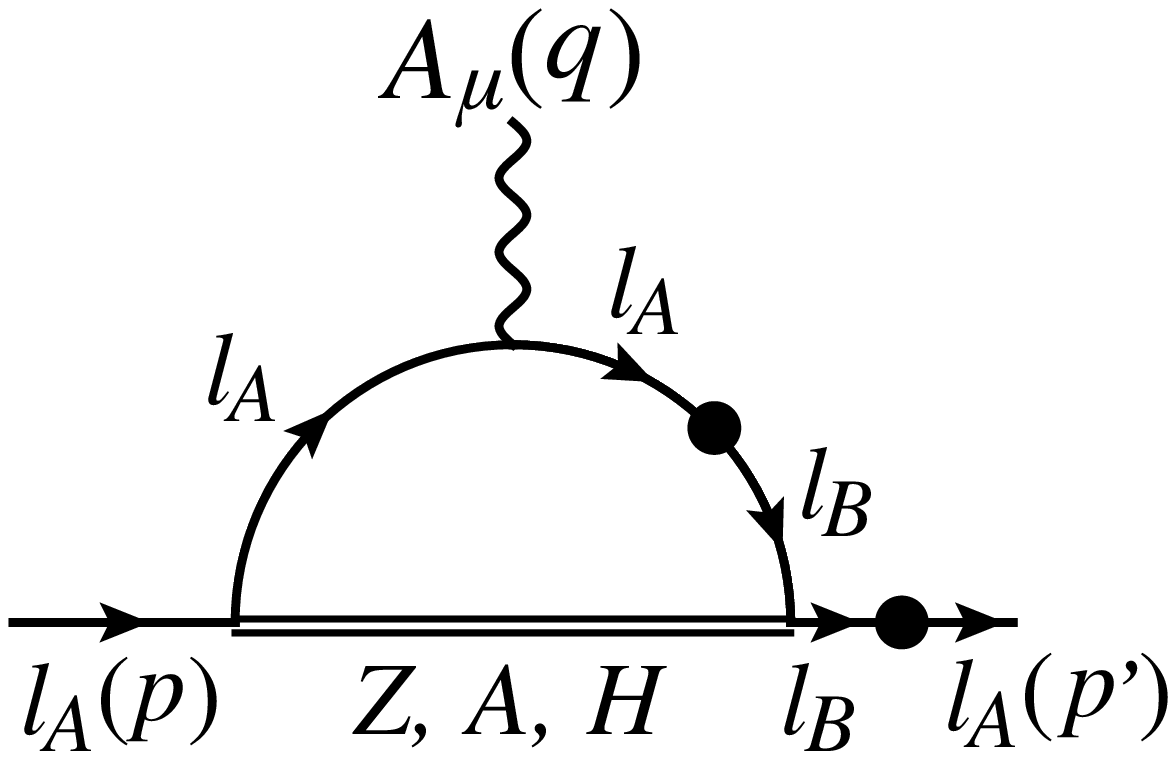}
\vspace{0.3cm}
\\
\center
\includegraphics[width=4cm]{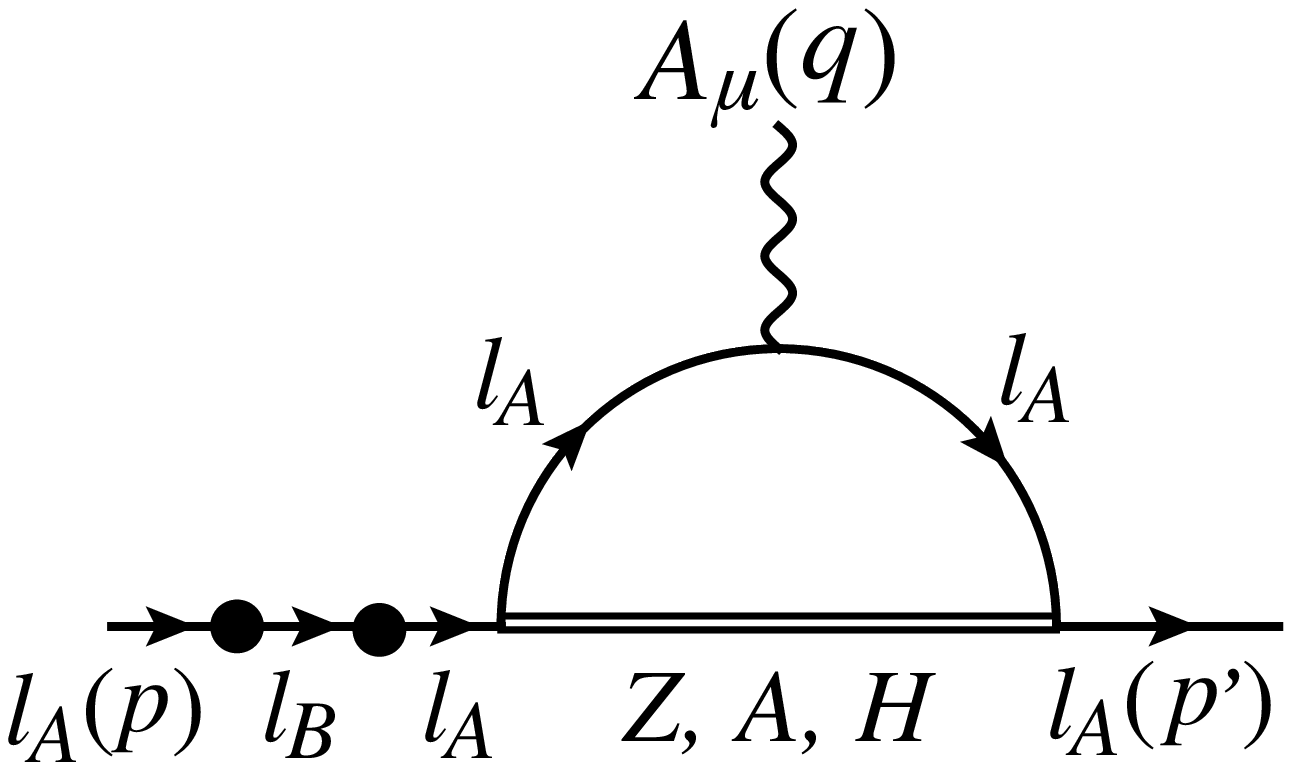}
\hspace{0.45cm}
\includegraphics[width=4cm]{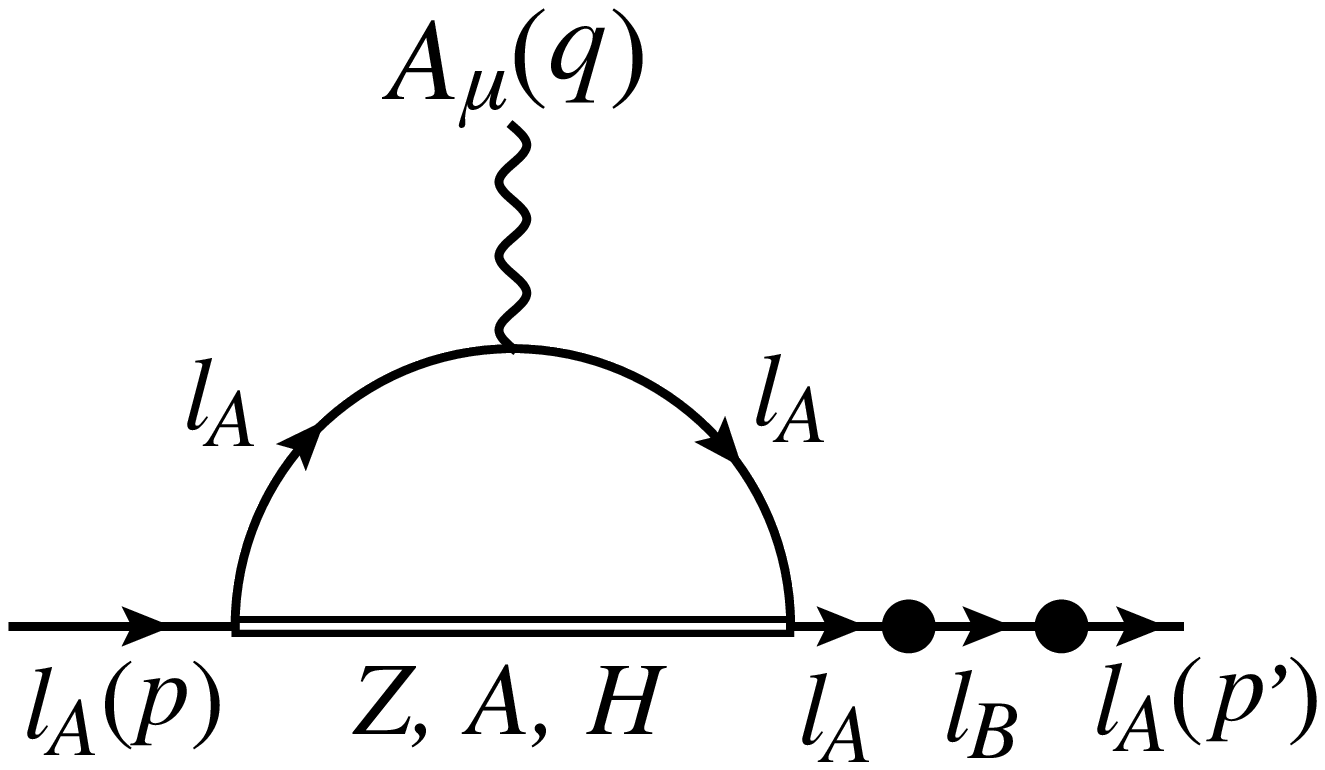}
\caption{\label{gbdiagrams} Feynman diagrams $A_\mu f_A f_A$ contributing to magnetic and electric form factors, with Lorentz-nonconservation effects entering exclusively through bilinear insertions $l_Al_B$, where either $A=B$ or $A\ne B$. Virtual double lines in loops stand for a $Z$ boson, a photon, or a Higgs boson.}
\end{figure}

\begin{figure}[ht]
\center
\includegraphics[width=3cm]{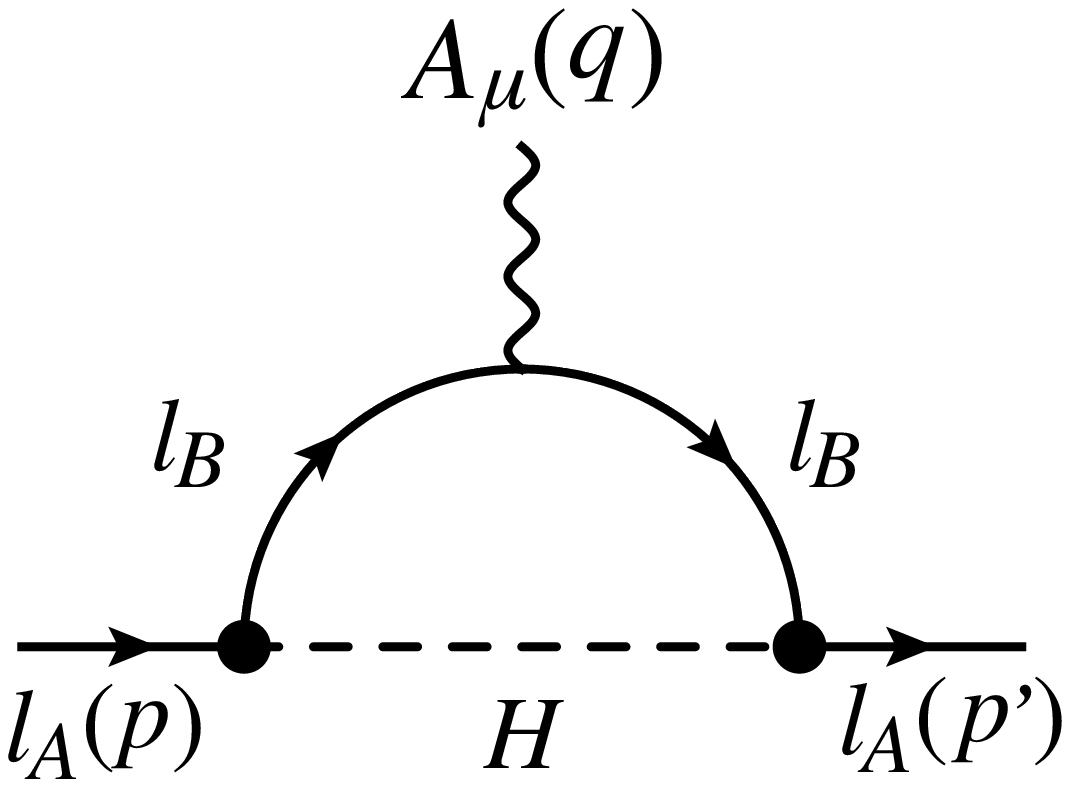}
\vspace{0.5cm}
\\
\center
\includegraphics[width=3.3cm]{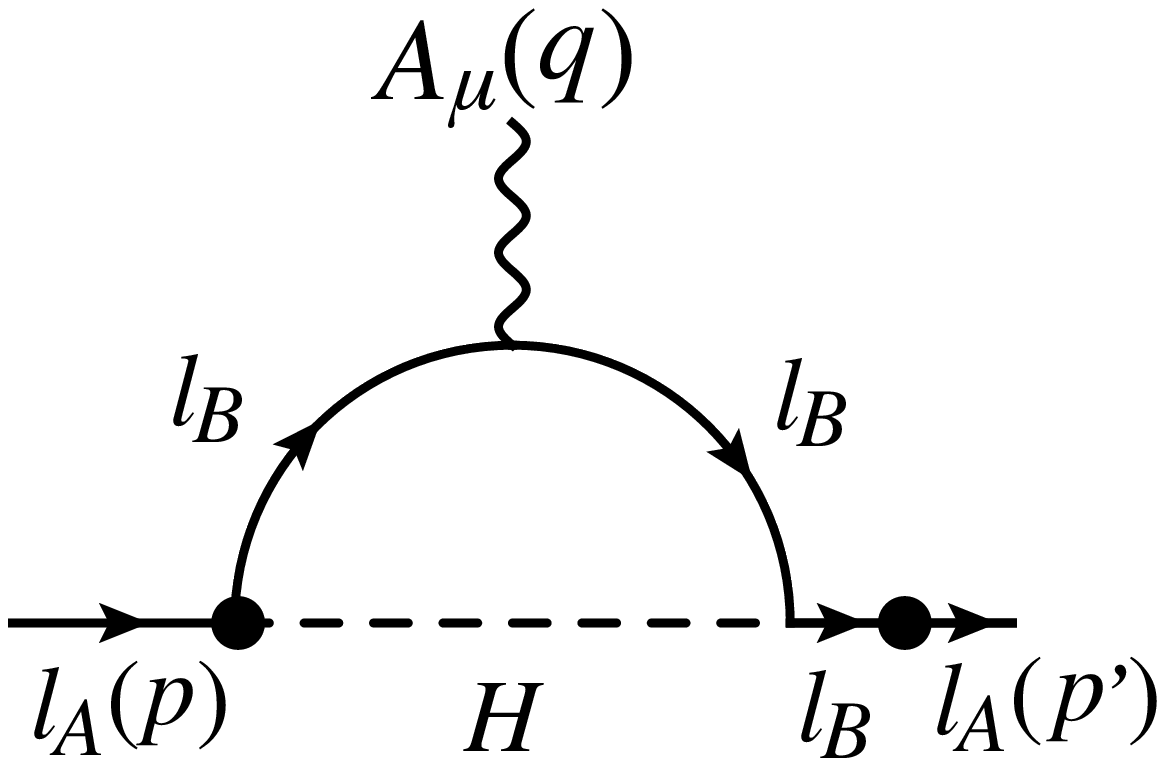}
\hspace{0.3cm}
\includegraphics[width=3.3cm]{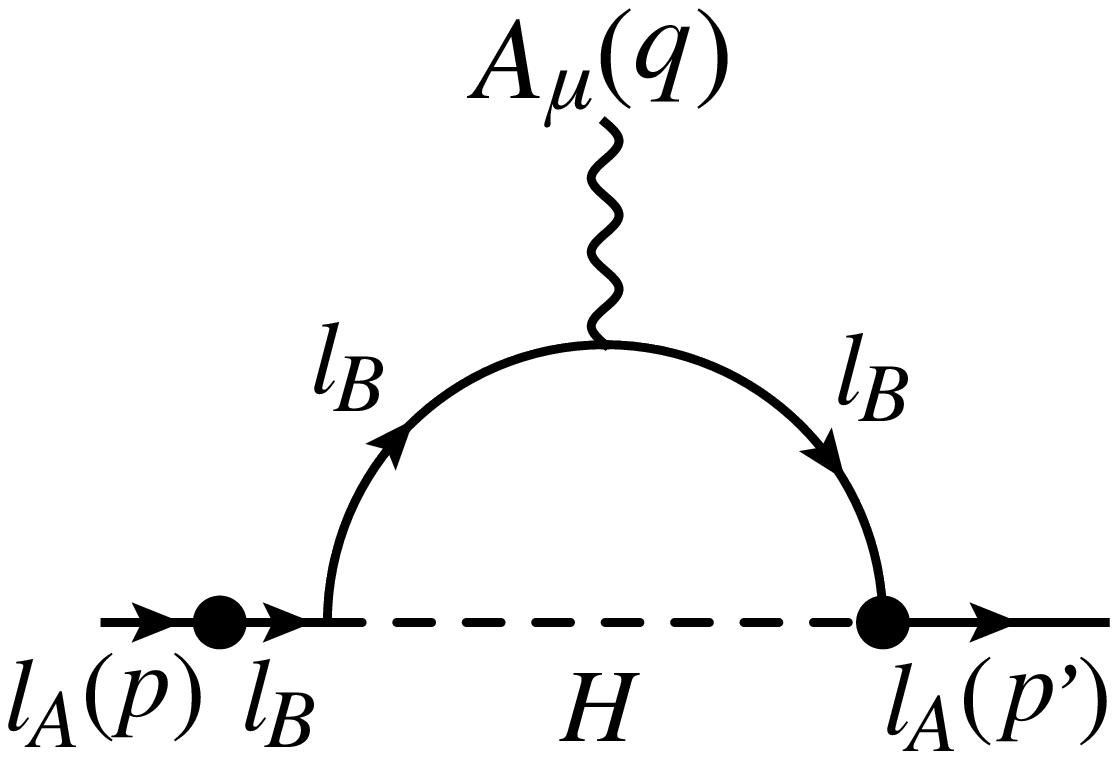}
\vspace{0.5cm}
\\
\center
\includegraphics[width=3cm]{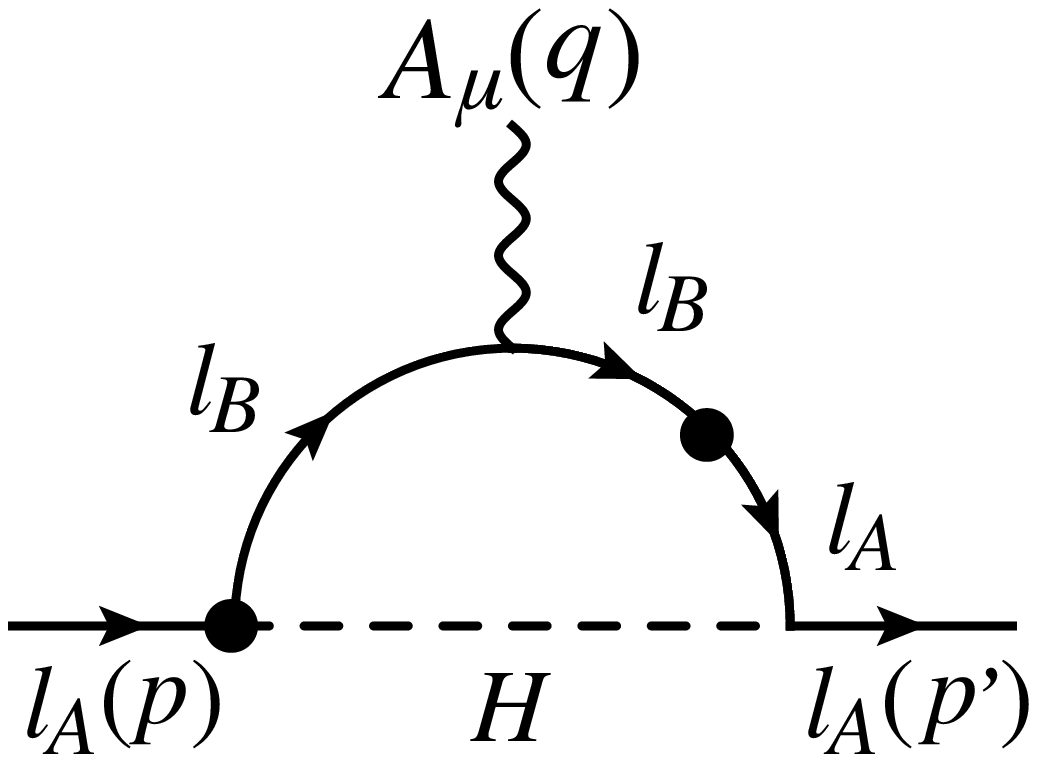}
\hspace{1cm}
\includegraphics[width=3cm]{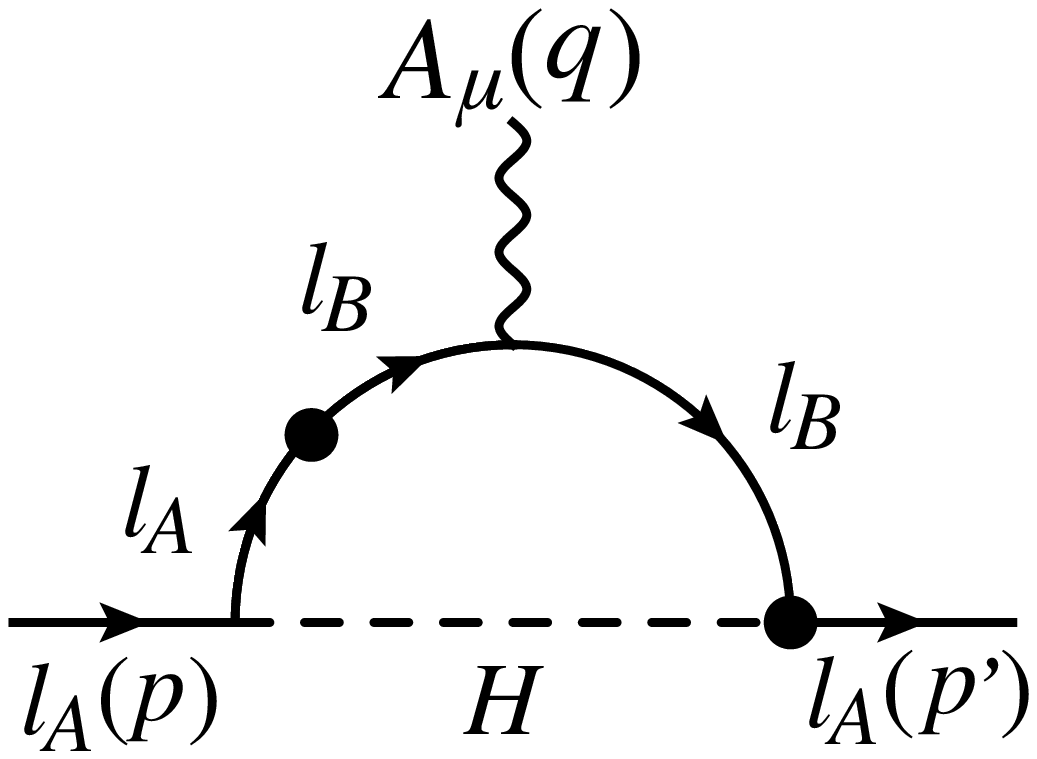}
\vspace{0.5cm}
\\
\center
\includegraphics[width=3cm]{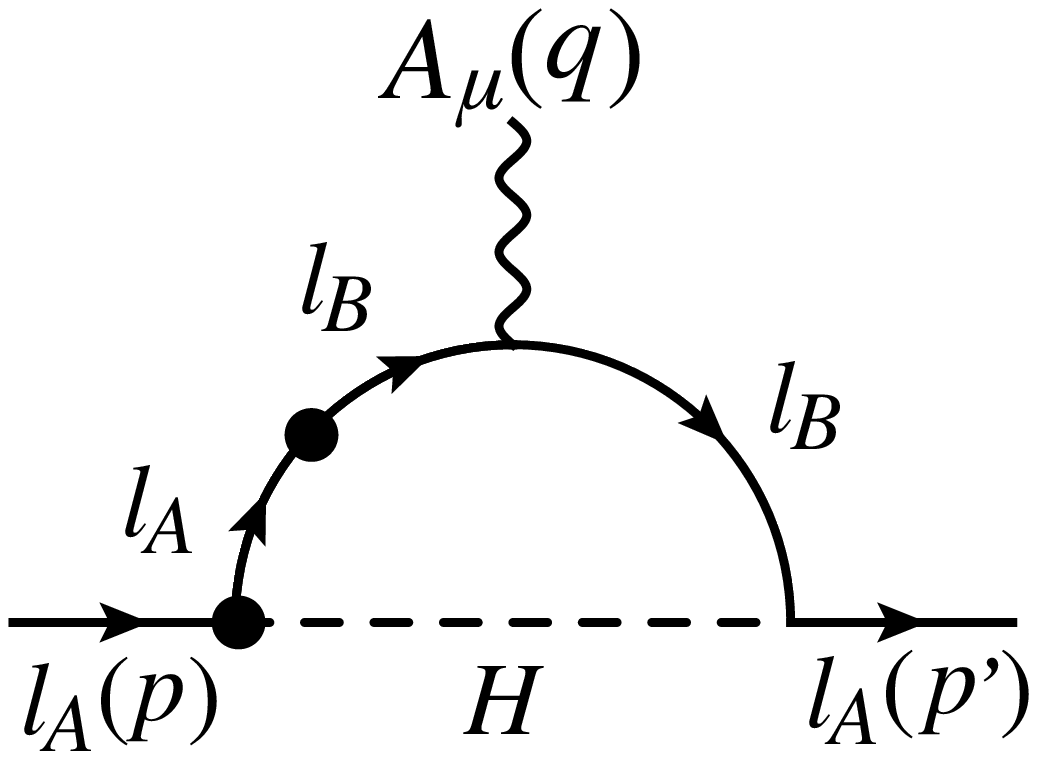}
\hspace{1cm}
\includegraphics[width=3cm]{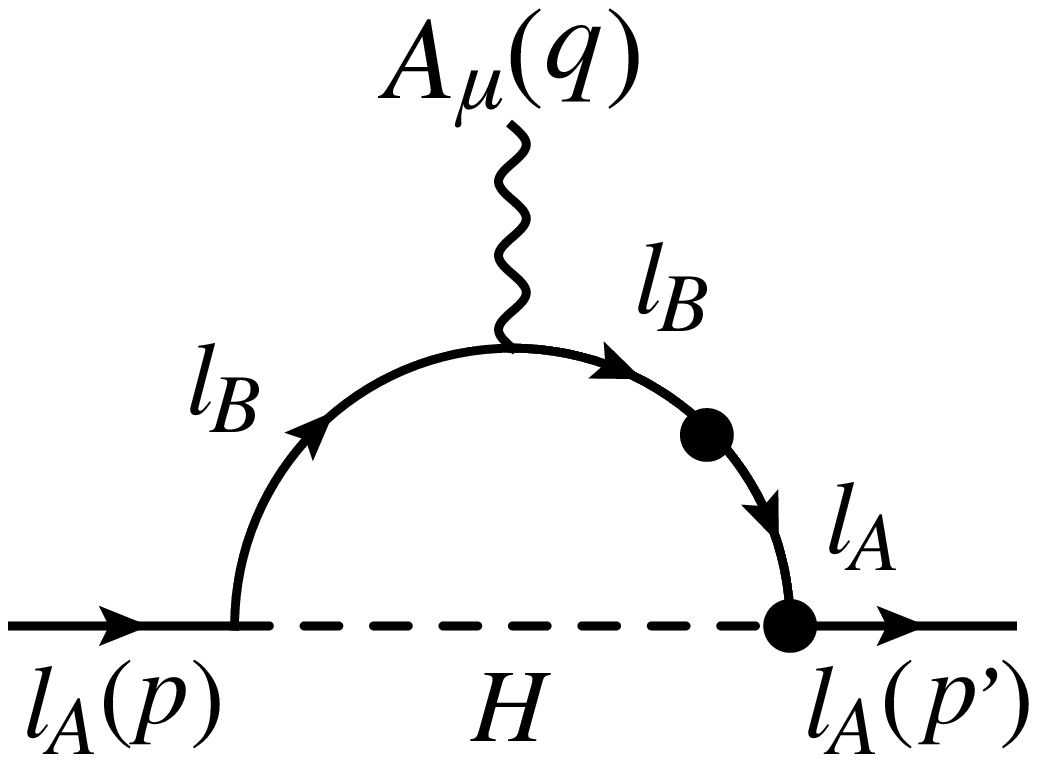}
\vspace{0.5cm}
\\
\center
\includegraphics[width=3.3cm]{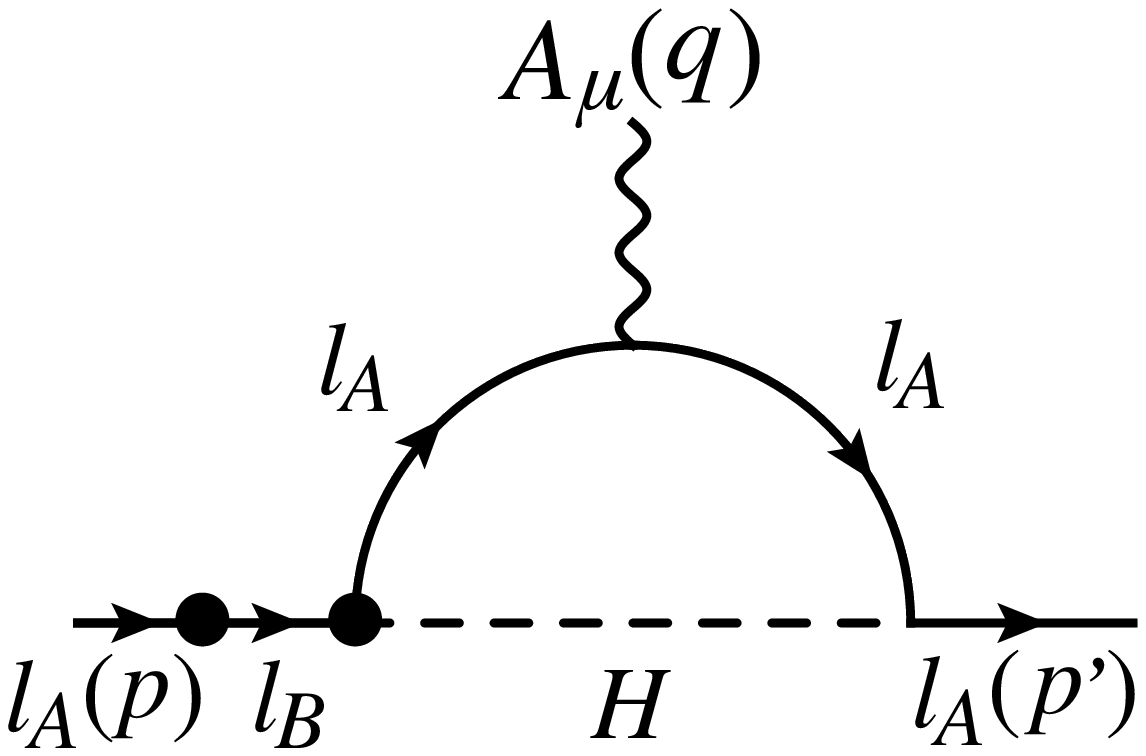}
\hspace{1cm}
\includegraphics[width=3.3cm]{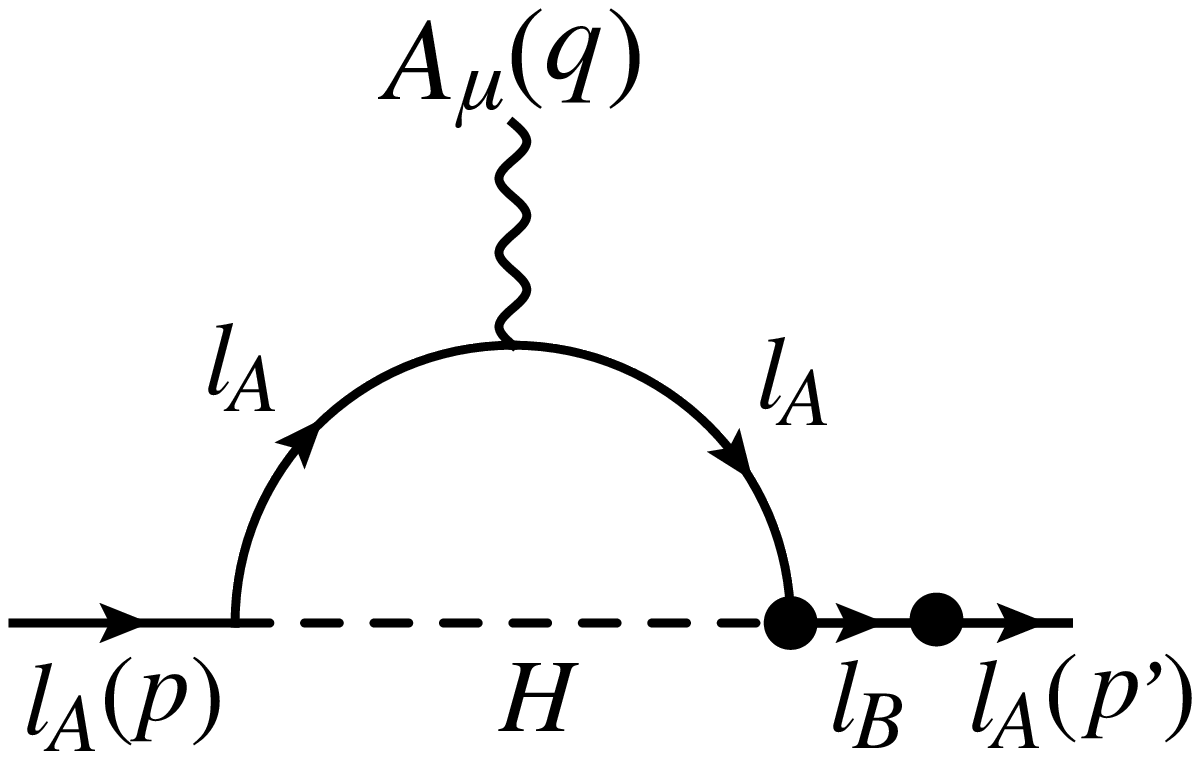}
\caption{\label{HFdiagrams} Feynman diagrams $A_\mu f_A f_A$ contributing to magnetic and electric form factors, with Lorentz-nonconservation effects entering through both bilinear insertions $l_Al_B$ and three-point vertices $Hf_Af_B$, where either $A=B$ or $A\ne B$. These diagrams require the presence of a virtual Higgs-boson line in order to exist.}
\end{figure}

With the previous discussion in mind, we consider the contributions to the electromagnetic vertex $A_\mu f_A f_A$ produced by the sum of the Feynman diagrams shown in Figs.~\ref{gbdiagrams}-\ref{HFdiagrams},
in which either two two-point insertions or two three-point vertices or simultaneously one two-point insertion and one three-point vertex appear, thus resulting in second-order contributions of mSME coefficients. The full set of contributing diagrams can be classified into three types: diagrams with a virtual $Z$-boson line; diagrams with a virtual photon line; and diagrams with a virtual Higgs-boson line. Neutrinos are assumed to be massless, so no couplings among them and the Higgs field arises, with the consequence that no contributing diagrams with virtual $W$-boson lines exist. Double lines in the loops of diagrams of Fig.~\ref{gbdiagrams} generically represent virtual-field lines which can be associated to either a $Z$ boson, a photon, or a Higgs boson. This figure displays the whole set of Feynman diagrams in which a virtual $Z$ boson or a virtual photon participate. Vertices $Hf_Af_B$ involving the coefficients $V^{AB}_{\mu\nu}$ and $A^{AB}_{\mu\nu}$ emerge from Eq.~(\ref{Yassb}), thus giving rise to the diagrams of Fig.~\ref{HFdiagrams}, which add together with diagrams comprising bilinear insertions, comprehended by Fig.~\ref{gbdiagrams}, to give the full set of diagrams with Higgs-boson loop lines.
We find it worth emphasizing that both two-point insertions and three-point vertices generated by Eq.~(\ref{Yassb}) are flavor changing, which enlarges the number of contributing diagrams. Also notice that the calculation is performed in the unitary gauge, so no diagrams with pseudo-Goldstone bosons exist. 
\\

We have performed all the calculations by following the Passarino-Veltman tensor-reduction method~\cite{PassVe}, for which the software Mathematica, by Wolfram, has been utilized, together with the packages Feyncalc~\cite{MBD} and Package X~\cite{Patel}. After carrying out these calculations and organizing the resulting expressions, we identified the contributions from each set of diagrams in Figs.~\ref{gbdiagrams}-\ref{HFdiagrams} to AMMs and EDMs. The occurrence of two-point insertions in contributing Feynman diagrams comes along with technical complications. The presence of Lorentz-violating coefficients, which cannot be simplified by on-shell conditions, diversifies the set of different Lorentz structures participating in the loop contributions. For the same reason, the expressions for the generated form factors are quite large, even though this is an on-shell calculation performed in a specific gauge. Under such circumstances, for us it made no sense to provide the explicit expressions of the AMMs and EDMs contributions. Nonetheless, using Package X, we were able to numerically check the consistent cancelation of ultraviolet (UV) divergences in all the contributions. Another practical complication arises because each bilinear insertion introduces an extra loop denominator, so loop integrals involve several propagator denominators, for which calculation strategies were realized and implemented. 
\\

\subsection{Dominant contributions to electromagnetic moments}
We classify contributing diagrams into two sets: (1) diagrams where bilinear insertions or three-point vertices involve change of lepton flavor, that is with $A\ne B$, which we refer to as the case of {\it virtual-lepton-flavor change}; and (2) diagrams in which no lepton-flavor change occurs due to two-point insertions or three-point vertices, that is $A=B$, which we call the case of {\it virtual-lepton-flavor conservation}. The calculation of the electromagnetic vertex $A_\mu l_Al_A$ is carried out by taking both external fermions on shell, but keeping the photon field off shell, so that $q^2\ne0$ (see Figs.~\ref{gbdiagrams}-\ref{HFdiagrams} for kinematical notation). The leading contributions to AMMs and EDMs are produced by diagrams with a virtual-photon line, provided in Fig.~\ref{gbdiagrams}. Quantitatively, the difference between such dominant contributions and those arising from other diagrams is of 10 orders of magnitude at least. While all diagrams, of all sorts, have been calculated and their contributions estimated, we explain the techniques involved in the calculations by specifically discussing the aforementioned leading contributions in some detail. As usual, virtual-photon diagrams come along with infrared (IR) divergences, which motivates us to introduce a fictitious photon mass, $m_\gamma$~\cite{PeSch,Schwartzbuch}. An aspect worth commenting is that diagrams with two two-point insertions on a single virtual-fermion line generate IR divergencies, whereas diagrams in which propagators involve exactly one of such insertions are IR finite. \\

For a variety of models, analytic expressions of triangle diagrams written in the unitary gauge bear three propagator denominators. Nonetheless, the use of two-point insertions in the present calculation yields expressions with up to five of such denominators. We have executed this calculation by following two different paths and we have verified that the results obtained through both approaches coincide with each other. One of such approaches consists in a direct calculation in which  four- and five-point Passarino-Veltman scalar functions~\cite{PassVe} emerge. In the other method, which we discuss in detail next, the number of propagator denominators in loop integrals is reduced to just three by implementing a trick involving squared-mass derivatives. Consider the first, third and fourth diagrams of Fig.~\ref{gbdiagrams} in the case of virtual-photon line. These diagrams, which have been identified to produce the leading contributions among the whole set of virtual-photon diagrams, have Lorentz-violating two-point insertions in loop lines, exclusively. After defining $\Delta_\gamma=k^2-m_\gamma^2$ and $\Delta_j(p)=(k+p)^2-m_j^2$, and implementing the Feynman parameters technique~\cite{PeSch,Feynman}, we write the analytical expressions of these diagrams as
\begin{eqnarray}
&&
\Gamma^{AB}_{1\mu}=\frac{i}{(4\pi)^2}\frac{(2\pi\mu)^{4-D}}{i\pi}
\int d^Dk
\nonumber \\ && \hspace{0.4cm}\times
\int_0^1dx\,dy\,
\frac{\partial^2}{\partial m_j^2\partial m_k^2}\frac{N_{1\mu}(m_A,m_B)}{\Delta_\gamma\,\Delta_j(l)\Delta_k(l')}
\bigg|_{\substack{m_j=m_A
\vspace{0.1cm}\\ m_k=m_B}},
\label{Gamma1}
\end{eqnarray}

\begin{eqnarray}
&&
\Gamma^{AB}_{2\mu}=\frac{i}{(4\pi)^2}\frac{(2\pi\mu)^{4-D}}{i\pi}\int d^Dk
\nonumber \\ &&\hspace{1cm}
\times\int_0^1dx\,\frac{\partial^2}{\partial (m_j^2)^2}\frac{x\,N_{2\mu}(m_A,m_B)}{\Delta_\gamma\,\Delta_j(l)\Delta_B(p)}\bigg|_{m_j=m_A},
\label{Gamma2}
\end{eqnarray}

\begin{eqnarray}
&&\Gamma^{AB}_{3\mu}=\frac{i}{(4\pi)^2}\frac{(2\pi\mu)^{4-D}}{i\pi}
\int d^Dk
\nonumber \\ && \hspace{0.2cm}
\times\int_0^1dx\,\frac{\partial^2}{\partial (m_j^2)^2}\frac{(1-x)\,N_{3\mu}(m_A,m_B)}{\Delta_\gamma\,\Delta_j(l)\Delta_B(p')}\bigg|_{m_j=m_A},
\label{Gamma3}
\end{eqnarray}
where $l=xp+(1-x)p'$ and $l'=yp+(1-y)p'$ have been defined. Keep in mind that the cases in which virual-lepton flavor is conserved and changed are both comprehended by Eqs.~(\ref{Gamma1})-(\ref{Gamma3}). The regularization of loop integrals in these equations is carried out within the approach of dimensional regularization~\cite{PeSch,BoGi}, in which case $\mu$ is a quantity with units $[\mu]={\rm mass}$, introduced to correct mass dimensions of amplitudes. In Eqs.~(\ref{Gamma1})-(\ref{Gamma3}), the factors $N_{j\mu}=N_{j\mu}(m_A,m_B)$, depending on charged-lepton masses and not affected by squared-mass derivatives, read 
\begin{eqnarray}
&&
N_{1\mu}=-\frac{e^3v^2}{4}
\gamma^\nu(\slashed{k}+\slashed{p}'+m_A)
\big( A^{BA*}_{\alpha\beta}\gamma_5+V^{AB}_{\alpha\beta}\big)
\nonumber \\ &&
\hspace{1.4cm}
\times
\sigma^{\alpha\beta}
(\slashed{k}+\slashed{p}'+m_B)\gamma_\mu(\slashed{k}+\slashed{p}+m_B)
\nonumber \\ &&
\hspace{1.4cm}
\times
\big(A^{BA*}_{\rho\lambda}\gamma_5+V^{AB}_{\rho\lambda}\big)
\sigma^{\rho\lambda}
(\slashed{k}+\slashed{p}+m_A)\gamma_\nu, 
\end{eqnarray}

\begin{eqnarray}
&&
N_{2\mu}=-\frac{e^3v^2}{4}\gamma^\nu(\slashed{k}+\slashed{p}'+m_A)\gamma_\mu(\slashed{k}+\slashed{p}+m_A)
\nonumber \\&&
\hspace{1.1cm}
\times
\big( A^{BA*}_{\alpha\beta}\gamma_5+V^{AB}_{\alpha\beta} \big)\sigma^{\alpha\beta}(\slashed{k}+\slashed{p}+m_B)
\nonumber \\&&
\hspace{1.1cm}
\times
\big( A^{BA*}_{\rho\lambda}\gamma_5+V^{AB}_{\rho\lambda} \big)
\sigma^{\rho\lambda}
(\slashed{k}+\slashed{p}+m_A)\gamma_\nu,
\end{eqnarray}

\begin{eqnarray}
&&
N_{3\mu}=-\frac{e^3v^2}{4}\gamma^\nu(\slashed{k}+\slashed{p}'+m_A)\big( A^{BA*}_{\alpha\beta}\gamma_5+V^{AB}_{\alpha\beta} \big)
\nonumber \\ &&
\hspace{1.4cm}
\times
\sigma^{\alpha\beta}(\slashed{k}+\slashed{p}'+m_B)\big( A^{BA*}_{\rho\lambda}\gamma_5+V^{AB}_{\rho\lambda} \big)\sigma^{\rho\lambda}
\nonumber \\ &&
\hspace{1.4cm}
\times
(\slashed{k}+\slashed{p}'+m_A)\gamma_\mu(\slashed{k}+\slashed{p}+m_A)\gamma_\nu.
\end{eqnarray}
\\

We add all the individual loop-diagram analytic expressions to get the total contribution $\Gamma^A_\mu=\sum_{B=e,\mu,\tau}\big(\Gamma^{AB}_{1\mu}+\Gamma^{AB}_{2\mu}+\Gamma^{AB}_{3\mu}\big)$. Through algebraic manipulations, we write this as 
\begin{equation}
\Gamma^A_\mu=\frac{e}{2m_A}f^m_A\,\sigma_{\mu\nu}q^\nu+if^d_A\,\sigma_{\mu\nu}q^\nu\gamma_5+\cdots,
\label{totcontana}
\end{equation}
where $f^m_A$ and $f^d_A$ are functions of field masses, squared photon momentum $q^2$ and  Lorentz-violating tensor coefficients $V^{AB}_{\alpha\beta}$ and $A^{AB}_{\alpha\beta}$. We emphasize that $f^m_A$ and $f^d_A$, which from Eq.~(\ref{emvpar}) are respectively recognized as magnetic and electric form factors, are invariant under particle Lorentz transformations. So, contributions to the AMM and to the EDM of the charged-lepton $l_A$ can be straightforwardly extracted from such coefficients. Ellipsis in Eq.~(\ref{totcontana}) represent a large set of terms, most of which involve violations of invariance under particle Lorentz transformations. 
\\

We define, in the space of matrix representations of Lorentz transformations, the $4\times4$ matrices $\kappa_1^{AB}$, $\kappa_2^{AB}$, $\kappa_3^{AB}$, $\kappa_4^{AB}$, and $\kappa_5^{AB}$, with entries
\begin{equation}
(\kappa^{AB}_1)_\alpha\hspace{0.000001cm}^\beta=V^{AB}_{\alpha\nu}V^{AB\nu\beta},
\label{k1AB}
\end{equation}
\begin{equation}
(\kappa^{AB}_2)_\alpha\hspace{0.000001cm}^\beta=A^{BA*}_{\alpha\nu}A^{BA\nu\beta*},
\label{k2AB}
\end{equation}
\begin{equation}
(\kappa^{AB}_3)_\alpha\hspace{0.000001cm}^\beta=V^{AB}_{\alpha\nu}A^{BA\nu\beta*},
\label{k3AB}
\end{equation}
\begin{equation}
(\kappa^{AB}_4)_\alpha\hspace{0.000001cm}^\beta=V^{AB}_{\alpha\nu}\tilde{V}^{AB\nu\beta},
\label{k4AB}
\end{equation}
\begin{equation} (\kappa^{AB}_5)_\alpha\hspace{0.000001cm}^\beta=A^{BA*}_{\alpha\nu}\tilde{A}^{BA\nu\beta*}, 
\label{k5AB}
\end{equation}
where $\tilde{V}^{AB}_{\alpha\beta}=\epsilon_{\alpha\beta\rho\lambda}V^{AB\rho\lambda}$ and $\tilde{A}^{BA}_{\alpha\beta}=\epsilon_{\alpha\beta\rho\lambda}A^{BA\rho\lambda}$. Since $A,B=e,\mu,\tau$, there are, in principle, 45 of these complex matrices, each of them bearing 32 parameters,
thus yielding a grand total of 1440 parameters, which, however, are not independent. In particular, the following relations hold: $(\kappa^{AB}_1)_{\alpha\beta}=(\kappa^{AB}_1)_{\beta\alpha}$ and $(\kappa^{AB}_1)_{\alpha\beta}=(\kappa^{BA}_1)^*_{\alpha\beta}$; $(\kappa^{AB}_2)_{\alpha\beta}=(\kappa^{AB}_2)_{\beta\alpha}$ and $(\kappa^{AB}_2)_{\alpha\beta}=(\kappa^{BA}_2)^*_{\alpha\beta}$;
$(\kappa_3^{AB})_{\alpha\beta}=-(\kappa_3^{BA})^*_{\alpha\beta}$; $(\kappa_4^{AB})_{\alpha\beta}=(\kappa_4^{BA})^*_{\alpha\beta}$; and $(\kappa_5^{AB})_{\alpha\beta}=(\kappa_5^{BA})^*_{\alpha\beta}$. As $(\kappa_j^{AB})_{\alpha\beta}=\pm(\kappa_j^{BA})^*_{\alpha\beta}$, for any fixed $j$, there are 9 real matrices, say ${\rm Re}\{ \kappa_j^{AB} \}$ and ${\rm Im}\{ \kappa_j^{AB} \}$, so the whole set of matrices comprises 45 elements with 16 real entries per matrix. Nonetheless, notice that in the cases $j=1,2$ a further reduction occurs due to symmetry in the space of matrix Lorentz representations. The numbers of parameters resulting from the complex quantities $(\kappa_j^{AB})_\alpha\hspace{0.00001cm}^\beta$, for each $j=1,2,3,4,5$ and after taking these relations into account, are given in Table~\ref{numpars},
\begin{table}[ht]
\center
\begin{tabular}{|c|c|c|c|c|c|}
\hline
Parameters & $(\kappa^{AB}_1)_\alpha\hspace{0.000001cm}^\beta$ & $(\kappa^{AB}_2)_\alpha\hspace{0.000001cm}^\beta$ & $(\kappa^{AB}_3)_\alpha\hspace{0.000001cm}^\beta$ & $(\kappa^{AB}_4)_\alpha\hspace{0.000001cm}^\beta$ & $(\kappa^{AB}_5)_\alpha\hspace{0.000001cm}^\beta$
\\ \hline
Number & 90 & 90 & 144 & 144 & 144
\\ \hline
\end{tabular}
\caption{\label{numpars} The number of parameters associated to Lorentz-violating factors $(\kappa_j^{AB})_\alpha\hspace{0.000001cm}^\beta$ for each $j=1,2,3,4,5$. Keep in mind that coefficients $(\kappa_j^{AB})_\alpha\hspace{0.000001cm}^\beta$ are complex quantities, so the numbers shown in this table take into account both independent real and imaginary parts.}
\end{table} 
from which a total of 612 parameters are counted. However, the Lorentz-violating contributions to AMMs and EDMs under consideration do not carry information on all these parameters. A major reason motivating the definitions given in Eqs.~(\ref{k1AB})-(\ref{k5AB}) is that, as we show below, all Lorentz-violating contributions from the Yukawa sector to AMMs and EDMs emerge as linear combinations of traces ${\rm tr}\,\kappa^{AB}_j=(\kappa^{AB}_j)_\alpha\hspace{0.000001cm}^{\alpha}$, which are the SME quantities to be bounded. 
\\

The total contributions from the Lorentz-violating Yukawa sector, Eq.~(\ref{Yassb}), to $f_A$-lepton magnetic and electric form factors, $f^{m}_{A}$ and $f^{d}_{A}$, are expressed as
\begin{equation}
f^{m}_A=\hat{f}^{m}_A+\sum_{B\ne A}\tilde{f}^{m}_{AB},
\label{AMMtot}
\end{equation}
\begin{equation}
f^{d}_A=\hat{f}^{d}_A+\sum_{B\ne A}\tilde{f}^{d}_{AB},
\label{EDMtot}
\end{equation}
where $\hat{f}^{m}_A$ and $\hat{f}^{d}_A$ are virtual-lepton-flavor conserving, whereas $\tilde{f}^{m}_{AB}$ and $\tilde{f}^{d}_{AB}$, with $A\ne B$, come from Feynman diagrams with virtual-lepton-flavor change. In terms of the kappa notation, defined by Eqs.~(\ref{k1AB})-(\ref{k5AB}) , we write the magnetic- and electric-form-factor contributions from Feynman diagrams that preserve virtual-lepton flavor, that is $A=B$, as $\hat{f}^m_A=-{\rm tr}\big\{\kappa_1^{AA}\,h_{a,1} +\kappa_2^{AA}\,h_{a,2} \big\}$ and $\hat{f}^d_A=-{\rm tr}\big\{ \kappa_3^{AA}\,h_{d,1}+\kappa_4^{AA}\,h_{d,2}+\kappa_5^{AA}\,h_{d,3} \big\}$, where the symbol ``tr'' denotes, as before, a trace operating on $4\times4$ matrices in the space of matrix representations of Lorentz transformations. The explicit expressions for the coefficients $h_{a,1}$, $h_{a,2}$, $h_{d,1}$, $h_{d,2}$, and $h_{d,3}$ are quite large and intricate functions of masses, with no practical use for the reader, so we have not included them in the present paper. UV divergences introduced by each contributing loop diagram lie exclusively within two-point scalar functions $B_0$. Any of such functions can be expressed, after dimensional regularization, as $B_0=\Delta_{\rm UV}+(\text{finite terms})$, where $\Delta_{\rm UV}=1/(4-D)-\gamma_{\rm E}+\log(4\pi/\mu^2)$ diverges as $D\rightarrow4$~\cite{HooVe}. All $B_0$ functions share the same UV-divergent term $\Delta_{\rm UV}$, so any difference of the form $B^j_0-B_0^k$, with $B_0^j$ and $B_0^k$ denoting different two-point scalar functions, is free of UV divergencies. We have been able to establish that all $B_0$ functions in both $\hat{f}^m_A$ and $\hat{f}^d_A$ appear in a difference like this, so we conclude that every UV divergence is cancelled from the contributions with $A=B$. In the next step, we perform derivatives with respect to squared masses, as indicated in Eqs.~(\ref{Gamma1})-(\ref{Gamma3}). Then parametric integrals, also shown in such equations, are carried out, and the on-shell condition $q^2\rightarrow0$, which defines the AMM and EDM contributions, is implemented. The resulting virtual-lepton-flavor-conserving expressions are
\begin{eqnarray}
&&
\hat{a}^{f_A}_A=\frac{e^3v^2}{4\pi^2m_A^2}{\rm tr}
\Big\{
-\kappa_1^{AA}\Big(\Delta_{\rm IR}+\log\frac{\mu^2}{m_A^2} \Big)
\nonumber \\&&
\hspace{0.8cm}
+\frac{3}{8}\big( \kappa_2^{AA}-5\kappa_1^{AA} \big)
\Big\},
\label{AMMd}
\end{eqnarray}
\begin{eqnarray}
&&
\hat{d}^{f_A}_A=\frac{e^3v^2}{64\pi^2m_A^3}{\rm tr}
\Big\{
-\big( \kappa_5^{AA}+8i\kappa_3^{AA}
+15\kappa_4^{AA} \big)
\nonumber \\ &&
\hspace{0.5cm}
\times
\Big( \Delta_{\rm IR}+\log\frac{\mu^2}{m_A^2} \Big)
+\big( 5\kappa_5^{AA}
+12i\kappa_3^{AA}-13\kappa_4^{AA} \big)
\Big\}.
\label{EDMd}
\nonumber \\
\end{eqnarray}
These contributions involve IR divergences, given by the factor $\Delta_{\rm IR}+\log(\mu^2/m_A^2)$, which we omit in what follows as such divergences are expected to eliminate through renormalization at the level of cross section~\cite{PeSch,Schwartzbuch}. For diagrams with virtual-lepton-flavor change, the same steps are followed. The corresponding contributions to AMMs are free of both UV and IR divergences, whereas the resulting EDMs turn out to be IR divergent. Once such IR divergences have been removed, the virtual-lepton-flavor-changing contributions to AMMs and EDMs are expressed as
\begin{widetext}
\begin{eqnarray}
&&
\tilde{a}^{f_A}_B=\frac{e^3v^2}{192\pi^2m_A^6(m_A^2-m_B^2)^2}{\rm tr}
\Big\{
\big(\kappa_1^{AB}-\kappa_2^{AB}\big)\Big( 34m_A^8\log\frac{m_A^2}{m_B^2}+(m_A^2-m_B^2)\Big( -13m_A^4m_B^2+18m_A^2m_B^4
\nonumber \\ &&
+2(-19m_A^4m_B^2+11m_A^2m_B^4+17m_A^6-9m_B^6)\log\frac{m_B^2}{m_B^2-m_A^2}-39m_A^6 \Big) \Big)+24m_A^3m_B\big(\kappa_1^{AB}+\kappa_2^{AB}\big)
\nonumber \\ &&
\times
\Big( m_A^4\log\frac{m_A^2}{m_B^2}
+(m_A^2-m_B^2)\Big( 2(m_A^2-m_B^2)\log\frac{m_B^2}{m_B^2-m_A^2}-m_A^2 \Big) \Big)
\Big\},
\label{AMMnd}
\end{eqnarray}
\begin{eqnarray}
&&
\tilde{d}^{f_A}_B=\frac{e^3v^2}{96\pi^2m_A^6(m_A^2-m_B^2)^2}{\rm tr}
\Big\{
m_A^4m_B^3 \Big( 24(i\kappa_3^{AB}+3\kappa_4^{AB}+3 \kappa_5^{AB})\log\frac{m_B^2}{m_B^2-m_A^2}-(\kappa_4^{AB}+\kappa_5^{AB}) \Big)
\nonumber \\ &&
-8m_A^2m_B^5\Big( 3(2\kappa_5^{AB}+i\kappa_3^{AB}+3\kappa_4^{AB})\log\frac{m_B^2}{m_B^2-m_A^2}+\kappa_4^{AB}+\kappa_5^{AB} \Big)
\nonumber \\ &&
-2m_A^7(\kappa_4^{AB}-\kappa_5^{AB})\Big( 10\log\frac{m_A^2}{m_B^2-m_A^2}-9 \Big)
-m_A^6m_B(\kappa_5^{AB}+\kappa_4^{AB})\Big( 32\log\frac{m_B^2}{m_B^2-m_A^2}+20\log\frac{m_A^2}{m_B^2}-9 \Big)
\nonumber \\ &&
-m_A^3m_B^4(\kappa_4^{AB}-\kappa_5^{AB})\Big( 18\log\frac{m_B^2}{m_B^2-m_A^2}-1 \Big)
+m_A^5m_B^2(\kappa_4^{AB}-\kappa_5^{AB})\Big( 39\log\frac{m_B^2}{m_B^2-m_A^2}-19 \Big)
\nonumber \\ &&
+8m_B^7(\kappa_5^{AB}+\kappa_4^{AB})\log\frac{m_B^2}{m_B^2-m_A^2}
-m_Am_B^6(\kappa_4^{AB}-\kappa_5^{AB})\log\frac{m_B^2}{m_B^2-m_A^2}
\Big\}.
\label{EDMnd}
\end{eqnarray}
\\
\end{widetext}


\section{Estimation of effects and discussion}
\label{numb}
Known to exist since the realization of the Stern-Gerlach experiment~\cite{StGe} and the ulterior theoretical explanation provided by Goudsmit and Uhlenbeck~\cite{UhGo1,UhGo2}, intrinsic magnetic moments of elementary particles, which gave birth to the concept of spin, receive quantum corrections, dubbed AMMs~\cite{Schwinger}. In the cases of the electron and the muon, the corresponding SM predictions have been calculated and estimated with remarkable precision~\cite{AHKNe,AHKNmu}, whereas experimental studies have reached exceptional sensitivity~\cite{muong2,HFG1,HFG2}. The disparity among the SM contribution to the AMM of some fermion $f$, $a_f^{\rm SM}$, and the current best experimental measurement, $a_f^{\rm exp}$, is conventionally characterized by the quantity $\Delta a_f=a_f^{\rm exp}-a_f^{\rm SM}$. For the electron AMM, 
\begin{equation}
\Delta a_e=-1.06(082)\times10^{-12}
\label{dae}
\end{equation}
has been reported~\cite{AHKNe} and
\begin{equation}
\Delta a_\mu=249(87)\times10^{-11}
\label{damu}
\end{equation}
is the current value for the case of the muon~\cite{AHKNmu}. These discrepancies, being so tiny, can be interpreted as suitable places to look for suppressed new physics, beyond the SM. The much-less known tau-lepton AMM was investigated by the authors of Ref.~\cite{GSV}, who performed an analysis of collider data and then established the model-independent limits
\begin{equation}
-0.007<a^{\rm NP}_\tau<0.005
\label{datau}
\end{equation}
on new-physics contributions to this quantity. \\

While theoretically plausible and phenomenologically relevant, the EDMs of elementary particles have not been measured ever, so our best experimental knowledge on the matter consists in bounds. Investigations regarding EDMs of elementary particles have found much motivation in their connection to the phenomenon of $CP$ violation, an essential ingredient behind baryonic assymetry~\cite{Sakharov}. Limits on the electron EDM, $d_e$, are particularly stringent. Experiments with thallium atoms and ytterbium fluoride molecules achieved high sensitivities, thus yielding upper bounds of order $10^{-27}e\cdot {\rm cm}$ on $|d_e|$~\cite{RCSD,HKSTH,KSHSTH}. Furthermore, the ACME Collaboration reported the improved upper limit~\cite{ACMEColl}
\begin{equation}
|d_e|<8.7\times10^{-29}\,e\cdot{\rm cm}, 
\label{deexp}
\end{equation}
at 90\% C. L. The SM prediction lies about 15 orders of magnitude below current experimental sensitivity~\cite{PoRi}, thus rendering the search for new physics presumably originating EDMs a promising task. Three analyses aimed at the observation of the muon EDM were performed and reported in Ref.~\cite{Muong2Coll}, by the Muon $g-2$ Collaboration. This group concluded that the lack of any signal yields the bound~\cite{Muong2Coll}
\begin{equation}
|d_\mu|<1.8\times10^{-19}\,e\cdot{\rm cm},
\label{dmuexp}
\end{equation}
at 95\% C. L. An experimental investigation, carried out by the Belle Collaboration, searched for $CP$ violation induced by the tau-lepton EDM, determining, at 95\% C. L., the limits~\cite{BelleColl}
\begin{equation}
-2.2\times10^{-17}\,e\cdot{\rm cm}<{\rm Re}(d_\tau)<4.5\times10^{-17}\,e\cdot{\rm cm},
\label{dtauexpRe}
\end{equation}
\begin{equation}
-2.5\times10^{-17}\,e\cdot{\rm cm}<{\rm Im}(d_\tau)<0.8\times10^{-17}\,e\cdot{\rm cm}.
\label{dtauexpIm}
\end{equation}
\\

An estimation of bounds on SME coefficients from the Yukawa sector given in Eq.~(\ref{Yassb}) is a main objective of the present section. Being an effective field theory characterized by coefficients with Lorentz indices, this Lagrangian sector bears a large amount of parameters. Aiming at the reduction of the number of such parameters, we consider scenarios defined by specific assumptions on Lorentz-nonconserving coefficients. This matter is addressed below, in a concrete manner. Moreover, the implementation of on-shell conditions on Eqs.~(\ref{AMMtot}) and (\ref{EDMtot}) defines the mSME Yukawa leading contributions to AMMs, as $f^m_A(q^2=0)=a^{\rm SME}_A$, and EDMs, as $f^d_A(q^2=0)=d^{\rm SME}_A$. Then, using Eqs.~(\ref{AMMd})-(\ref{EDMnd}), the resulting electromagnetic-factors contributions can be rearranged as
\begin{equation}
a^{\rm SME}_{A}=\sum_{j=1}^2\sum_{B=e,\mu,\tau}a_j^{AB}\,{\rm tr}\,\kappa_j^{AB},
\label{afA}
\end{equation}
\begin{equation}
d^{\rm SME}_{A}=\sum_{k=3}^5\sum_{B=e,\mu,\tau}d_k^{AB}\,{\rm tr}\,\kappa_k^{AB},
\label{dfA}
\end{equation}
which is convenient in order to determine bounds. Following Ref.~\cite{HMNSTV}, we assume that matrices $V_{\alpha\beta}$ and $A_{\alpha\beta}$ are symmetric in flavor space. Since these matrices are Hermitian and antiHermitian, respectively, $V_{\alpha\beta}$ are real and $A_{\alpha\beta}$ are imaginary. Therefore, according to Eqs.~(\ref{k1AB})-(\ref{k5AB}), traces ${\rm tr}\,\kappa_1^{AB}$, ${\rm tr}\,\kappa_2^{AB}$, ${\rm tr}\,\kappa_4^{AB}$, and ${\rm tr}\,\kappa_5^{AB}$ are real quantities, but ${\rm tr}\,\kappa_3^{AB}$ is imaginary. Furthermore, these assumptions ensure that virtual-lepton-flavor-conserving contributions $\hat{a}^{f_A}_A$ and $\hat{d}^{f_A}_A$, respectively given by Eqs.~(\ref{AMMd}) and (\ref{EDMd}), are real. \\

Consider, in general, some sort of new physics generating contributions to magnetic and/or electric form factors of fermions. The resulting set of magnetic and electric form factors can be classified into~\cite{NPR,BGS} diagonal electromagnetic form factors, in which external fermions coincide with each other, and transition electromagnetic form factors, characterized by different external fermions. If transitions connecting leptons to quarks are forbidden, each of these types of fermions yields nine magnetic moments and nine electric moments, with each set arranged as a $3\times3$ matrix. All such matrices, whose diagonal entries are the diagonal moments and with the transition moments playing the roles of nondiagonal components, are conventionally assumed to be Hermitian, meaning that diagonal moments are real, whereas transition moments might be complex. Notice, however, that working with the vertex $A_\mu f_Af_A$ off shell may introduce thresholds, beyond which imaginary parts of diagonal moments might be induced. It turns out that, even though AMMs and EDMs are on-shell quantities, electromagnetic moments of unstable particles may have imaginary parts. This issue was adressed by the authors of Ref.~\cite{AvKa}, who asseverated that AMMs and EDMs are ensured to be real only as long as calculations are performed in the context of Lorentz-conserving quantum electrodynamics and suggested that {\it ad hoc} definitions of these electromagnetic properties should be given in more general situations. Their discussion included a two-loop calculation, which showed that even the SM produces complex AMMs and EDMs. The emergence of complex electromagnetic moments has been pointed out in Refs.~\cite{GaSi,BiPa} as well. \\

\begin{figure}[ht]
\center
\includegraphics[width=8cm]{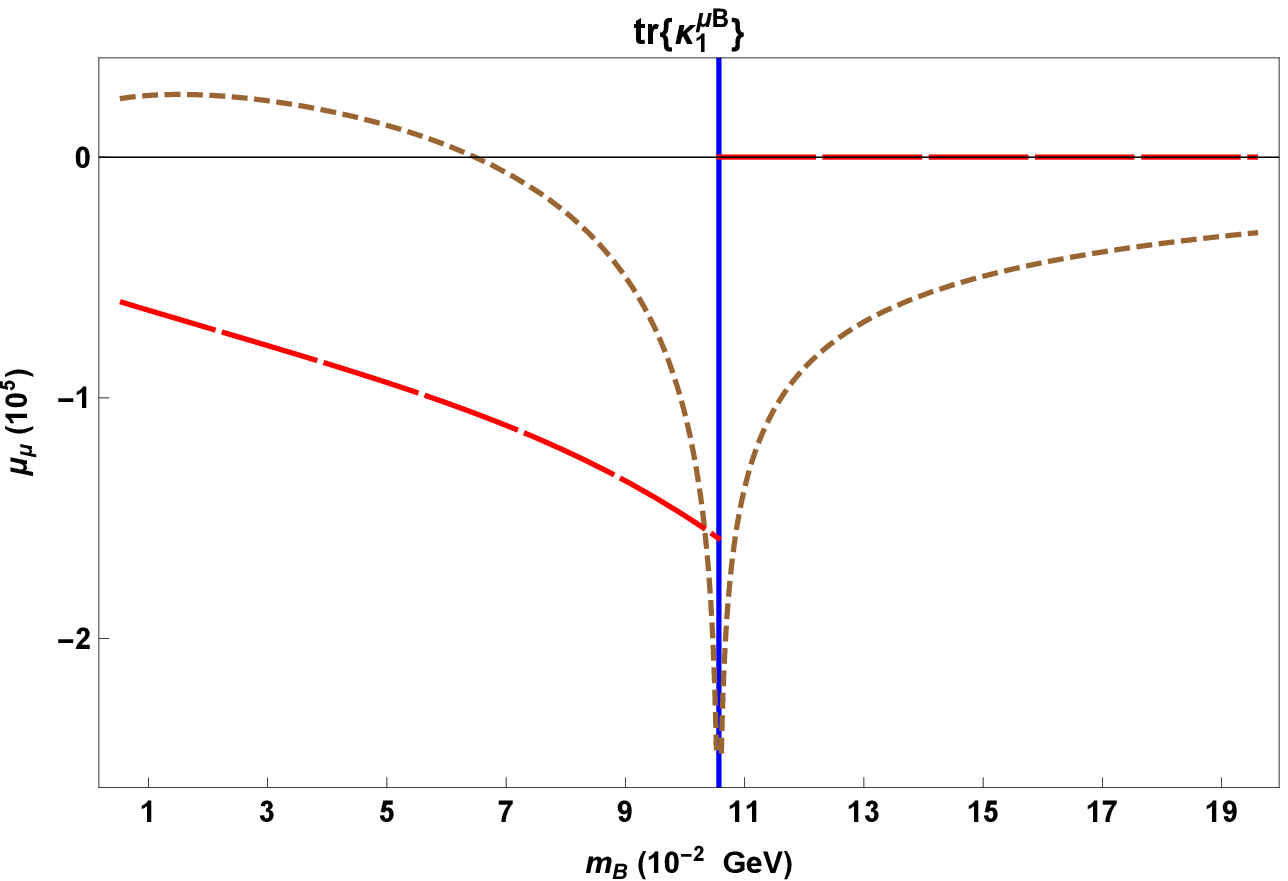}
\vspace{0.2cm}
\\
\includegraphics[width=8cm]{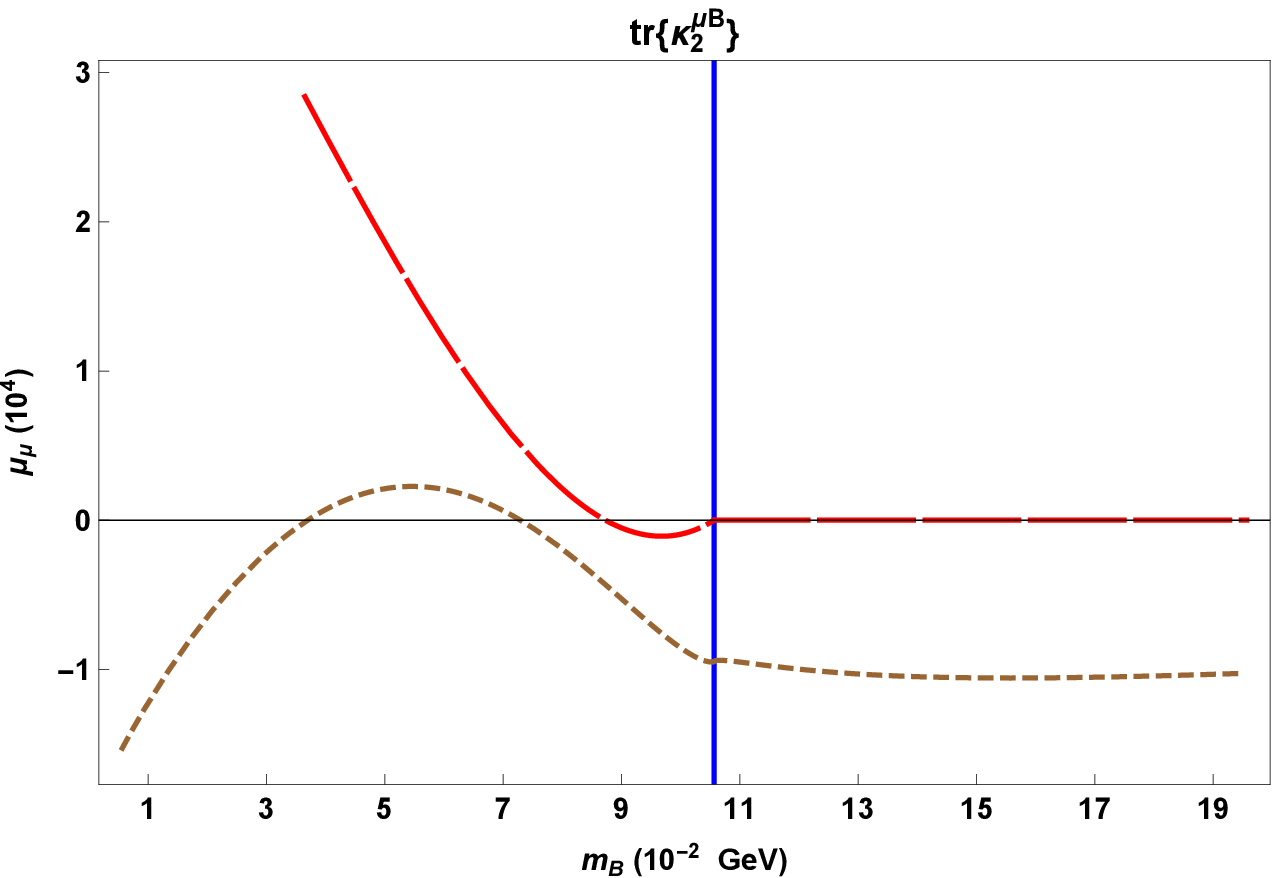}
\caption{\label{AMMtrs} Factors of traces in $a^{f_\mu}$, Eq.~(\ref{afA}), as functions of the virtual-lepton mass $m_B$. The first graph displays $a^{\mu B}_1$, the coefficient of ${\rm tr}\,\kappa^{\mu B}_1$, while $a^{\mu B}_2$, the coefficient of ${\rm tr}\,\kappa^{\mu B}_2$, is shown in the second graph. Short dashed plots stand for real parts of the factors, whereas long-dashed curves display imaginary parts of them. Vertical solid lines indicate where the threshold value $m_B=m_\mu$ lies.}
\end{figure}

\begin{figure}[ht]
\center
\includegraphics[width=8cm]{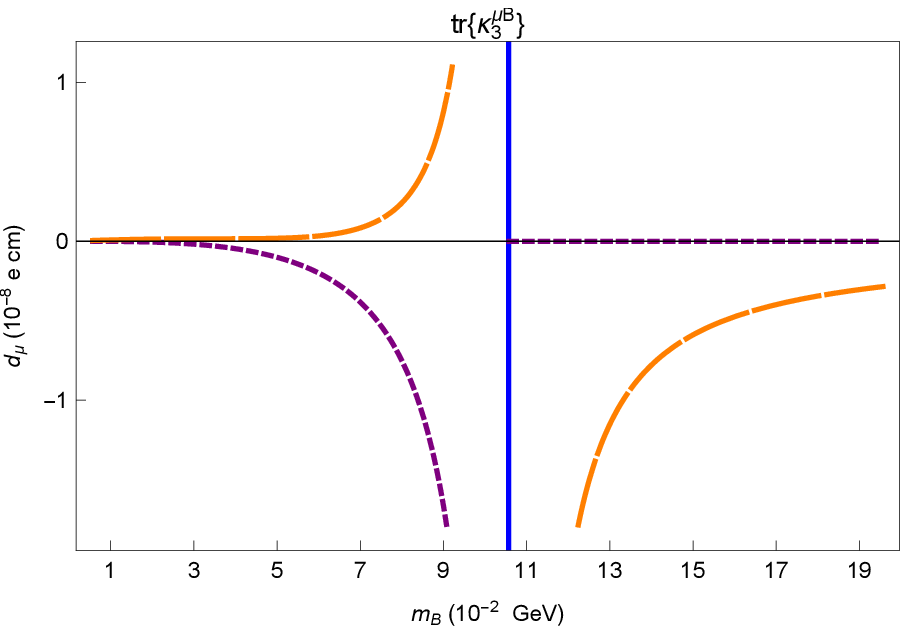}
\vspace{0.2cm}
\\
\includegraphics[width=8cm]{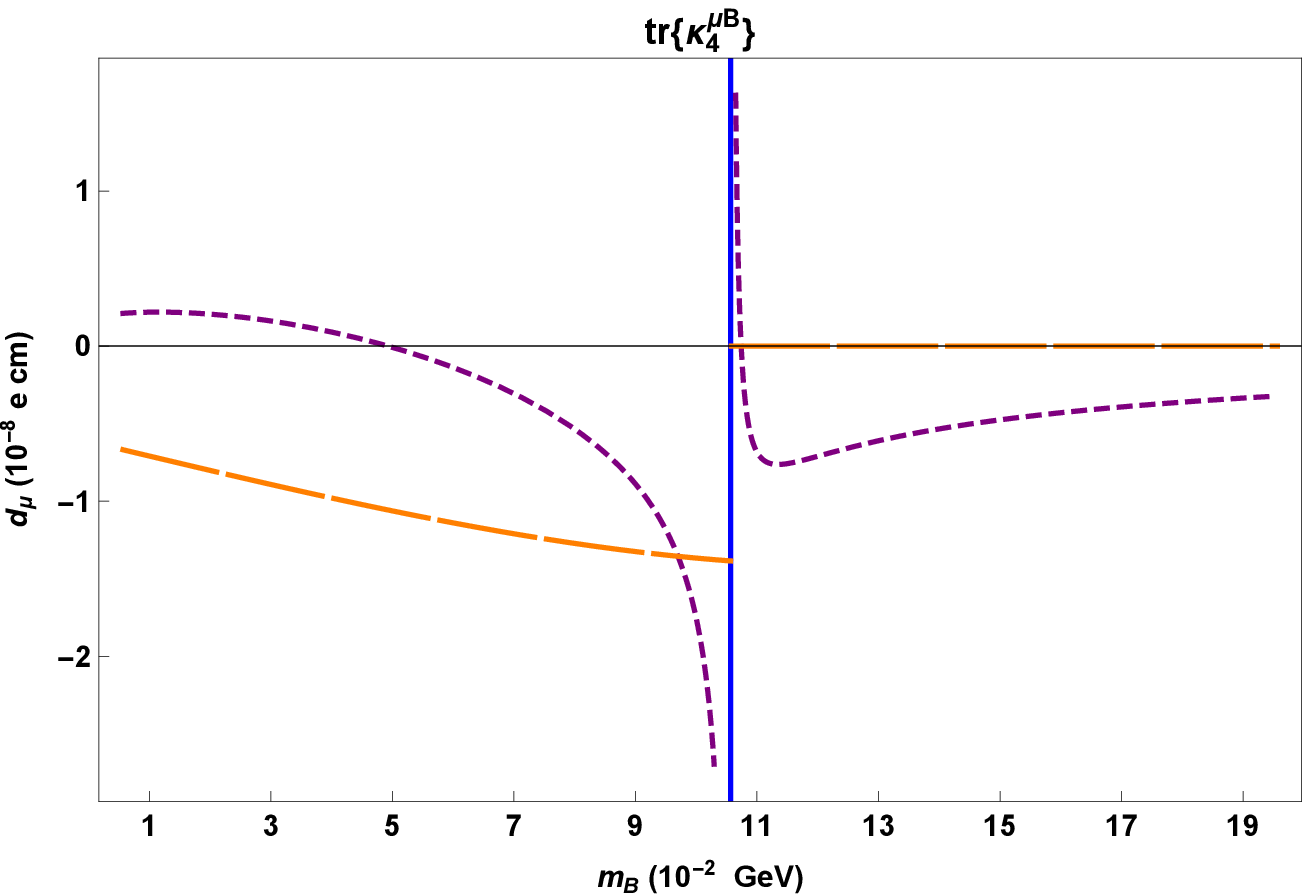}
\vspace{0.2cm}
\\
\includegraphics[width=8cm]{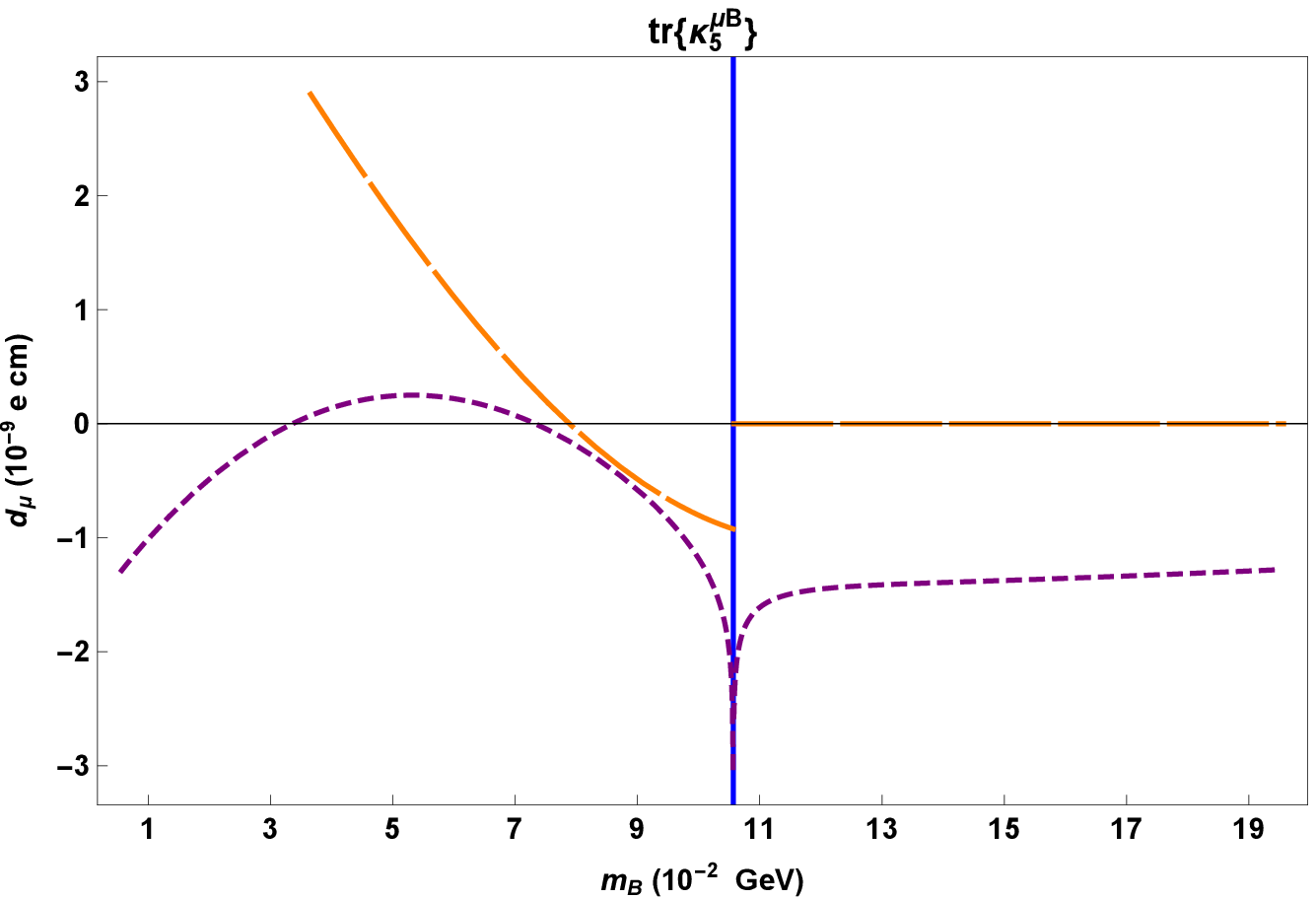}
\caption{\label{EDMtrs} Factors of traces in $d^{f_\mu}$, Eq.~(\ref{afA}), as functions of the virtual-lepton mass $m_B$. The first graph displays $d^{\mu B}_3$, the coefficient of ${\rm tr}\,\kappa^{\mu B}_3$, while $d^{\mu B}_4$, the coefficient of ${\rm tr}\,\kappa^{\mu B}_4$, is shown in the second graph, and $d^{\mu B}_5$, the coefficient of ${\rm tr}\,\kappa^{\mu B}_5$, is depicted by the third graph. Short dashed plots stand for real parts of the factors, whereas long-dashed curves display imaginary parts of them. Vertical solid lines indicate where the threshold value $m_B=m_\mu$ lies.}
\end{figure}

From the explicit expressions provided in Eqs.~(\ref{AMMnd})-(\ref{EDMnd}), notice that Lorentz-nonconserving contributions to charged-lepton AMMs and EDMs are complex, even though these electromagnetic moments are not transition-like, but diagonal moments instead, and even though they have been calculated on shell. In a general context, analogous imaginary parts of one-loop amplitudes, if present, usually emerge when some external field is connected to loop lines corresponding to particles which together are lighter than the external particle. In the case of the Lorentz-violating theory considered in the present investigation, the insertion of bilinear vertices connecting some external-field line to a lighter virtual-field line produces an alike effect, which can be appreciated in Eqs.~(\ref{AMMnd}) and (\ref{EDMnd}), where logarithmic factors $\log\frac{m^2_B}{m_B^2-m_A^2}$ are real or imaginary depending on whether $m_B>m_A$ or $m_A>m_B$ holds. This is illustrated by the graphs in Figs.~\ref{AMMtrs}-\ref{EDMtrs}, 
which display the behavior of real and imaginary parts of coefficients $a^{AB}_j$ and $d^{AB}_k$, defined by Eqs.~(\ref{afA}) and (\ref{dfA}), as functions of the virtual-lepton mass $m_B$, for the case $A=\mu$, corresponding to external muons. The upper graph of Fig.~\ref{AMMtrs} corresponds to $a^{\mu B}_1$ whereas the lower graph of this figure respresents $a^{\mu B}_2$, both quantities being factors within the muon AMM contribution. In these graphs, short-dashed curves represent the real parts of coefficients $a^{\mu B}_j$, while long-dashed plots depict imaginary parts of such quantities. Moreover, horizontal solid lines represent the values $a^{\mu B}_j=0$. Vertical solid lines, located at value of the muon mass $m_\mu$, correspond to a $m_B$ threshold, beyond which factors $a^{\mu B}_j$ are either complex or real quantities. Both graphs make it evident that values of $l_B$ mass such that $m_B<m_\mu$ yield imaginary-part contributions, but these contributions are real as long as $m_B>m_\mu$. Just as in Fig.~\ref{AMMtrs}, short-dashed plots represent real parts of factors $d^{\mu B}_k$ in all the graphs of Fig.~\ref{EDMtrs}, whereas long-dashed curves are associated to their imaginary parts. Factors $d^{\mu B}_4$ and $d^{\mu B}_5$, whose real and imaginary parts, as functions of $m_B$, are displayed in the second and third graphs of Fig~\ref{EDMtrs}, behave as the AMM factors $a^{\mu B}_j$. In the case of $d^{\mu B}_3$, shown in the first graph of Fig.~\ref{EDMtrs}, roles are inverted, as its real part vanishes for $m_B>m_\mu$ but its imaginary part remains nonzero. This is correct, since the trace ${\rm tr}\,\kappa^{\mu B}_3$ is purely imaginary, so a global imaginary factor from this trace should get things right. We have made sure that analogous behaviors occur if external leptons $f_e$ or $f_\tau$ are considered.\\

Now consider the $3\times3$ matrices
\begin{equation}
\mathscr{X}
_j
=
\left(
\begin{array}{ccc}
\mathscr{X}_j^{ee} & \mathscr{X}_j^{e\mu} & \mathscr{X}_j^{e\tau}
\vspace{0.1cm}
\\
\mathscr{X}_j^{\mu e} & \mathscr{X}_j^{\mu\mu} & \mathscr{X}_j^{\mu\tau}
\vspace{0.1cm}
\\
\mathscr{X}_j^{\tau e} & \mathscr{X}_j^{\tau\mu} & \mathscr{X}_j^{\tau\tau}
\end{array}
\right)
\equiv
\left(
\begin{array}{ccc}
{\rm tr}\,\kappa^{ee}_j & {\rm tr}\,\kappa^{e\mu}_j & {\rm tr}\,\kappa^{e\tau}_j
\vspace{0.1cm}
\\
{\rm tr}\,\kappa^{\mu e}_j & {\rm tr}\,\kappa^{\mu\mu}_j & {\rm tr}\,\kappa^{\mu\tau}_j
\vspace{0.1cm}
\\
{\rm tr}\,\kappa^{\tau e}_j & {\rm tr}\,\kappa^{\tau\mu}_j & {\rm tr}\,\kappa^{\tau\tau}_j
\end{array}
\right).
\end{equation}
Keep in mind that these matrices are not, by any means, related to transition electromagnetic moments, in which external fermions have different flavors. They exclusively correspond to diagonal electromagnetic moments, and rather characterize the terms, of such quantities, in which virtual-lepton flavor is preserved or changed. For instance, from Eq.~(\ref{afA}), the contribution from Lorentz violation to the tau AMM is expressed as $a^{\rm SME}_{\tau}=\sum_{j=1}^2\big( a_j^{\tau e}\,\mathscr{X}_j^{\tau e}+a_j^{\tau\mu}\,\mathscr{X}_j^{\tau\mu}+a_j^{\tau\tau}\,\mathscr{X}_j^{\tau\tau} \big)$. Then notice that the third rows of matrices $\mathscr{X}_1$ and $\mathscr{X}_2$ comprise all the SME traces $\mathscr{X}_j^{\tau B}={\rm tr}\,\kappa_j^{\tau B}$ necessary to determine this AMM contribution. 
In general, $\mathscr{X}_j^{\dag}=\mathscr{X}_j$ holds for $j=1,2,4,5$, while $\mathscr{X}_3^{\dag}=-\mathscr{X}_3$, so not all the traces defining the entries of matrices $\mathscr{X}_j$ are independent. Moreover, our previous assumption that $V^{AB}_{\alpha\beta}=V^{BA}_{\alpha\beta}$ and $A^{AB}_{\alpha\beta}=A^{BA}_{\alpha\beta}$ ensures that $\mathscr{X}_1$, $\mathscr{X}_2$, $\mathscr{X}_4$, and $\mathscr{X}_5$ are symmetric and real, whereas $\mathscr{X}_3$ is symmetric and imaginary. Thus, each matrix $\mathscr{X}_j$ is determined by 6 independent parameters $\mathscr{X}^{AB}_j$, yielding a total of 30 independent parameters. Since AMM contributions are given exclusively in terms of $\mathscr{X}_1$ and $\mathscr{X}_2$, these quantities are determined by 12 real traces, whereas EDMs, expressed as combinations of traces in $\mathscr{X}_3$, $\mathscr{X}_4$, and $\mathscr{X}_5$, involve 18 independent traces, in total. With these definitions at hand, we now consider scenarios, distinguished by the textures of $\mathscr{X}_j$.


\subsection{Quasi-diagonal textures}
The scenario of quasi-diagonal textures is defined by the assumption that the diagonal entries of $\mathscr{X}_j$ are by far dominant, whereas off-diagonal components of such matrices practically equal zero, namely $\mathscr{X}_j^{AB}\approx0$ for $A\ne B$. Then $\mathscr{X}_j$ matrices look like
\begin{equation}
\mathscr{X}_j\approx
\left(
\begin{array}{ccc}
\mathscr{X}^{e}_j & 0 & 0
\\
0 & \mathscr{X}^{\mu}_j & 0
\\
0 & 0 & \mathscr{X}^{\tau}_j
\end{array}
\right),
\end{equation}
where we have denoted $\mathscr{X}^{AA}_j=\mathscr{X}^A_j$. So, Eqs.~(\ref{afA}) and (\ref{dfA}) are expressed as
\begin{eqnarray}
a_{A}^{\rm SME}&\approx&a_1^{AA}\,\mathscr{X}^{A}_1+a_2^{AA}\,\mathscr{X}^{A}_2,
\label{diagtextAMM}
\\ \nonumber \\
d_{A}^{\rm SME}&\approx&d_3^{AA}\,\mathscr{X}^{A}_3+d_4^{AA}\,\mathscr{X}^{A}_4+d_5^{AA}\,\mathscr{X}^{A}_5,
\label{diagtextEDM}
\end{eqnarray}
where repeated flavor indices do not indicate sums. \\

\begin{figure}[ht]
\center
\includegraphics[width=5.35cm]{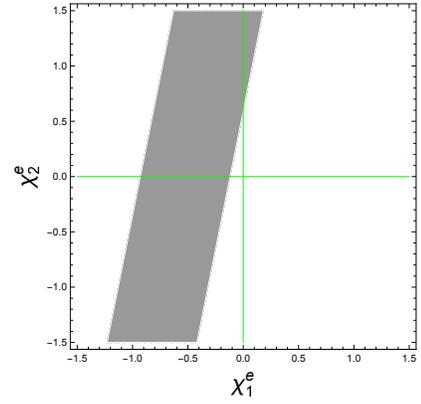}
\caption{\label{diagtexelectron} For quasi-degenerate texture, allowed region within the parameter space $\big(\mathscr{X}^{e}_1,\mathscr{X}^{e}_2\big)\times10^{21}$, for $|\mathscr{X}^{e}_1|<1.5\times10^{-21}$ and $|\mathscr{X}^{e}_2|<1.5\times10^{-21}$, in accordance with the AMM constraint displayed in Eq.~(\ref{dae}).}
\end{figure}

\begin{figure}[ht]
\center
\includegraphics[width=5.35cm]{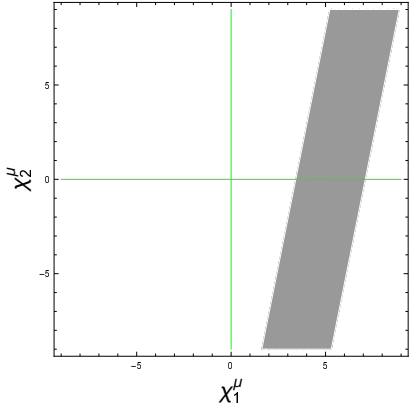}
\caption{\label{diagtexmuon} For quasi-degenerate texture, allowed region within the parameter space $\big(\mathscr{X}^{\mu}_1,\mathscr{X}^{\mu}_2\big)\times10^{14}$, for $|\mathscr{X}^{\mu}_1|<9\times10^{-14}$ and $|\mathscr{X}^{\mu}_2|<9\times10^{-14}$, in accordance with the AMM constraint displayed in Eq.~(\ref{damu}).}
\end{figure}

\begin{figure}[ht]
\center
\includegraphics[width=5.35cm]{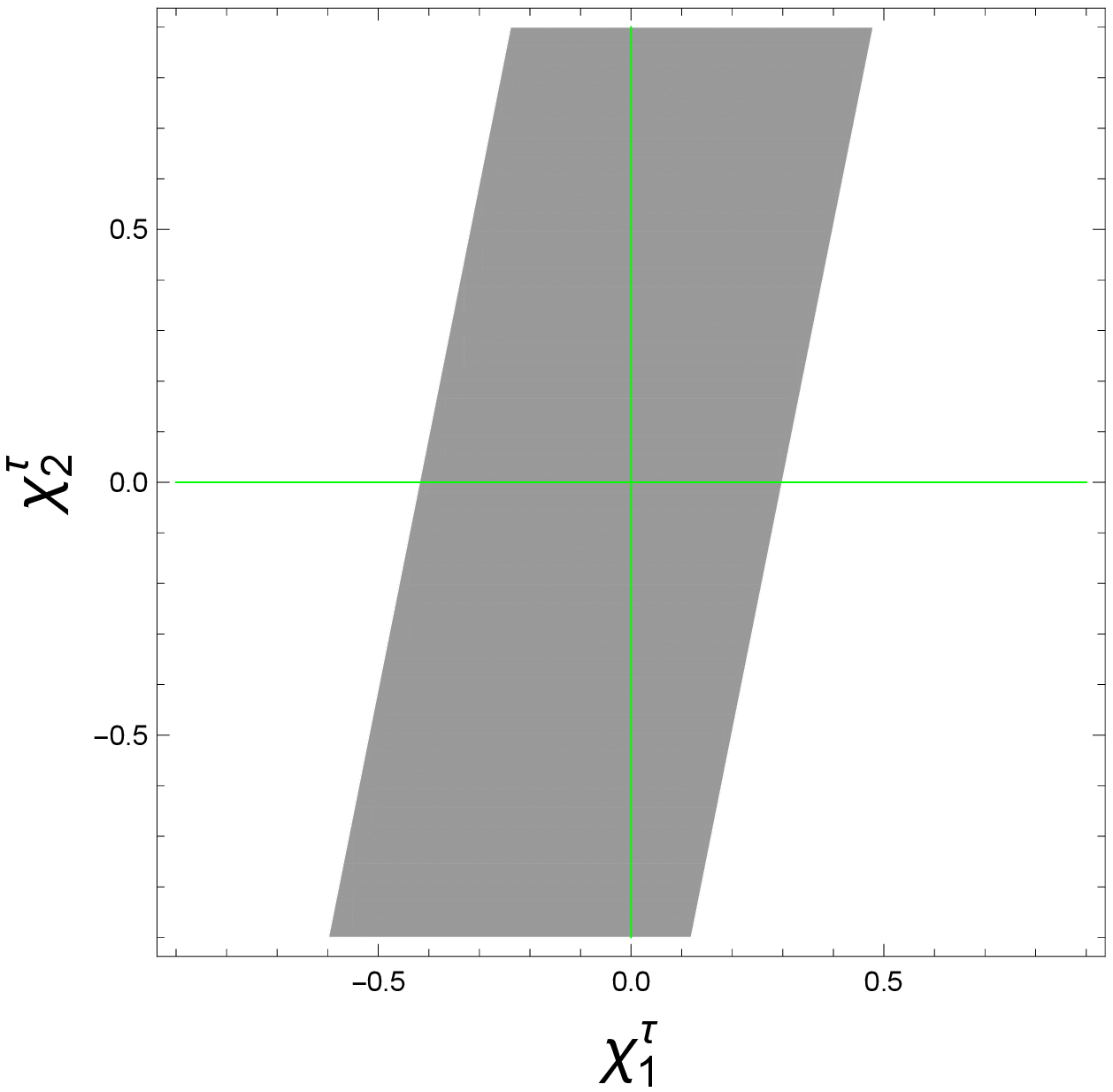}
\caption{\label{diagtextau} For quasi-degenerate texture, allowed region within the parameter space $\big(\mathscr{X}^{\tau}_1,\mathscr{X}^{\tau}_2\big)\times10^{4}$, for $|\mathscr{X}^{\tau}_1|<1.5\times10^{-4}$ and $|\mathscr{X}^{\tau}_2|<1.5\times10^{-4}$, in accordance with the AMM constraint displayed in Eq.~(\ref{datau}).}
\end{figure}

As shown by Eq.~(\ref{diagtextAMM}), the assumption of quasi-degenerate textures yields, for each lepton flavor $A$, a mSME contribution to AMM $a^{\rm SME}_A$ determined by only two parameters. We provide Figs.~\ref{diagtexelectron}-\ref{diagtextau}, which show parameter regions in $\big(\mathscr{X}^{A}_1,\mathscr{X}^{A}_2\big)$ spaces, allowed by current constraints from beyond-SM physics on AMM, given in Eqs.~(\ref{dae})-(\ref{datau}). Fig.~\ref{diagtexelectron} displays the allowed region for the case of SME contributions to the electron AMM, within $|\mathscr{X}^{e}_1|<1.5\times10^{-21}$ and $|\mathscr{X}^{e}_2|<1.5\times10^{-21}$. 
Aiming at a cleaner representation, the graph has been suitably rescaled by a factor $10^{21}$. An examination of this figure makes it clear that the Lorentz-violation coefficient $\mathscr{X}_1^{e}$ is more restricted than $\mathscr{X}_2^{e}$. For any value of $\mathscr{X}_2^{e}$, the trace $\mathscr{X}_1^{e}$ lies within a narrow interval of width $\approx8.1\times10^{-22}$. Furthermore, as long as Lorentz-violation traces of order $\lesssim10^{-21}$ are assumed, the value of $\mathscr{X}_1^{e}$ is more likely to be negative. Similar explanations for Figs.~\ref{diagtexmuon} and \ref{diagtextau} apply, but with the corresponding graphs respectively rescaled by factors $10^{14}$ and $10^{4}$. From Fig.~\ref{diagtexmuon}, $\mathscr{X}^{\mu}_1>0$ if muon-flavor traces of Lorentz-violating matrices are of order $\lesssim10^{-14}$. 
To better illustrate the values of the parameters $\mathscr{X}_1^{A}$ and $\mathscr{X}_2^{A}$ considered for the realization of Figs.~\ref{diagtexelectron}-\ref{diagtextau}, we refer the reader to Tables~\ref{ditexfle}-\ref{ditexfltau},
\begin{table}[ht]
\begin{tabular}{|c|c|c|}
\hline
 & Bottom & Top
\\ \hline
$\mathscr{X}^{e}_2$ fixed & $-15\times10^{-22}$ & $+15\times10^{-22}$
\\
$\Rightarrow\mathscr{X}^{e}_1$ & $-8.25(405)\times10^{-22}$ & $-2.25(405)\times10^{-22}$
\\ \hline
$\mathscr{X}^{e}_5$ fixed & $-10^{-26}$ & $+10^{-26}$
\\
$\Rightarrow\mathscr{X}^{e}_4$ & $-3.76(503)\times10^{-28}$ & $+3.76(503)\times10^{-28}$
\\ \hline
$-i\mathscr{X}^{e}_3$ fixed & $-10^{-26}$ & $+10^{-26}$
\\
$\Rightarrow\mathscr{X}^{e}_5$ & $+9.60(536)\times10^{-28}$ & $-9.60(536)\times10^{-28}$
\\ \hline
$\mathscr{X}^{e}_4$ fixed & $-10^{-26}$ & $+10^{-26}$
\\
$\Rightarrow-i\mathscr{X}^{e}_3$ & $-3.42(176)\times10^{-27}$ & $+3.42(176)\times10^{-27}$
\\ \hline
\end{tabular}
\caption{\label{ditexfle} Values of Lorentz-violation parameters $\mathscr{X}_j^e$ from electron AMM (1st row) and EDM (2nd to 4th rows) constraints. Fixed parameters $\mathscr{X}^e_j$, in each row, correspond to either the bottom (2nd column) or top (3rd column) values of a graph in Figs.~\ref{diagtexelectron} or \ref{eEDMdiagtexc1EDM}. In the case of EDM-allowed regions, the values $\delta^e_{34},\delta^e_{45},\delta^e_{53}=10$ have been used.}
\end{table}
\begin{table}[ht]
\begin{tabular}{|c|c|c|}
\hline
 & Bottom & Top
\\ \hline
$\mathscr{X}^{\mu}_2$ fixed & $-9\times10^{-14}$ & $+9\times10^{-14}$
\\
$\Rightarrow\mathscr{X}^{\mu}_1$ & $3.46(184)\times10^{-14}$ & $7.06(184)\times10^{-14}$
\\ \hline
$\mathscr{X}^\mu_5$ fixed & $-5\times10^{-10}$ & $+5\times10^{-10}$
\\
$\Rightarrow\mathscr{X}^\mu_4$ & $-7.52(972)\times10^{-12}$ & $+7.52(972)\times10^{-12}$
\\ \hline
$-i\mathscr{X}^\mu_3$ fixed & $-5\times10^{-10}$ & $+5\times10^{-10}$
\\
$\Rightarrow\mathscr{X}^\mu_5$ & $+1.92(103)\times10^{-11}$ & $-1.92(103)\times10^{-11}$
\\ \hline
$\mathscr{X}^\mu_4$ fixed & $-5\times10^{-10}$ & $+5\times10^{-10}$
\\ 
$\Rightarrow-i\mathscr{X}^\mu_3$ & $-6.84(340)\times10^{-11}$ & $+6.84(340)\times10^{-11}$
\\ \hline
\end{tabular}
\caption{\label{ditexflmu} Values of Lorentz-violation parameters $\mathscr{X}_j^\mu$ from muon AMM (1st row) and EDM (2nd to 4th rows) constraints. Fixed parameters $\mathscr{X}^\mu_j$, in each row, correspond to either the bottom (2nd column) or top (3rd column) values of a graph in Figs.~\ref{diagtexmuon} or \ref{muEDMdiagtexc1EDM}. In the case of EDM-allowed regions, the values $\delta^\mu_{34},\delta^\mu_{45},\delta^\mu_{53}=10$ have been used.}
\end{table}
\begin{table}[ht]
\begin{tabular}{|c|c|c|}
\hline
 & Bottom & Top
\\ \hline
$\mathscr{X}^{\tau}_2$ fixed & $-9\times10^{-5}$ & $+9\times10^{-5}$
\\
$\Rightarrow\mathscr{X}^{\tau}_1$ & $-2.39(358)\times10^{-5}$ & $1.20(358)\times10^{-5}$
\\ \hline
$\mathscr{X}^\tau_5$ fixed & $-1.5\times10^{-4}$ & $+1.5\times10^{-4}$
\\
$\Rightarrow\mathscr{X}^\tau_4$ & $-2.84(814)\times10^{-6}$ & $+8.44(814)\times10^{-6}$
\\ \hline
$-i\mathscr{X}^\tau_3$ fixed & $-1.5\times10^{-4}$ & $+1.5\times10^{-4}$
\\
$\Rightarrow\mathscr{X}^\tau_5$ & $+1.73(86)\times10^{-5}$ & $-1.14(86)\times10^{-5}$
\\ \hline
$\mathscr{X}^\tau_4$ fixed & $-1.5\times10^{-4}$ & $+1.5\times10^{-4}$
\\
$\Rightarrow-i\mathscr{X}^\tau_3$ & $-6.11(285)\times10^{-5}$ & $+4.15(285)\times10^{-5}$
\\ \hline
\end{tabular}
\caption{\label{ditexfltau} Values of Lorentz-violation parameters $\mathscr{X}_j^\tau$ from tau-lepton AMM (1st row) and EDM (2nd to 4th rows) constraints. Fixed parameters $\mathscr{X}^\tau_j$, in each row, correspond to either the bottom (2nd column) or top (3rd column) values of a graph in Figs.~\ref{diagtextau} or \ref{tauEDMdiagtexc1EDM}. In the case of EDM-allowed regions, the values $\delta^\tau_{34},\delta^\tau_{45},\delta^\tau_{53}=10$ have been used.}
\end{table}
where Table~\ref{ditexfle} carries data for lepton flavor $A=e$,  whereas Tables~\ref{ditexflmu} and Table~\ref{ditexfltau} do it for the cases $A=\mu$ and $A=\tau$, respectively. The first rows of these tables display the minimum (``Bottom'' column) and maximum (``Top'' column) $\mathscr{X}^A_2$ values considered for the graphs of Figs.~\ref{diagtexelectron}-\ref{diagtextau}. The same rows include intervals of $\mathscr{X}^A_1$ values within allowed regions in the graphs, determined by the fixed upper and lower values of parameters $\mathscr{X}^A_2$. \\

\begin{figure}[ht]
\center
\includegraphics[width=5.35cm]{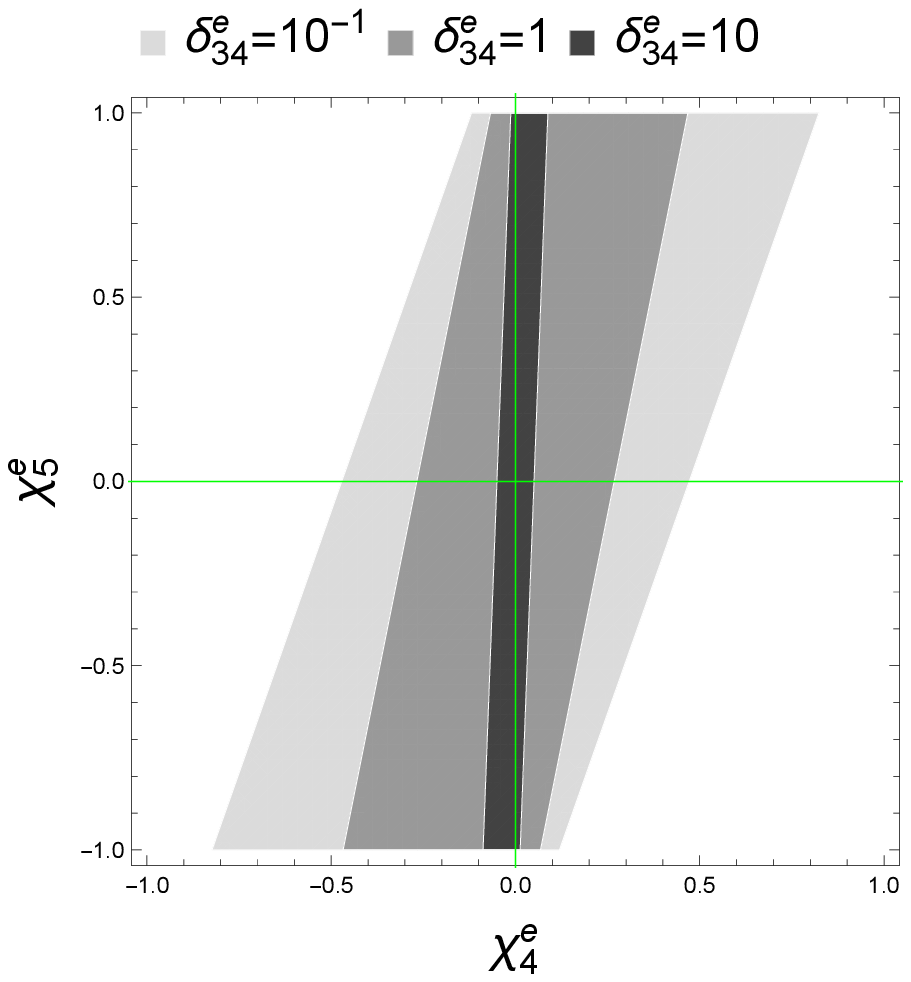}
\\
\vspace{0.2cm}
\includegraphics[width=5.35cm]{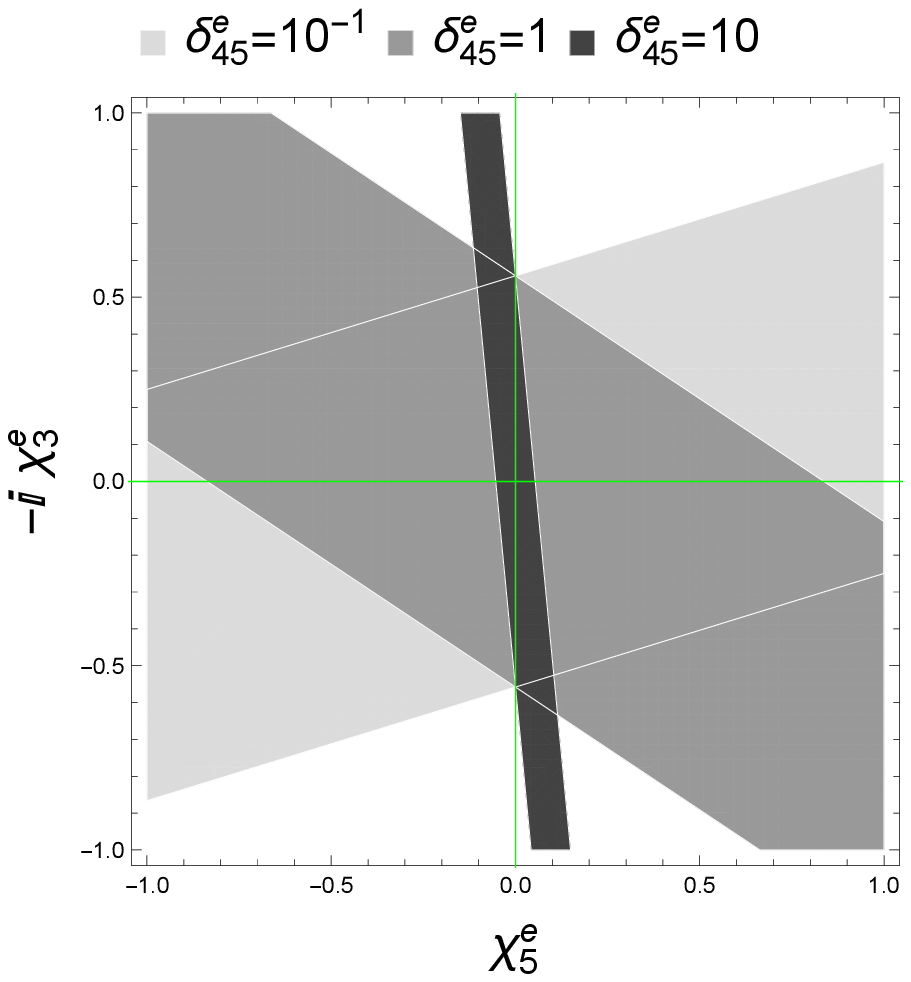}
\\
\vspace{0.2cm}
\includegraphics[width=5.35cm]{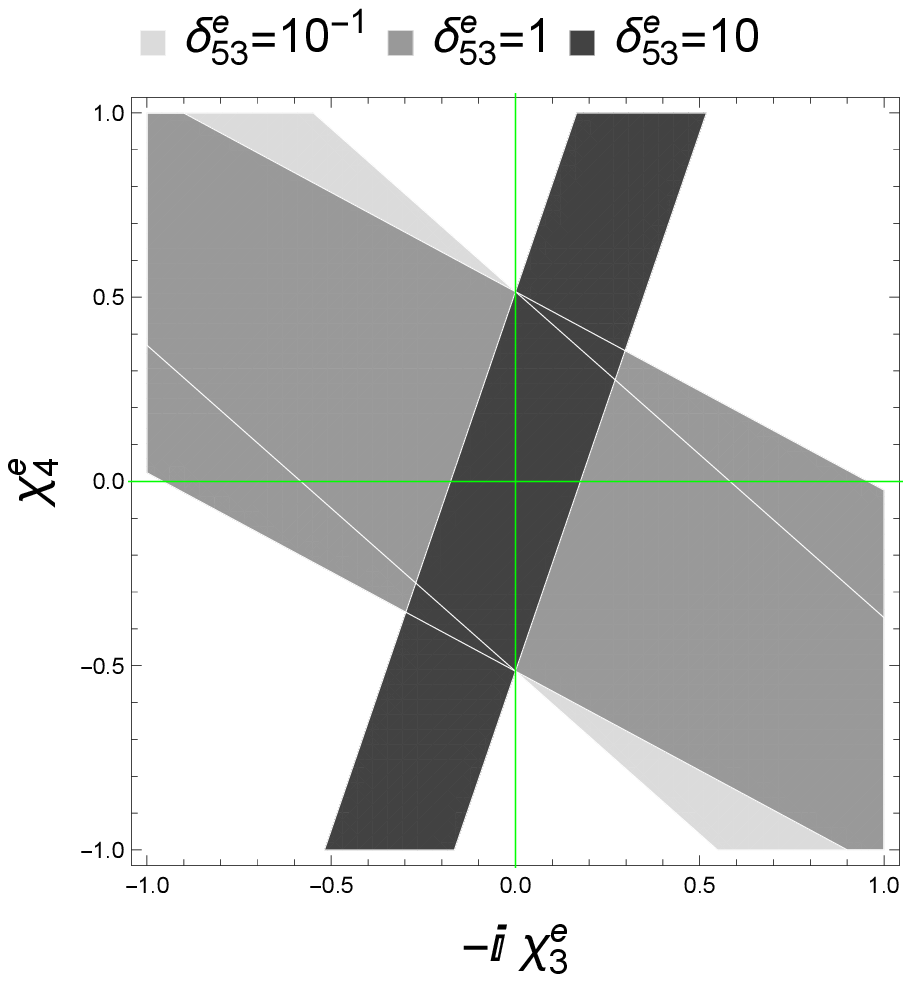}
\caption{\label{eEDMdiagtexc1EDM} Allowed regions, in scale $10^{-26}$, for mSME coefficients from electron EDM. {\it 1st graph}: Parameter space $(\mathscr{X}^e_4,\mathscr{X}^e_5)$ for fixed $\delta^e_{34}$ values. {\it 2nd graph}: Parameter space $(\mathscr{X}^e_5,-i\mathscr{X}^e_3)$ for fixed $\delta^e_{45}$ . {\it 3rd graph}: Parameter space $(-i\mathscr{X}^e_3,\mathscr{X}^e_4)$ for fixed $\delta^e_{34}$ values.}
\end{figure}

\begin{figure}[ht]
\center
\includegraphics[width=5.35cm]{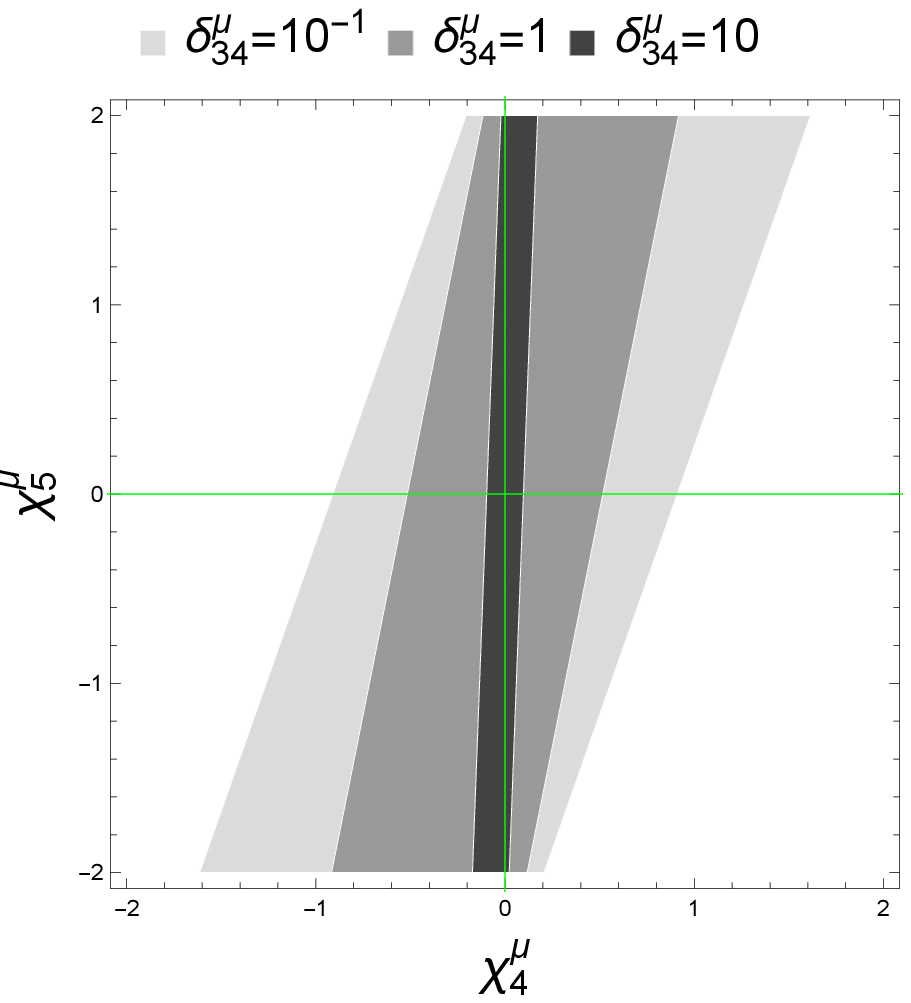}
\\
\vspace{0.2cm}
\includegraphics[width=5.35cm]{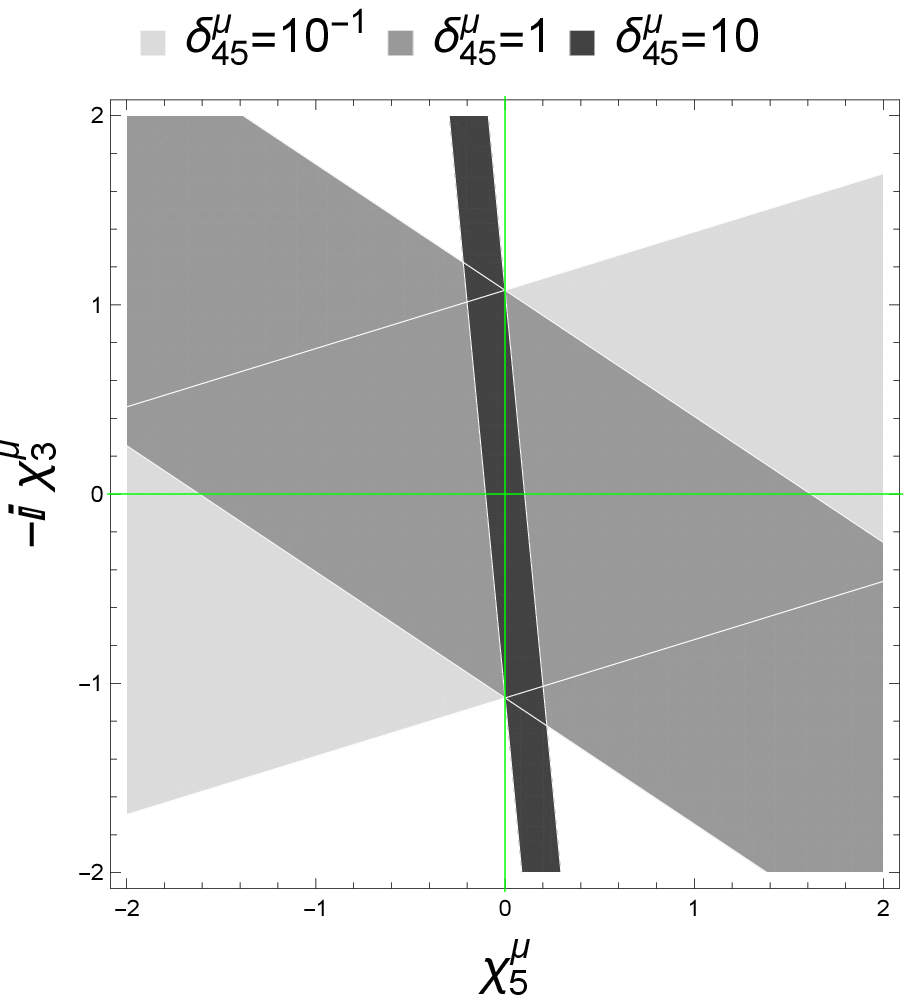}
\\
\vspace{0.2cm}
\includegraphics[width=5.35cm]{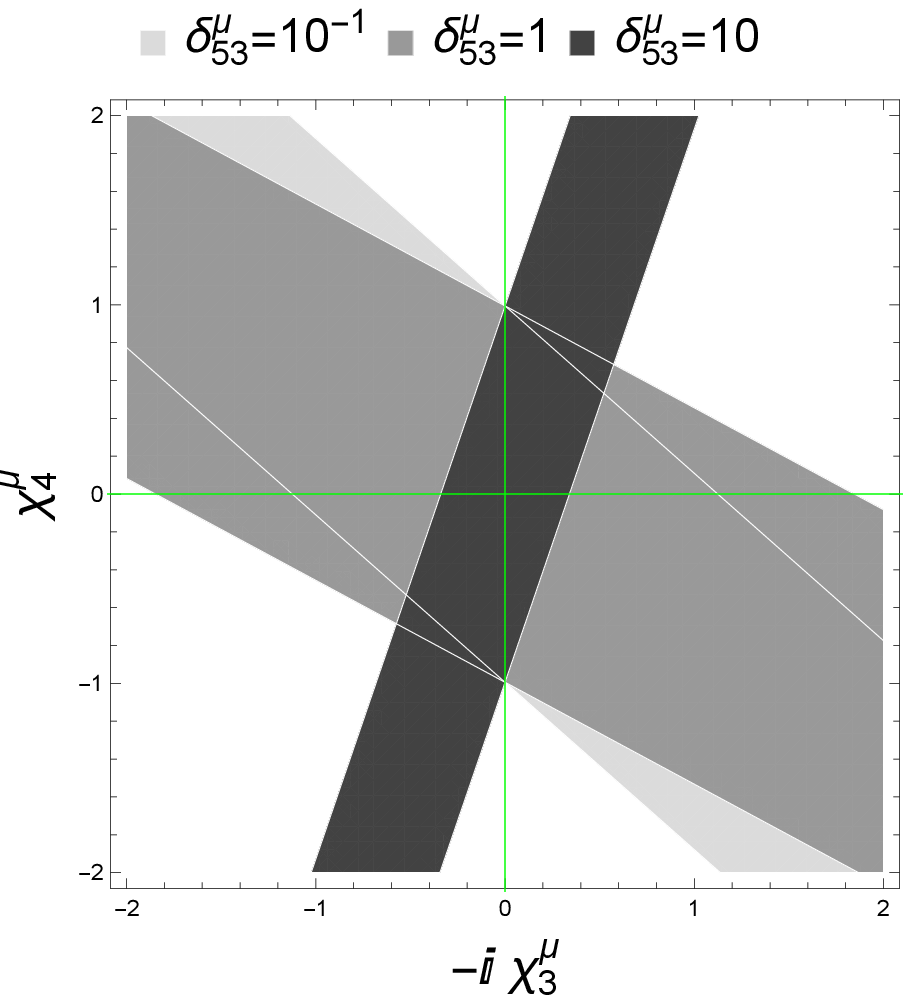}
\caption{\label{muEDMdiagtexc1EDM} Allowed regions, in scale $10^{-10}$, for mSME coefficients from muon EDM. {\it 1st graph}: Parameter space $(\mathscr{X}^\mu_4,\mathscr{X}^\mu_5)$ for fixed $\delta^\mu_{34}$ values. {\it 2nd graph}: Parameter space $(\mathscr{X}^\mu_5,-i\mathscr{X}^\mu_3)$ for fixed $\delta^\mu_{45}$ . {\it 3rd graph}: Parameter space $(-i\mathscr{X}^\mu_3,\mathscr{X}^\mu_4)$ for fixed $\delta^\mu_{34}$ values.}
\end{figure}

\begin{figure}[ht]
\center
\includegraphics[width=5.35cm]{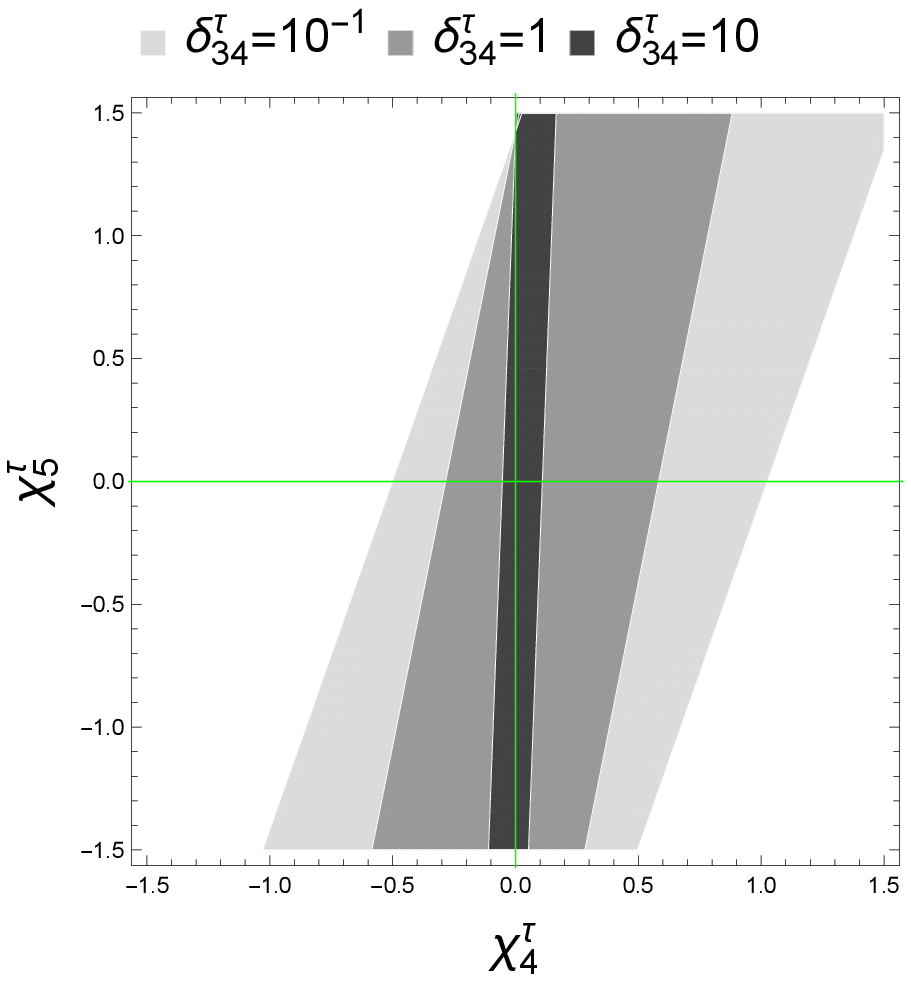}
\\
\vspace{0.2cm}
\includegraphics[width=5.35cm]{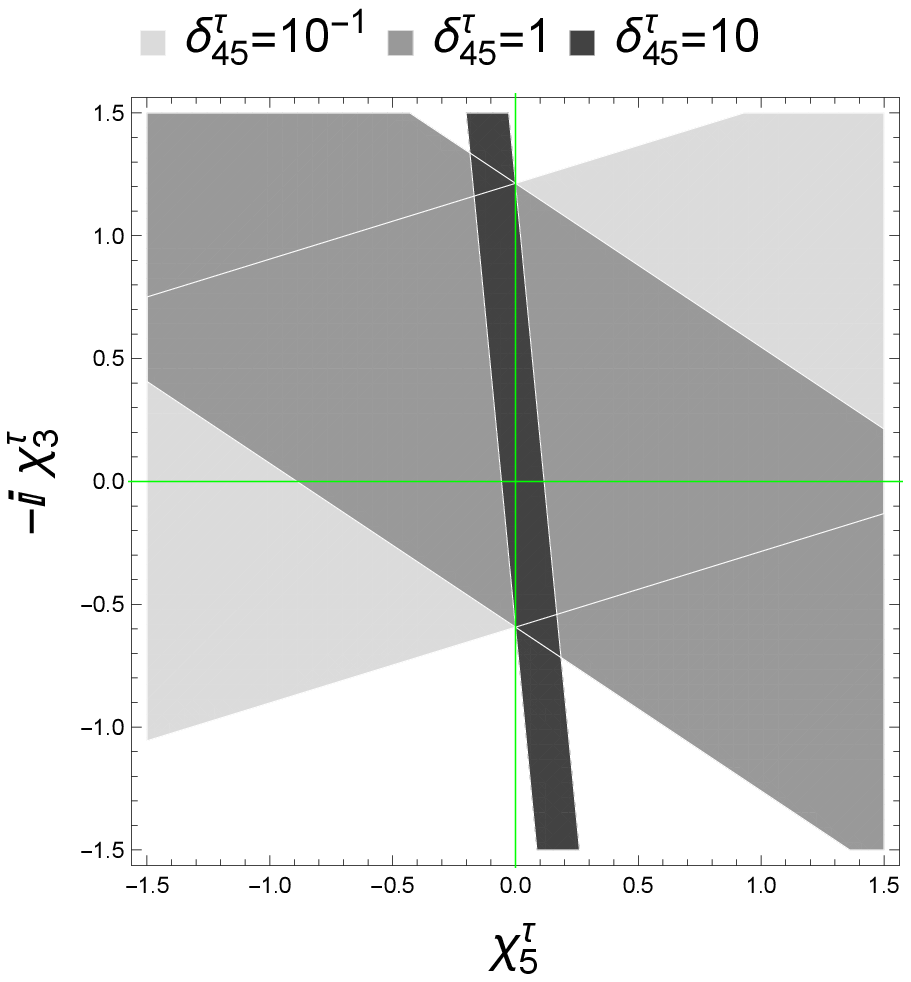}
\\
\vspace{0.2cm}
\includegraphics[width=5.35cm]{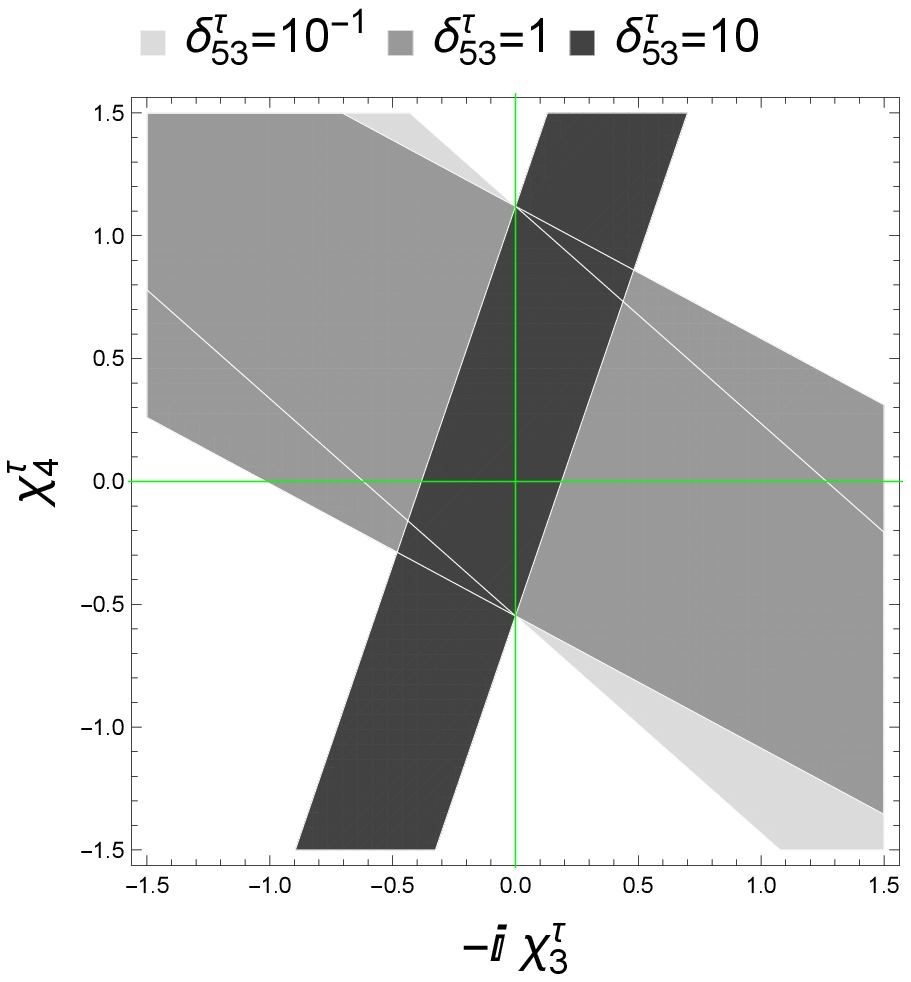}
\caption{\label{tauEDMdiagtexc1EDM} Allowed regions, in scale $10^{-4}$, for mSME coefficients from muon EDM. {\it 1st graph}: Parameter space $(\mathscr{X}^\tau_4,\mathscr{X}^\tau_5)$ for fixed $\delta^\tau_{34}$ values. {\it 2nd graph}: Parameter space $(\mathscr{X}^\tau_5,-i\mathscr{X}^tau_3)$ for fixed $\delta^\tau_{45}$ . {\it 3rd graph}: Parameter space $(-i\mathscr{X}^\tau_3,\mathscr{X}^\tau_4)$ for fixed $\delta^\tau_{34}$ values.}
\end{figure}

From Eq.~(\ref{diagtextEDM}), each new-physics EDM contribution $d^{\rm SME}_A$ is expressed in terms of three traces if quasi-degenerate textures are assumed. In order to provide an illustrative representation of effects from the mSME, we define the real quantities
\begin{equation}
\delta^A_{34}=\frac{-i\,\mathscr{X}^{A}_3}{\mathscr{X}^{A}_4},
\hspace{0.5cm}
\delta^{A}_{45}=\frac{\mathscr{X}^{A}_4}{\mathscr{X}^{A}_5},
\hspace{0.5cm}
\delta^A_{53}=\frac{\mathscr{X}^{A}_5}{-i\,\mathscr{X}^{A}_3},
\end{equation}
for $A=e,\mu,\tau$. With the aid of these expressions, allowed regions for Lorentz-violation parameters, determined by EDM limits given in Eqs.~(\ref{deexp})-(\ref{dtauexpRe}), have been established and depicted by Figs.~\ref{eEDMdiagtexc1EDM}-\ref{tauEDMdiagtexc1EDM}. In particular, Fig.~\ref{eEDMdiagtexc1EDM} 
displays regions for the case of lepton flavor $A=e$, for fixed $\delta^e_{34}$, $\delta^{e}_{45}$, or $\delta^{e}_{53}$ values and within intervals $-10^{-26}<-i\mathscr{X}^{e}_3<10^{-26}$, $-10^{-26}<\mathscr{X}^{e}_4<10^{-26}$, and $-10^{-26}<\mathscr{X}^{e}_5<10^{-26}$. These graphs have been rescaled by a factor $10^{26}$. Regarding the first graph of this figure, plotted in the parameter space $(\mathscr{X}^e_4,\mathscr{X}^e_5)$, three regions are shown, defined by the values $\delta^e_{34}=10^{-1},1,10$. The narrowest strip, which is also the darkest region (dark gray), is associated to $\delta^{e}_{34}=10$, thus corresponding to $-i\mathscr{X}^e_3$ larger, by one order of magnitude, than $\mathscr{X}^e_4$. A lighter and wider strip (regular gray) corresponds to a scenario in which $-i\mathscr{X}^e_3=\mathscr{X}^e_4$, that is, one in which $\delta^e_{34}=1$. The assumption that $\delta^e_{34}=10^{-1}$, in which case $\mathscr{X}^e_4>-i\mathscr{X}^e_3$ by one order of magnitude, defines the largest region in this graph, and is displayed in the lightest tone of gray. From this graph, notice that the scenario $\delta^e_{34}=10$ yields the most stringent restrictions over $\mathscr{X}^e_4$, though notice that even if $\delta^e_{34}=1,10^{-1}$, this parameter is still more constrained than $\mathscr{X}^e_5$. Similarly, the second and third graphs of Fig.~\ref{eEDMdiagtexc1EDM}, respectively plotted in the parameter spaces $\big( \mathscr{X}^e_5,-i\mathscr{X}^e_3 \big)$ and $\big( -i\mathscr{X}^e_3,\mathscr{X}^e_4 \big)$, show that the choices $\delta^e_{45}=10$ and $\delta^e_{53}=10$ produce the most strict limits on traces $\mathscr{X}^e_5$ and $-i\mathscr{X}^e_3$, respectively. Allowed values intended to quantitatively illustrate the scenarios $\delta^e_{34}, \delta^e_{45}, \delta^e_{53}=10$ are given in the second, third, and fourth rows of Table~\ref{ditexfle}. Figs~\ref{muEDMdiagtexc1EDM} and \ref{tauEDMdiagtexc1EDM} display EDM-allowed regions for Lorentz-violation parameters in the cases of lepton flavors $A=\mu$ and $A=\tau$, respectively.
The graphs of Fig.~\ref{muEDMdiagtexc1EDM}, plotted within intervals $-2\times10^{-10}<-i\mathscr{X}^{\mu}_3<2\times10^{-10}$, $-2\times10^{-10}<\mathscr{X}^{\mu}_4<2\times10^{-10}$, and $-2\times10^{-10}<\mathscr{X}^{\mu}_5<10^{-10}$, have been rescaled by a factor $10^{10}$. Meanwhile, a rescaling factor $10^{4}$ was utilized to carry out the graphs of Fig.~\ref{tauEDMdiagtexc1EDM}, within intervals $-1.5\times10^{-4}<-i\mathscr{X}^{\tau}_3<1.5\times10^{-4}$, $-1.5\times10^{-4}<\mathscr{X}^{\tau}_4<1.5\times10^{-4}$, and $-1.5\times10^{-4}<\mathscr{X}^{\tau}_5<1.5\times10^{-4}$. With the proper implementation of these rescalings and with the plotting intervals under consideration, the graphs of Figs.~\ref{muEDMdiagtexc1EDM} and \ref{tauEDMdiagtexc1EDM} look quite similar to those of Fig.~\ref{eEDMdiagtexc1EDM}, so alike explanations apply. For lepton flavors $A=\mu,\tau$ and taking $\delta^A_{34},\delta^A_{45},\delta^A_{53}=10$, the second, third and fourth rows of Tables~\ref{ditexflmu} and \ref{ditexfltau} provide representative values of parameters $-i\mathscr{X}^A_3$, $\mathscr{X}^A_4$, $\mathscr{X}^A_5$ included by the narrowest allowed regions in graphs of Figs.~\ref{muEDMdiagtexc1EDM}-\ref{tauEDMdiagtexc1EDM}, which are displayed in dark gray. The aforementioned qualitative resemblance among Figs.~\ref{eEDMdiagtexc1EDM}-\ref{tauEDMdiagtexc1EDM} is less accurate in the case of the graphs of Fig~\ref{tauEDMdiagtexc1EDM}, associated to $A=\tau$, as the corresponding regions are slightly non-symmetric with respect to the origin, which is a direct consequence of the tau EDM bounds given in Eq.~(\ref{dtauexpRe}). Such a disparity is quantitatively appreciated by an examination of the values provided in Table~\ref{ditexfltau}, when comparing them to the values consituting Tables~\ref{ditexfle} and \ref{ditexflmu}.


\subsection{Hermitian matrices $Y_{\alpha\beta}$}
Consider a scenario in which $Y^\dag_{\alpha\beta}=Y_{\alpha\beta}$ holds. As previously pointed out, this assumption yields, according to Eqs.~(\ref{VABab}) and (\ref{AABab}), an exact cancellation of Lorentz-violating coefficients $A^{AB}_{\alpha\beta}$ while leaving nonzero factors $V^{AB}_{\alpha\beta}$. Under such circumstances, only matrices $\mathscr{X}_1$ and $\mathscr{X}_4$ remain nonzero, so only 12 traces $\mathscr{X}^{AB}_j$ are still independent. The whole set of AMM contributions are then written in terms of six $\mathscr{X}_1$ parameters, whereas the six independent entries of $\mathscr{X}_4$ are the only ones defining the contributions to EDMs. \\

\begin{figure}[ht]
\center
$\Delta^{\rm H}_1=10$
\vspace{0.2cm} \\
\includegraphics[width=5.35cm]{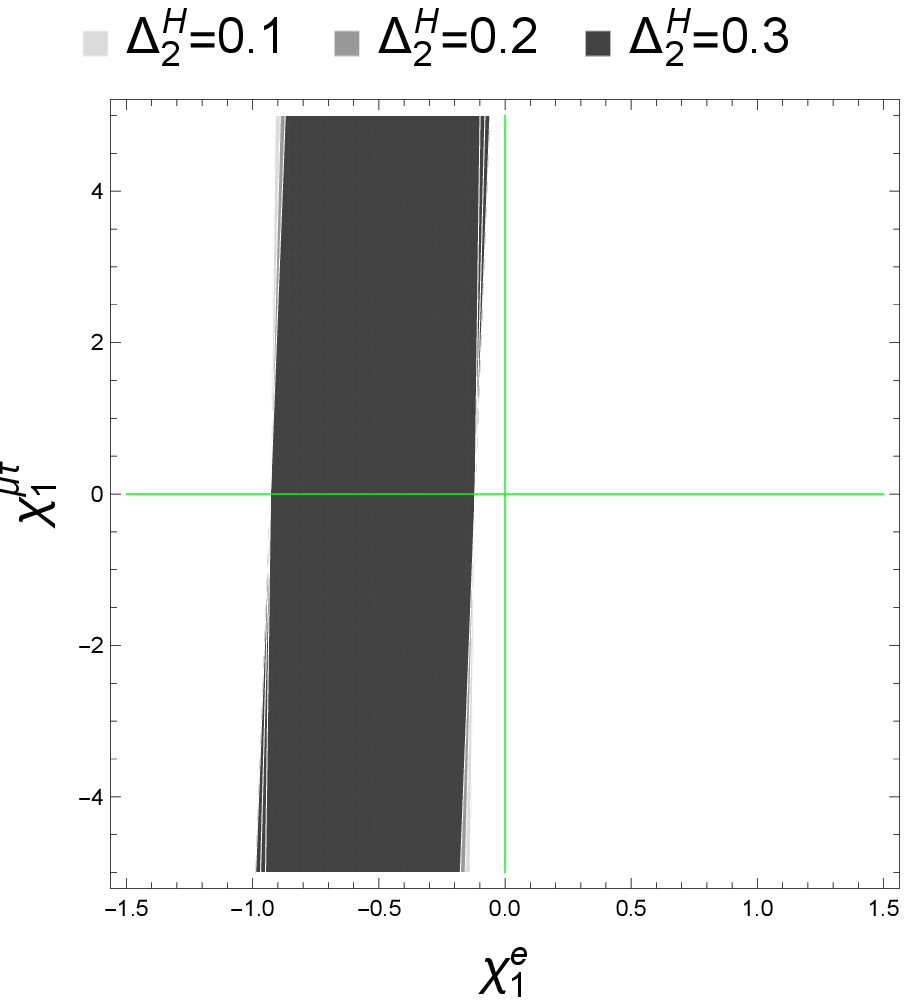}
\\
\vspace{0.4cm}
$\Delta^{\rm H}_1=10^2$
\vspace{0.2cm} \\
\includegraphics[width=5.35cm]{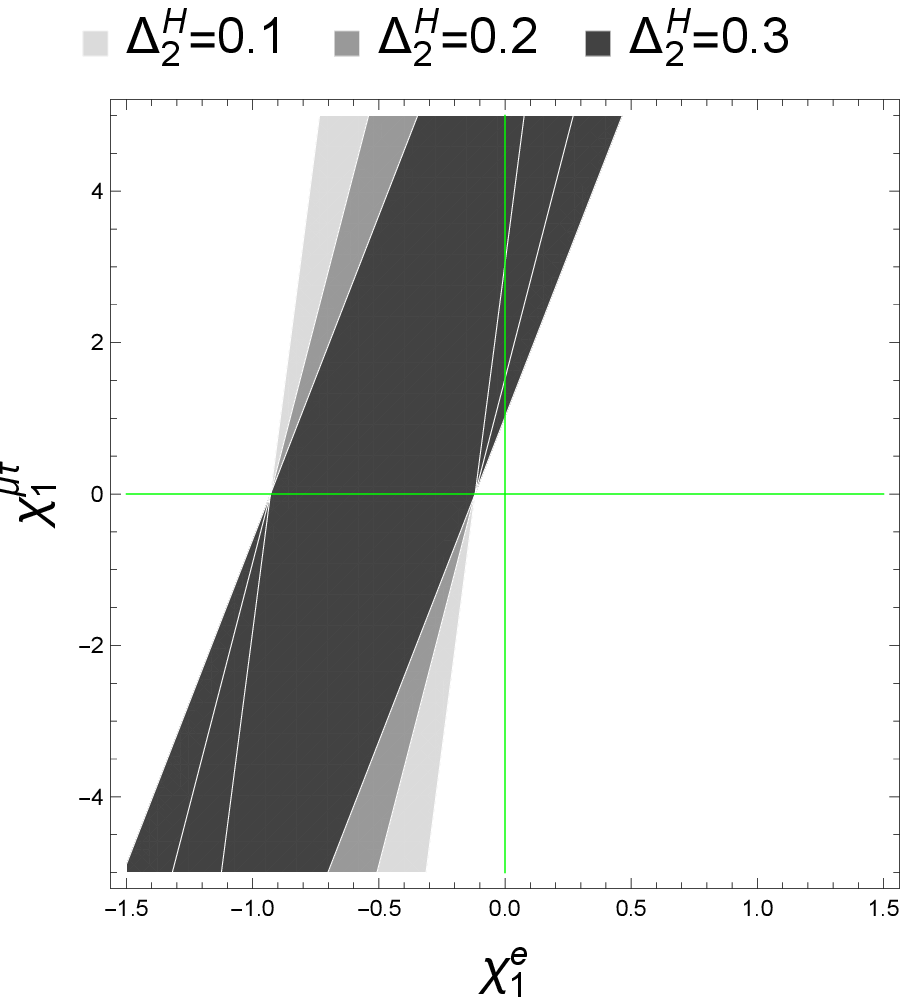}
\caption{\label{eHtextgph} Allowed regions in the parameter space $(\mathscr{X}_1^e,\mathscr{X}_1^{\mu\tau})$, within $|\mathscr{X}_1^e|<1.5\times10^{-26}$ and $|\mathscr{X}_1^{\mu\tau}|<5\times10^{-26}$, with both graphs rescaled by $10^{26}$. We consider values $\Delta^{\rm H}_1=10$ (upper graph) and $\Delta^{\rm H}_1=10^2$ (lower graph), whereas $\Delta^{\rm H}_2=0.1,0.2,0.3$ are used in both cases.}
\end{figure}

\begin{figure}[ht]
\center
$\Delta^{\rm H}_2=0.3$
\vspace{0.2cm} \\
\includegraphics[width=5.35cm]{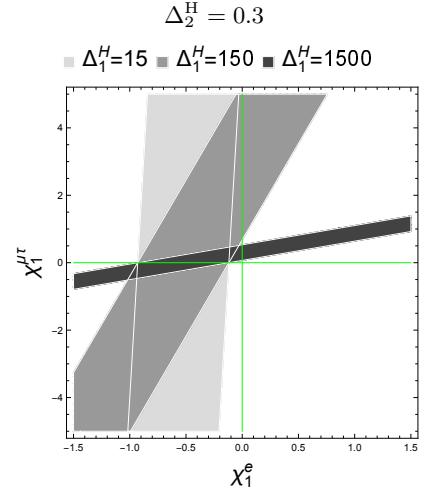}
\caption{\label{eHtextgphoth} Allowed regions in the parameter space $(\mathscr{X}_1^e,\mathscr{X}_1^{\mu\tau})$, within $|\mathscr{X}_1^e|<1.5\times10^{-26}$ and $|\mathscr{X}_1^{\mu\tau}|<5\times10^{-26}$, with the graph rescaled by $10^{26}$. We consider values $\Delta^{\rm H}_1=15,150,1500$, whereas $\Delta^{\rm H}_2=0.3$ is taken.}
\end{figure}

\begin{figure}[ht]
\center
$\Delta^{\rm H}_1=10$
\vspace{0.2cm} \\
\includegraphics[width=5.35cm]{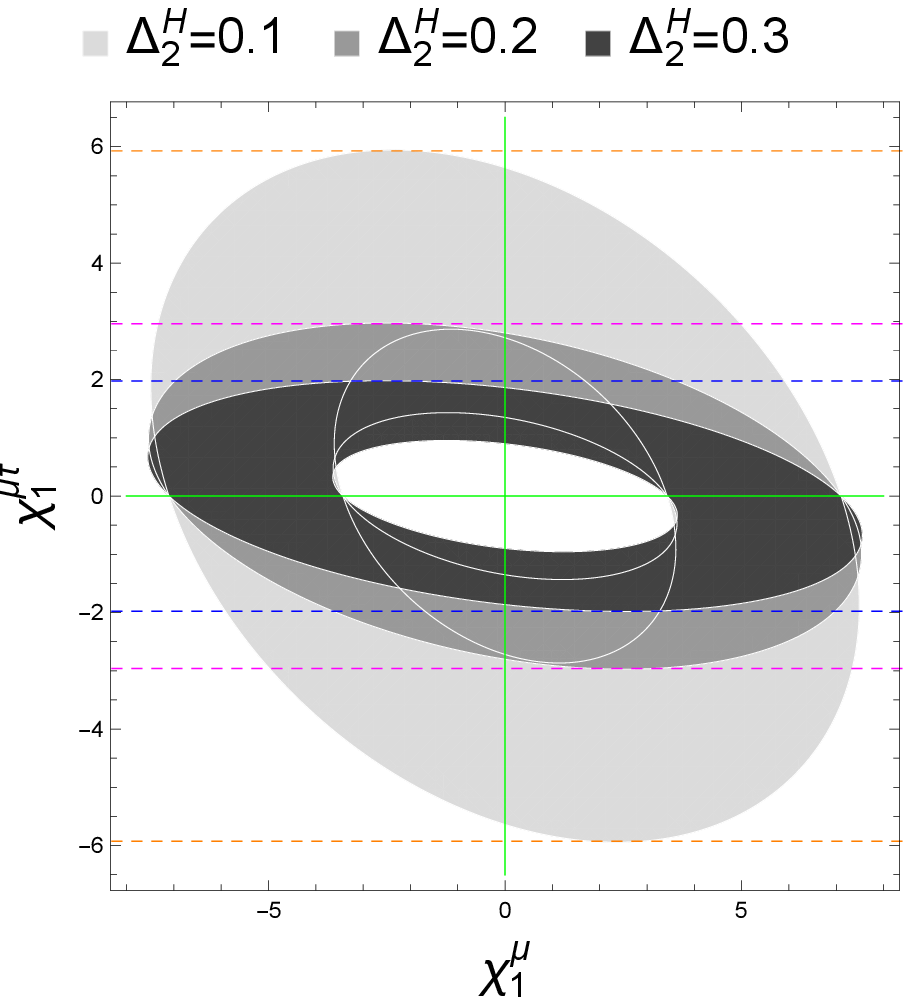}
\\
\vspace{0.4cm}
$\Delta^{\rm H}_1=10^2$
\vspace{0.2cm} \\
\includegraphics[width=5.35cm]{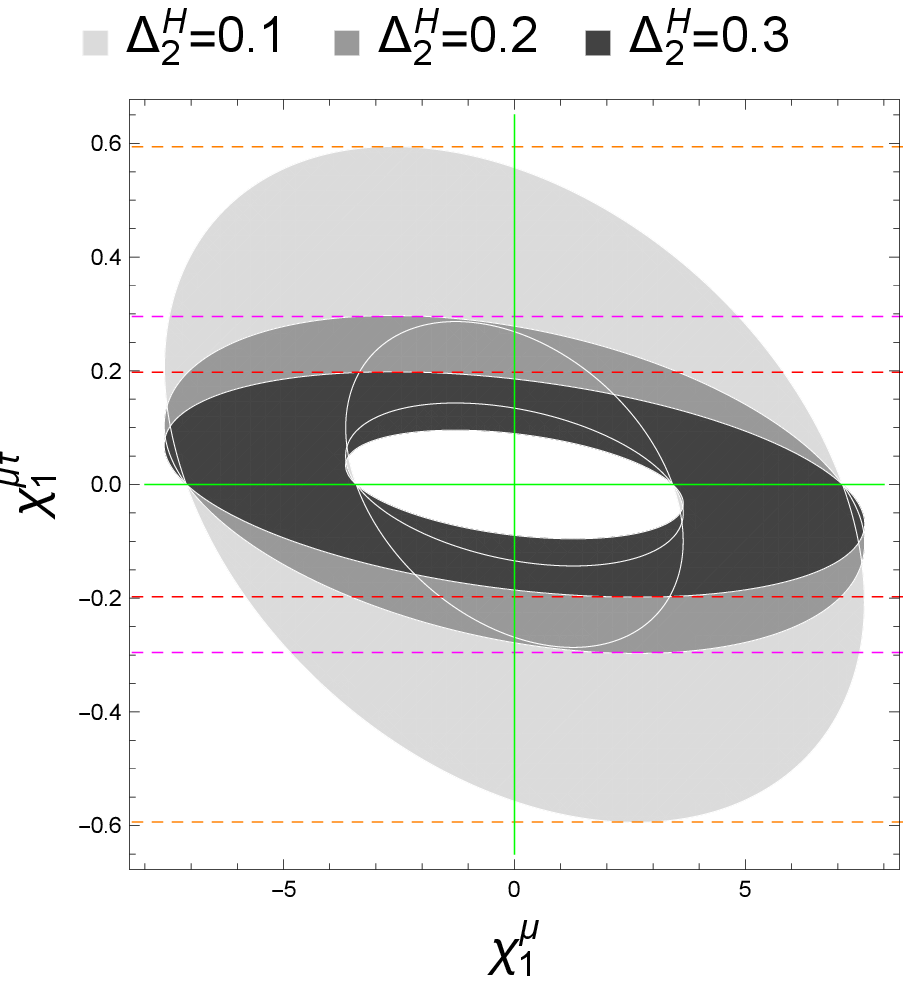}
\caption{\label{muHtextgph} Allowed regions in the parameter space $(\mathscr{X}_1^\mu,\mathscr{X}_1^{\mu\tau})$, within $|\mathscr{X}_1^\mu|<8\times10^{-14}$ and either $|\mathscr{X}_1^{\mu\tau}|<6\times10^{-14}$ (upper graph) or $|\mathscr{X}_1^{\mu\tau}|<0.6\times10^{-14}$ (lower graph), with both graphs rescaled by $10^{14}$. We consider values $\Delta^{\rm H}_1=10$ (upper graph) and $\Delta^{\rm H}_1=10^2$ (lower graph), whereas $\Delta^{\rm H}_2=0.1,0.2,0.3$ are used in both cases. Pairs of dashed horizontal lines define $\mathscr{X}^{\mu\tau}_1$-allowed intervals around $\mathscr{X}^{\mu\tau}_1=0$.}
\end{figure}

\begin{figure}[ht]
\center
\includegraphics[width=5.35cm]{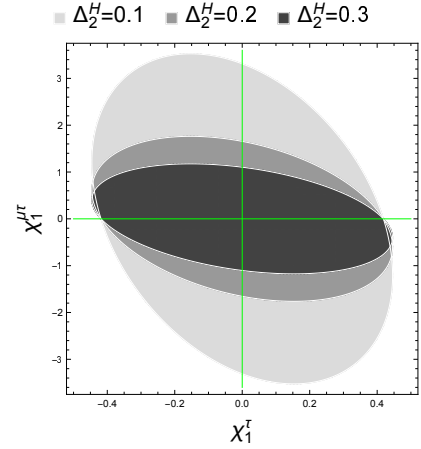}
\caption{\label{tauHtextgph} Allowed regions in the parameter space $(\mathscr{X}_1^\tau,\mathscr{X}_1^{\mu\tau})$, within $|\mathscr{X}_1^\tau|<0.35\times10^{-4}$ and $|\mathscr{X}_1^{\mu\tau}|<2.6\times10^{-4}$, with the graph rescaled by $10^{4}$. We consider values $\Delta^{\rm H}_2=0.1,0.2,0.3$, for any $\Delta^{\rm H}_1$.}
\end{figure}

To discuss the AMM contributions, we define the factors
\begin{equation}
\Delta^{\rm H}_1=\frac{\mathscr{X}_1^{e\mu}}{\mathscr{X}_1^{\tau e}},\hspace{0.3cm}\Delta^{\rm H}_2=\frac{\mathscr{X}_1^{\tau e}}{\mathscr{X}_1^{\mu\tau}},
\end{equation}
in terms of which the new-physics contributions are written as
\begin{equation}
a_e^{\rm SME}=a_1^{ee}\mathscr{X}^{e}_1+\Delta^{\rm H}_2(a^{e\mu}_1\Delta^{\rm H}_1+a^{e\tau}_1)\mathscr{X}^{\mu\tau}_1,
\label{aeHtext}
\end{equation}
\begin{equation}
a_\mu^{\rm SME}=a^{\mu\mu}_1\mathscr{X}^{\mu}_1+(a^{\mu e}_1\Delta^{\rm H}_1\Delta^{\rm H}_2+a^{\mu\tau}_1)\mathscr{X}^{\mu\tau}_1,
\label{amuHtext}
\end{equation}
\begin{equation}
a_\tau^{\rm SME}=a^{\tau\tau}_1\mathscr{X}^\tau_1+(a^{\tau e}_1\Delta^{\rm H}_2+a^{\tau\mu}_1)\mathscr{X}^{\mu\tau}_1.
\label{atauHtext}
\end{equation}
In this manner, each contribution, corresponding to any lepton flavor $A$, is expressed in terms of four parameters. For any flavor $A$, three of such quantities are the factors $\Delta^{\rm H}_1$, $\Delta^{\rm H}_2$, and the trace $\mathscr{X}^{\mu\tau}_1$, while the parameter $\mathscr{X}^A_1$, which is the only one distinguishing the specific $A$-flavor contribution, defines the expression as well. The fact that Eqs.~(\ref{aeHtext})-(\ref{atauHtext}) share three Lorentz-violation parameters is a feature to bear in mind, for the contributions $a^{\rm SME}_e$, $a^{\rm SME}_\mu$, $a^{\rm SME}_\tau$ are, in part, simultaneously determined by such parameters. To provide a qualitative description of the SME contributions $a^{\rm SME}_A$ in this scenario, the graphs of Figs.~\ref{eHtextgph} to \ref{tauHtextgph} have been plotted. \\

The two graphs of Fig.~\ref{eHtextgph},
displaying allowed regions in the space of parameters $(\mathscr{X}_1^e,\mathscr{X}_1^{\mu\tau})$ and determined by bounds on contributions from new physics to the electron AMM, Eq.~(\ref{dae}), have been realized within $|\mathscr{X}^{e}_1|<1.5$ and $|\mathscr{X}^{\mu\tau}_1|<5$, after a proper rescaling by the factor $10^{21}$. 
Two graphs have been included in order to compare allowed regions for scenarios characterized by different choices of the parameter $\Delta^{\rm H}_1$. The values $\Delta^{\rm H}_1=10$, in the upper graph, and $\Delta^{\rm H}_1=10^2$, in the lower graph, have been considered, whereas for each of such graphs the values $\Delta^{\rm H}_2=0.1,0.2,0.3$ have been taken into account. 
The three allowed regions shown by each graph are straight strips whose widths are similar to each other for each considered $\Delta^{\rm H}_2$ value. In the case $\Delta^{\rm H}_1=10$, the allowed regions barely distinguish from each other, whereas the shapes of the regions seem to be more sensitive to changes in $\Delta^{\rm H}_2$ as long as the value $\Delta^{\rm H}_1=10^2$, larger by one order of magnitude, is considered. These graphs illustrate how orientations of allowed regions are, in general, different for different values of $\Delta^{\rm H}_2$, with fixed $\Delta^{\rm H}_1$. In this context, the SME trace $\mathscr{X}^e_1$ is more stringently constrained than $\mathscr{X}^{\mu\tau}_1$ in both scenarios. Nonetheless, the last statement, realized for particular choices of factors $\Delta^{\rm H}_1$ and $\Delta^{\rm H}_2$, should not be understood to be valid in general. To illustrate this, we provide Fig.~\ref{eHtextgphoth},
which has been realized within same parameter region $(\mathscr{X}^e_1,\mathscr{X}^{\mu\tau}_1)$ as that of the graphs of Fig.~\ref{eHtextgph}, and with the same rescaling as well. In this case, the values $\Delta^{\rm H}_2=0.3$ and $\Delta^{\rm H}_1=15,150,1500$ have been utilized. Then notice that the largest value of $\Delta^{\rm H}_1$, among those considered for the realization of this graph, yields an allowed-region strip which, in comparison with the allowed regions of Fig.~\ref{eHtextgph}, is narrower with a clockwise-rotated orientation, thus corresponding to a parameter $\mathscr{X}^{\mu\tau}_1$ more stringently restricted than $\mathscr{X}^e_1$, as opposite to the allowed regions of Fig.~\ref{eHtextgph}. \\

Regarding the contributions from the SME to the AMM of the muon, the graphs of Fig.~\ref{muHtextgph}
provide a depiction of allowed regions in the parameter space $(\mathscr{X}^\mu_1,\mathscr{X}^{\mu\tau}_1)$, with a convenient rescaling by a factor $10^{14}$ implemented in the two graphs. In the present scenario, the contribution $a^{\rm SME}_\mu$ is complex valued, so its modulus, $|a^{\rm SME}_\mu|$, has been rather considered to compare it with the bound from new physics on the muon AMM, Eq.~(\ref{damu}), which corresponds to an interval of positive values. Thus, the allowed regions shown in Fig.~\ref{muHtextgph} are not straight strips, but rings instead. The values of the factors $\Delta^{\rm H}_1$ and $\Delta^{\rm H}_2$ utilized to plot the graphs of Fig.~\ref{muHtextgph} are the same as those of Fig.~\ref{eHtextgph}, that is, $\Delta^{\rm H}_1=10$ and $\Delta^{\rm H}_1=10^2$ were respectively used to plot the upper graph and the lower graph of Fig.~\ref{muHtextgph}, whereas the values $\Delta^{\rm H}_2=0.1,0.2,0.3$ were all considered in both graphs. With this in mind, notice that, for fixed $\Delta^{\rm H}_1$, increasing values of the parameter $\Delta^{\rm H}_2$ flatten the ring along the $\mathscr{X}^{\mu\tau}_1$ axis, in which case larger values of $\Delta^{\rm H}_2$ correspond to more restricted allowed regions. While values within $|\mathscr{X}^\mu_1|<8\times10^{-14}$ have been considered for both graphs, vertical axes range along different intervals: the upper-graph vertical axis runs over $|\mathscr{X}^{\mu\tau}_1|<6\times10^{-14}$; the lower graph, on the other hand, displays values of the the vertical axis within $|\mathscr{X}^{\mu\tau}_1|<0.6\times10^{-14}$. So, notice that the choice $\Delta^{\rm H}_1=10^2$, for the lower graph, yields more constrained regions than those corresponding to $\Delta^{\rm H}_1=10$. \\

Next we focus on the SME contributions to the tau AMM, which we illustrate by means of Fig.~\ref{tauHtextgph}.
This figure displays one sole graph, plotted within $|\mathscr{X}^\tau_1|<0.35\times10^{-4}$ and $|\mathscr{X}^{\mu\tau}_1|<2.6\times10^{-4}$, and normalized by the factor $10^{4}$.  Notice, from Eq.~(\ref{atauHtext}), that $a^{\rm SME}_\tau$ is $\Delta^{\rm H}_1$-independent, so there is no need to include more graphs to compare regions associated to different values of the factor $\Delta^{\rm H}_1$. As it occurred in the case $A=\mu$, previously discussed, the SME contribution $a^{\rm SME}_\tau$, to the tau AMM, is a complex-valued quantity. Thus the modulus $|a^{\rm SME}_\tau|$ was considered and, since the bounds on this AMM define an interval which includes both positive and negative values, the resulting allowed regions do not involve holes. Again, the plotted allowed regions correspond to the values $\Delta^{\rm H}_2=0.1,0.2,0.3$. This graph shows that the larger the value of $\Delta^{\rm H}_2$, the flattest the ellipse along the $\mathscr{X}^{\mu\tau}_1$ axis and, thus, the smaller the allowed region in the parameter space $(\mathscr{X}^\tau_1,\mathscr{X}^{\mu\tau}_1)$.\\

Among the three mSME AMM contributions, the most stringent constraints on $\mathscr{X}_1^{\mu\tau}$, of orders $10^{-15}$ to $10^{-14}$, are set by $|a^{\rm SME}_\mu|$. The corresponding allowed intervals are displayed in the graphs of Fig.~\ref{muHtextgph} through pairs of horizontal dashed lines which are equidistant from the horizontal axes. Each graph then shows three of such intervals. The precise numerical values of these limits are given in the fourth column of Table~\ref{tabtext2}. \\

\begin{table}[ht]
\center
\begin{tabular}{|c|c|c|c|}
\hline
& $\Delta^{\rm H}_1$ & $\Delta^{\rm H}_2$ & $|\mathscr{X}^{\mu\tau}_1|<$
\\ \hline
 & 10 & 0.1 & $5.93\times10^{-14}$
\\ 
& 10 & 0.2 & $2.96\times10^{-14}$
\\ 
$a^{\rm SME}_A$ & 10 & 0.3 & $1.98\times10^{-14}$
\\
& $10^2$ & 0.1 & $0.59\times10^{-14}$
\\
& $10^2$ & 0.2 & $0.29\times10^{-14}$
\\
& $10^2$ & 0.3 & $0.19\times10^{-14}$
\\ \hline
& $\Delta^{\rm H}_3$ & $\Delta^{\rm H}_4$ & $|\mathscr{X}^{\mu\tau}_4|<$
\\ \hline
 & 10 & 0.1 & $2.90\times10^{-11}$
\\ 
& 10 & 0.2 & $1.45\times10^{-11}$
\\ 
$d^{\rm SME}_A$ & 10 & 0.3 & $0.97\times10^{-11}$
\\
& $10^2$ & 0.1 & $0.29\times10^{-11}$ 
\\
& $10^2$ & 0.2 & $0.14\times10^{-11}$
\\
& $10^2$ & 0.3 & $0.10\times10^{-11}$
\\ \hline
\end{tabular}
\caption{\label{tabtext2} Allowed intervals of $\mathscr{X}^{\mu\tau}_1$ and $\mathscr{X}^{\mu\tau}_4$ values for different choices of parameters $\Delta^{\rm H}_1$, $\Delta^{\rm H}_2$, $\Delta^{\rm H}_3$, and $\Delta^{\rm H}_4$, and determined from lepton AMMs and EDMs bounds on new-physics effects, implemented to mSME Yukawa-sector contributions $a^{\rm SME}_A$ and $d^{\rm SME}_A$ in the scenario of Hermitian matrices $Y_{\alpha\beta}$.}
\end{table}

Next we address the contributions from the mSME Yukawa sector to EDMs, which we express as
\begin{equation}
d^{\rm SME}_e=d^{ee}_4\mathscr{X}^e_4+\Delta^{\rm H}_4\big( d^{e\mu}_4\Delta^{\rm H}_3+d_4^{e\tau} \big)\mathscr{X}^{\mu\tau}_4,
\label{desme}
\end{equation}
\begin{equation}
d^{\rm SME}_\mu=d_4^{\mu\mu}\mathscr{X}^\mu_4+\big( d_4^{\mu e}\Delta^{\rm H}_4\Delta^{\rm H}_3+d_4^{\mu\tau} \big)\mathscr{X}_4^{\mu\tau},
\label{dmusme}
\end{equation}
\begin{equation}
d^{\rm SME}_\tau=d_4^{\tau\tau}\mathscr{X}^\tau_4+\big( d_4^{\tau e}\Delta_4^{\rm H}+d_4^{\tau\mu} \big)\mathscr{X}^{\mu\tau}_4,
\label{dtausme}
\end{equation}
for which
\begin{equation}
\Delta^{\rm H}_3=\frac{\mathscr{X}_4^{e\mu}}{\mathscr{X}_4^{\tau e}},\hspace{0.3cm}\Delta^{\rm H}_4=\frac{\mathscr{X}_4^{\tau e}}{\mathscr{X}_4^{\mu\tau}},
\end{equation}
have been defined. So, these EDM contributions are given in terms of six Lorentz-violation parameters, with the three contributions sharing the parameters $\Delta^{\rm H}_3$, $\Delta^{\rm H}_4$, and $\mathscr{X}^{\mu\tau}_4$, while being distinguished from each other by the traces $\mathscr{X}^e_4$, $\mathscr{X}^\mu_4$, or $\mathscr{X}^\tau_4$. \\

\begin{figure}[ht]
\center
$\Delta^{\rm H}_3=10$
\vspace{0.2cm} \\
\includegraphics[width=5.35cm]{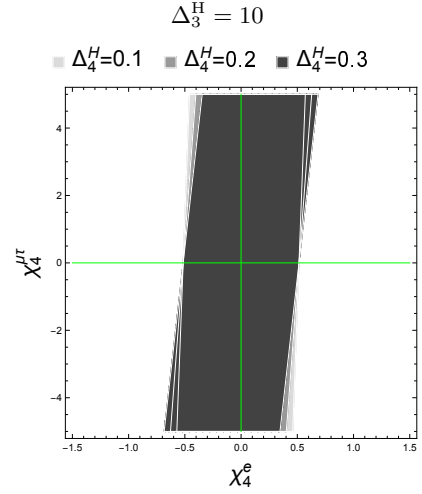}
\vspace{0.4cm}
\\
$\Delta^{\rm H}_3=10^2$
\vspace{0.2cm} \\
\includegraphics[width=5.35cm]{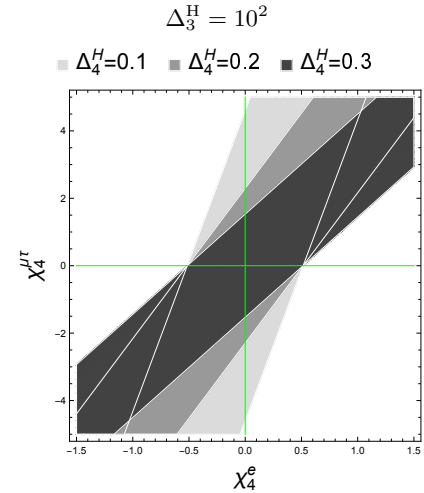}
\caption{\label{chachacha1} Allowed regions in the parameter space $(\mathscr{X}_4^e,\mathscr{X}_4^{\mu\tau})$, within $|\mathscr{X}_4^e|<1.5\times10^{-26}$ and $|\mathscr{X}_4^{\mu\tau}|<5\times10^{-26}$, with both graphs rescaled by $10^{26}$. We consider values $\Delta^{\rm H}_3=10$ (upper graph) and $\Delta^{\rm H}_3=10^2$ (lower graph), whereas $\Delta^{\rm H}_4=0.1,0.2,0.3$ are used in both cases.}
\end{figure}

\begin{figure}[ht]
\center
$\Delta^{\rm H}_3=10$
\vspace{0.2cm} \\
\includegraphics[width=5.35cm]{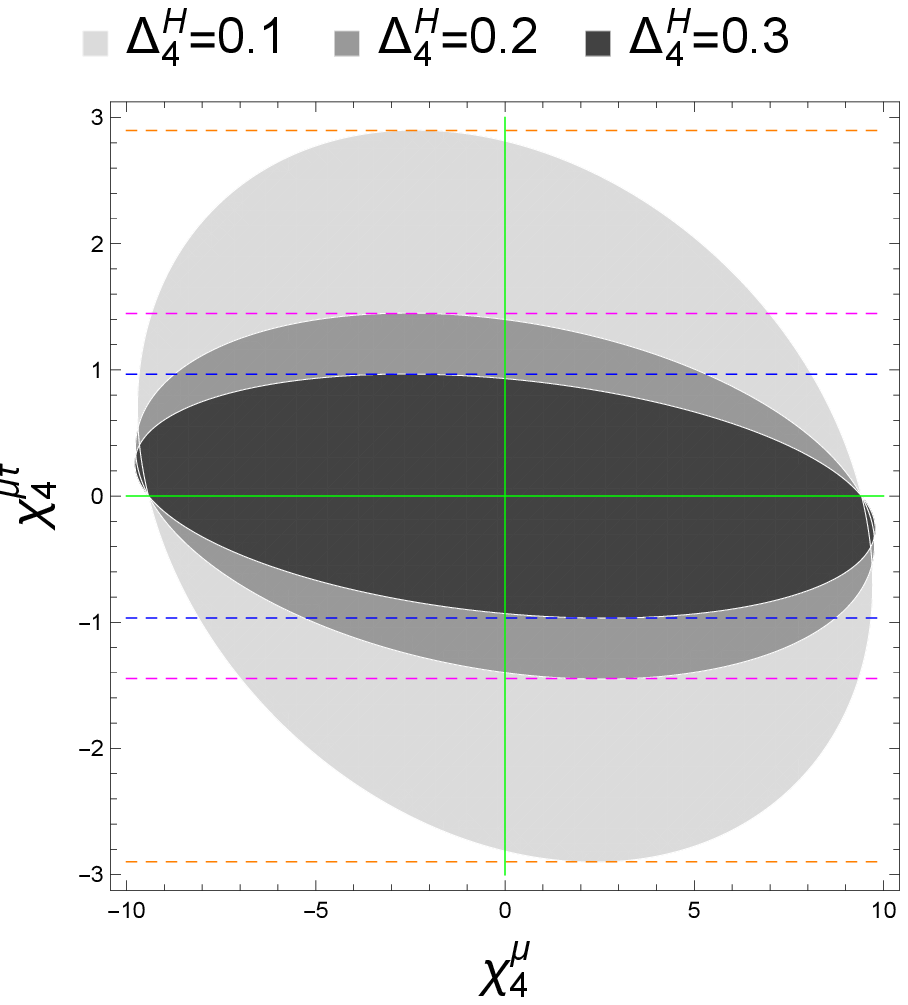}
\vspace{0.4cm}
\\
$\Delta^{\rm H}_3=10^2$
\vspace{0.2cm} \\
\includegraphics[width=5.35cm]{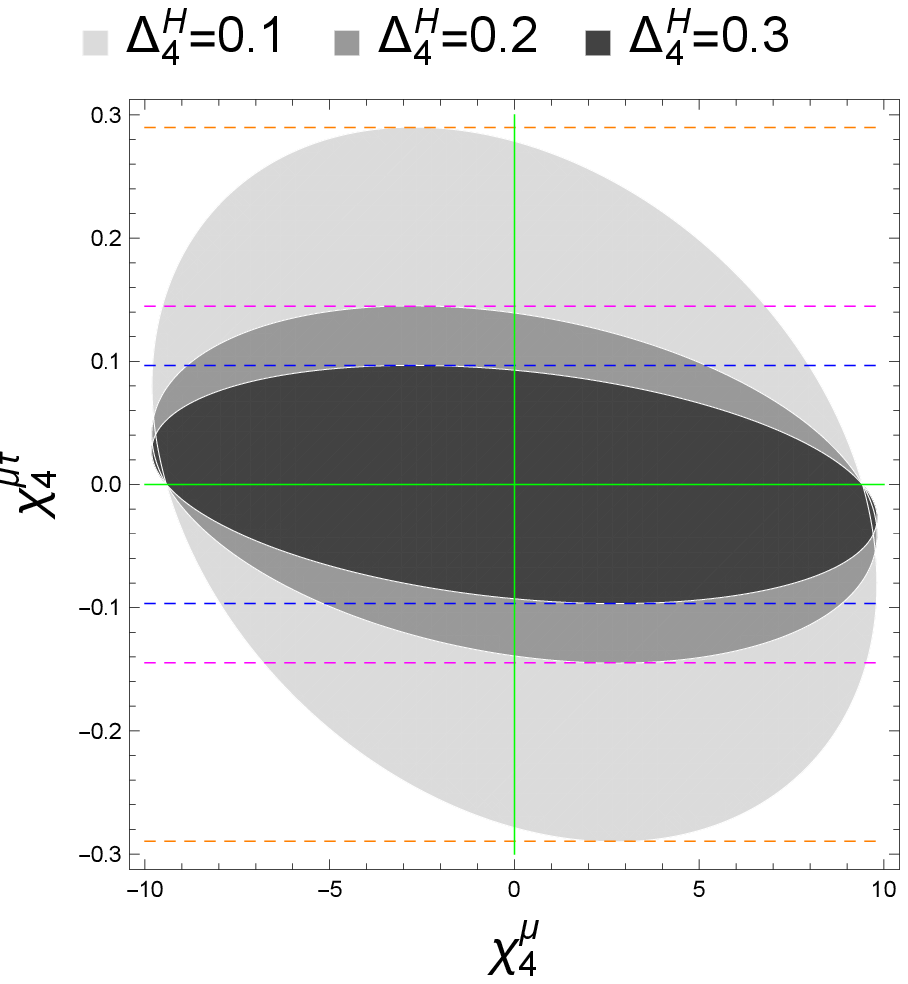}
\caption{\label{chachacha2} Allowed regions in the parameter space $(\mathscr{X}_4^\mu,\mathscr{X}_4^{\mu\tau})$, within $|\mathscr{X}_4^\mu|<10^{-10}$ and either $|\mathscr{X}_4^{\mu\tau}|<3\times10^{-11}$ (upper graph) or $|\mathscr{X}_4^{\mu\tau}|<0.3\times10^{-11}$ (lower graph), with both graphs rescaled by $10^{11}$. We consider values $\Delta^{\rm H}_3=10$ (upper graph) and $\Delta^{\rm H}_3=10^2$ (lower graph), whereas $\Delta^{\rm H}_4=0.1,0.2,0.3$ are used in both cases. Pairs of dashed horizontal lines define $\mathscr{X}^{\mu\tau}_4$-allowed intervals around $\mathscr{X}^{\mu\tau}_4=0$.}
\end{figure}

\begin{figure}[ht]
\center
\includegraphics[width=5.35cm]{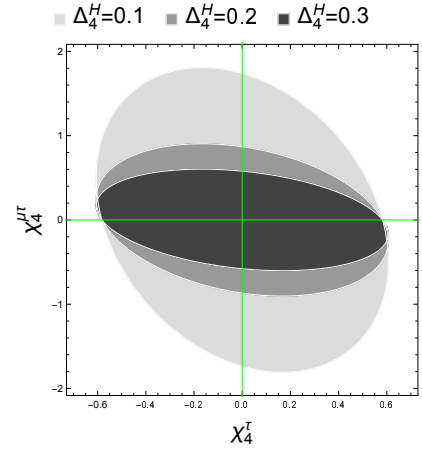}
\caption{\label{chachacha3} Allowed regions in the parameter space $(\mathscr{X}_4^\tau,\mathscr{X}_4^{\mu\tau})$, within $|\mathscr{X}_4^\tau|<0.7\times10^{-4}$ and $|\mathscr{X}_1^{\mu\tau}|<2\times10^{-4}$, with the graph rescaled by $10^{4}$. We consider values $\Delta^{\rm H}_4=0.1,0.2,0.3$, for any $\Delta^{\rm H}_3$.}
\end{figure}

Since the expressions given in Eqs.~(\ref{desme})-(\ref{dtausme}), for the mSME Yukawa-sector contributions to EDMs, have the very same structure as Eqs.~(\ref{aeHtext})-(\ref{atauHtext}), which correspond to AMMs Lorentz-violation effects, the discussion of bounds established by EDMs is quite similar to the one previously carried out for AMMs. Allowed regions in parameter spaces $(\mathscr{X}_4^e,\mathscr{X}_4^{\mu\tau})$, $(\mathscr{X}_4^\mu,\mathscr{X}_4^{\mu\tau})$, and $(\mathscr{X}_4^\tau,\mathscr{X}_4^{\mu\tau})$, for different choices of $\Delta^{\rm H}_3$ and $\Delta^{\rm H}_4$, are respectively displayed in Figs.~\ref{chachacha1}, \ref{chachacha2}, and \ref{chachacha3}. Again, the most restrictive constraints on the parameter $\mathscr{X}^{\mu\tau}_4$, of order $10^{-11}$, come from the muon EDM contribution $d^{\rm SME}_\mu$. Allowed intervals for this parameter have then been indicated, in the graphs of Fig.~\ref{chachacha2}, by horizontal dashed lines. The values characterizing such intervals are given in Table~\ref{tabtext2}.

\section{Summary}
\label{conc}
The present investigation has been performed in the context established by the Lorentz- and $CPT$-violating Standard-Model Extension, an effective field theory which sets a quite general framework to quantify, at relatively-low energies, effects to be expected from a higher-energy formulation incorporating violation of Lorentz invariance. While this model of new physics extends every sector of the Standard Model, by the inclusion of both renormalizable and nonrenormalizable lagrangian terms, our discussion has been restricted to couplings occurring in the Yukawa sector of the renormalizable part of the Standard-Model extension. This sort of Lorentz nonconservation is characterized by Yukawa-like couplings endowed with spacetime Lorentz indices, which come along with noninvariance under particle transformations, even though observer transformations remain a symmetry of the theory. \\

In a perturbative approach, the Lorentz-violating interactions resulting from this extended Yukawa sector, after implementation of the Brout-Englert-Higgs mechanism, yield two-point insertions and three-point vertices which induce one-loop corrections to the electromagnetic vertex $A_\mu l_Al_A$. In the presence of Lorentz nonpreservation, $A_\mu l_Al_A$ is characterized by a tensor structure with several form factors adding to those already known to define the ordinary Lorentz-invariant parametrization of such interactions. In this context, the aforementioned loop corrections involve contributions to both magnetic and electric form factors. For these quantities to consistently be Lorentz invariant, the contributions from this new physics have been argued to emerge for the first time at the second order in Lorentz-violating coefficients. With this in mind, leading contributions from Lorentz violation to anomalous magnetic moments and electric dipole moments have been identified, with the corresponding expressions being quantified by flavor-mixing coefficients ${\rm tr}\,\kappa^{AB}_1=V^{AB}_{\alpha\nu}V^{AB\nu\alpha}$, ${\rm tr}\,\kappa^{AB}_2=A^{BA*}_{\alpha\nu}A^{BA\nu\alpha*}$, ${\rm tr}\,\kappa^{AB}_3=V^{AB}_{\alpha\nu}A^{BA\nu\alpha*}$, ${\rm tr}\,\kappa^{AB}_4=V^{AB}_{\alpha\nu}\tilde{V}^{AB\nu\alpha}$, and ${\rm tr}\,\kappa^{AB}_5=A^{BA*}_{\alpha\nu}\tilde{A}^{BA\nu\alpha*}$. The resulting contributions from Lorentz-violation to anomalous magnetic moments and electric dipole moments were found to be ultraviolet finite. Imaginary parts of the muon and tau-lepton electromagnetic moments emerge, even though the calculation has been carried out on shell and has included only diagonal moments. This happens in the perturbative approach, followed here, because bilinear insertions allow for Feynman diagrams in which external lines connect with virtual lines associated to lighter particles, thus defining thresholds which are surpassed by the contributions after on-shell conditions are implemented. \\

Defining $3\times3$ matrices $\mathscr{X}_j$ with lepton-flavor components $\mathscr{X}_j^{AB}={\rm tr}\,\kappa_j^{AB}$, where $j=1,2,3,4,5$ and where the trace ``tr'' operates on $4\times4$ matrices $\kappa_j^{AB}$ given in the space of matrix representations of the Lorentz group, we have considered and explored two scenarios. One of them, which we named ``scenario of quasi-diagonal textures'', is defined by the conditions $\mathscr{X}^{AB}_j\approx0$ for $A\ne B$. In this context, each electromagnetic moment is determined by Lorentz-violation parameters not shared by any other moment, so the comparison between each electromagnetic-moment contribution and its corresponding bound, as taken from the literature, is carried out independently of any other contribution. Moreover, in this scenario all the contributions from Lorentz violation are real. A summary of the bounds determined in this scenario is provided in Table~\ref{finaltab}, where constraints derived within this scenario can be found in rows including the assumption ``QDT'', which is an acronym for quasi-diagonal textures. From this table, note that the most stringent bounds, restricting SME coefficients ${\rm tr}\,\kappa^{ee}_j$, are established by the electron EDM, whereas the most stringent bound given by AMMs limits on new-physics effects, of order $10^{-22}$, also correspond to the case of the electron. As expected, for both electromagnetic-moment Lorentz-violation contributions the weakest constraints are determined by the restrictions on the tau-lepton electromagnetic moments. A scenario of Hermitian Yukawa matrices, which in Table~\ref{finaltab} has been referred to by the acronym ``HYM'', is defined by the condition $Y^\dag_{\mu\nu}=Y_{\mu\nu}$, which yields the exact cancellation of coefficients $A_{\mu\nu}^{BA*}$. In this scenario, the Lorentz-violation contributions to the electromagnetic moments of the muon and the tau lepton, which are unstable particles, turned out to be complex quantities. The comparison of SME contributions with current bounds on the electromagnetic moments and the corresponding analysis is more intricate than the one executed in the other scenario, the reason being that all the anomalous-magnetic-moment contributions share SME parameters, together with the fact that bounds on these quantities are quite different from each other. And the same applies for the case of contributions to electric dipole moments. Under such circumstances, the parameters $|{\rm tr}\,\kappa^{\mu\tau}_1|$ and $|{\rm tr}\,\kappa^{\mu\tau}_4|$ are the ones which have been bounded. The most restrictive limits, which constrain $|{\rm tr}\,\kappa^{\mu\tau}_1|$, have been placed by contributions to the muon anomalous magnetic moments, being as restrictive as $10^{-15}$, while the effects from the mSME Yukawa sector on electric dipole moments yielded constraints of order $10^{-12}$ on $|{\rm tr}\,\kappa^{\mu\tau}_4|$.

\begin{widetext}

\begin{table}[ht]
\center
\begin{tabular}{cccc}
{\bf Assumptions} & {\bf EMMs} & {\bf Combinations} & {\bf Bounds}
\\ \hline \hline
QDT, ${\rm tr}\,\kappa^{ee}_2=\pm1.5\times10^{-21}$ & $a^{\rm SME}_e$ & ${\rm tr}\,\kappa^{ee}_1$ & $-2.25(405)\times10^{-22}$ 
\\
&&& $-8.25(405)\times10^{-22}$
\\ \hline
QDT, ${\rm tr}\,\kappa^{\mu\mu}_2=\pm5.0\times10^{-10}$ & $a^{\rm SME}_\mu$ & ${\rm tr}\,\kappa^{\mu\mu}_1$ & $+7.06(184)\times10^{-14}$
\\
&&& $+3.46(184)\times10^{-14}$
\\ \hline
QDT, ${\rm tr}\,\kappa^{\tau\tau}_2=\pm9.0\times10^{-5}$ & $a^{\rm SME}_\tau$ & ${\rm tr}\,\kappa^{\tau\tau}_1$ & $+1.20(358)\times10^{-5}$
\\
&&& $-2.39(358)\times10^{-5}$
\\ \hline
QDT, $\delta^e_{34}=10$, ${\rm tr}\,\kappa^{ee}_5=\pm10^{-26}$ & $d^{\rm SME}_e$ & ${\rm tr}\,\kappa^{ee}_4$ & $\pm3.76(503)\times10^{-28}$
\\ 
QDT, $\delta^e_{45}=10$, $-i{\rm tr}\,\kappa^{ee}_3=\pm10^{-26}$ & $d^{\rm SME}_e$ & ${\rm tr}\,\kappa^{ee}_5$ & $\mp9.60(536)\times10^{-28}$
\\ 
QDT, $\delta^e_{53}=10$, ${\rm tr}\,\kappa^{ee}_4=\pm10^{-26}$ & $d^{\rm SME}_e$ & $-i\,{\rm tr}\,\kappa^{ee}_3$ & $\pm3.42(176)\times10^{-27}$
\\ \hline
QDT, $\delta^\mu_{34}=10$, ${\rm tr}\,\kappa^{\mu\mu}_5=\pm5.0\times10^{-10}$ & $d^{\rm SME}_\mu$ & ${\rm tr}\,\kappa^{\mu\mu}_4$ & $\pm7.52(972)\times10^{-12}$
\\ 
QDT, $\delta_{45}^\mu=10$, $-i{\rm tr}\,\kappa^{\mu\mu}_3=\pm5.0\times10^{-10}$ & $d^{\rm SME}_\mu$ & ${\rm tr}\,\kappa^{\mu\mu}_5$ & $\mp1.92(103)\times10^{-11}$
\\ 
QDT, $\delta^\mu_{53}=10$, ${\rm tr}\,\kappa^{\mu\mu}_4=\pm5.0\times10^{-10}$ & $d^{\rm SME}_\mu$ & $-i{\rm tr}\,\kappa^{\mu\mu}_3$ & $\pm6.84(340)\times10^{-11}$
\\ \hline
QDT, $\delta^\tau_{34}=10$, ${\rm tr}\,\kappa^{\tau\tau}_5=\pm1.5\times10^{-4}$ & $d^{\rm SME}_\tau$ & ${\rm tr}\,\kappa^{\tau\tau}_4$ & $+8.44(814)\times10^{-6}$
\\
&&& $-2.84(814)\times10^{-6}$
\\ 
QDT, $\delta^\tau_{45}=10$, $-i{\rm tr}\,\kappa^{\tau\tau}_3=\pm1.5\times10^{-4}$ & $d^{\rm SME}_\tau$ & ${\rm tr}\,\kappa^{\tau\tau}_5$ & $-1.14(86)\times10^{-5}$
\\
&&& $+1.73(86)\times10^{-5}$
\\ 
QDT, $\delta^\tau_{53}=10$, ${\rm tr}\,\kappa^{\tau\tau}_4=\pm1.5\times10^{-4}$ & $d^{\rm SME}_\tau$ & $-i{\rm tr}\,\kappa^{\tau\tau}_3$ & $+4.15(285)\times10^{-5}$
\\
&&& $-6.11(285)\times10^{-5}$
\\ \hline
HYM, $\Delta^{\rm H}_1=10$, $\Delta^{\rm H}_2=0.3$ & $|a^{\rm SME}_\mu|$ & $|{\rm tr}\,\kappa^{\mu\tau}_1|$ & $<1.98\times10^{-14}$
\\
HYM, $\Delta^{\rm H}_1=10^2$, $\Delta^{\rm H}_2=0.3$ & $|a^{\rm SME}_\mu|$ & $|{\rm tr}\,\kappa^{\mu\tau}_1|$ & $<0.19\times10^{-14}$
\\ \hline
HYM, $\Delta^{\rm H}_3=10$, $\Delta^{\rm H}_4=0.3$ & $|d^{\rm SME}_\mu|$ & $|{\rm tr}\,\kappa^{\mu\tau}_4|$ & $<0.97\times10^{-11}$
\\
HYM, $\Delta^{\rm H}_3=10^2$, $\Delta^{\rm H}_4=0.3$ & $|d^{\rm SME}_\mu|$ & $|{\rm tr}\,\kappa^{\mu\tau}_4|$ & $<0.10\times10^{-11}$
\\ \hline \hline
\end{tabular}
\caption{\label{finaltab} Bounds on SME coefficients from the Lorentz-violating Yukawa sector.}
\end{table}

\end{widetext}

\acknowledgements
The authors acknowledge financial support from CONACYT and SNI (M\'exico).

\section*{References}


\end{document}